\documentclass[journal]{IEEEtran}

\usepackage{amsmath,amsfonts}
\usepackage{amssymb,amscd,amsthm,hyperref}
\usepackage{mathtools,enumerate}
\usepackage[caption=false,font=normalsize,labelfont=sf,textfont=sf]{subfig}
\usepackage{textcomp}
\usepackage{stfloats}
\usepackage{url}
\usepackage{graphicx}
\usepackage{cite}
\usepackage{color}
\usepackage{multirow,booktabs}

\newtheorem{Theorem}{Theorem}
\newtheorem{Lemma}[Theorem]{Lemma}
\newtheorem{Proposition}[Theorem]{Proposition}
\newtheorem{Corollary}[Theorem]{Corollary}
\theoremstyle{definition}
\newtheorem{definition}{Definition}
\newtheorem{example}{Example}
\theoremstyle{remark}

\DeclareMathOperator*{\argmin}{\arg\!\min}

\begin{document}
\bstctlcite{IEEEexample:BSTcontrol}

\title{On the Structure of Information}

\author{Sebastian~Gottwald and Daniel~A.~Braun
\thanks{The authors are with the Institute of Neural Information Processing, Ulm University, 89081 Ulm, Germany. (e-mail: sebastian.gottwald@uni-ulm.de)}
}

\maketitle

\begin{abstract}
We characterize information as risk reduction between knowledge states represented by partitions of the underlying probability space. Entropy is risk reduction from no (or partial) knowledge to full knowledge about a random variable, information is risk reduction to partial knowledge. This applies to any information measure that is based on expected loss minimization, with Shannon information and variance as prominent examples. In each case, core properties like the chain rule, non-negativity, and the relationship between information and divergence are preserved. The partition induced by an observation $Y$ encodes what it knows about the sample $\omega\in\Omega$. Its knowledge about another random variable $X$ is the coarser partition induced by the conditional law of $X$ given $Y$. The lattice of these distribution partitions fixes how information about $X$ decomposes into redundant, synergistic, and complementary terms. Redundancy is the risk reduced by their meet, which is Aumann's common knowledge about $X$. Synergy is the gap between their join and the distribution partition of the joint observation, which, unlike for observation partitions, can be non-zero. Complementary information further decomposes into unique contributions that depend on what is already known. Every term is the risk change along one transition between knowledge states, and none is obtained from the others by subtraction. Amounts of information do not overlap or add up like parts of a whole, contrary to common intuition and to existing decompositions. What combines are the knowledge states, through meets and joins, and information quantifies what such combinations gain.
\end{abstract}

\begin{IEEEkeywords}
Information, entropy, uncertainty, knowledge, partition, lattice, partial information decomposition, sub-$\sigma$-algebra, conditional expectation, Bregman divergence.
\end{IEEEkeywords}

\section{Introduction}

\IEEEPARstart{O}{bservations} matter because they change what we know. Their significance rarely lies in the raw values recorded but in what those values imply about other variables of interest, such as unobserved states of the world, future outcomes, or quantities we aim to predict. For example, the temperature shown by a living-room thermometer is not meaningful in itself, but through what it reveals about comfort, heating decisions, or related physical variables such as humidity or pressure. The purpose of an information measure is to assign a numerical value to such changes in knowledge, capturing how an observation improves our ability to learn, predict, and decide.

This viewpoint agrees with the classical foundations of information theory introduced by Shannon \cite{Shannon1948}, in which information is defined as the minimal expected length of a bit string required to encode a given message. More precisely, the total information, or uncertainty, of a discrete signal $X$ can be quantified by $H_S(X) = \mathbb E[-\log \mathbb P_X(X)]$, known as the \textit{Shannon entropy} of $X$. Shannon \cite{Shannon1959} and others \cite{Gallager1968} showed that this entropy plays a central role in coding theorems, advancing Shannon's original goal of characterizing the limits of communication. Specifically, $H_S(X)$ determines the minimum achievable average codeword length for encoding $X$ using binary strings \cite{Huffman1952}. Mutual information, $I_S(X;Y)= H_S(X) - H_S(X|Y)$, also admits interpretations in communication theory, for instance as the minimum message length required to simulate conditional distributions $\mathbb P_{X|Y=y}$ given $\mathbb P_X$ \cite{Harsha2010}.

Although Shannon originally conceived information theory in the context of channel coding and communication, its impact has extended to many other disciplines, including statistical mechanics \cite{Jaynes1957}, complexity theory \cite{Ladyman2020}, economics \cite{Stiglitz2002}, decision theory \cite{Tishby2011}, and machine learning \cite{MacKay2002}. Numerous alternative formulations of uncertainty and information have been developed for situations in which the length of binary strings transmitted over a channel is not the most appropriate quantifier, including generalized entropies, such as R\'enyi entropy \cite{Renyi1961}, Tsallis entropy \cite{Tsallis1988}, and Csisz\'ar's $f$-divergences \cite{Csiszar1972}, as well as Kolmogorov's algorithmic approach \cite{Kolmogorov1968}, and, in parallel to Shannon's coding-theoretic formulation, Wiener \cite{Wiener1948} arrived at an equivalent measure from the standpoint of prediction and filtering. All of these approaches share the assumption that uncertainty is quantified by a single number.

However, information also carries an inherent order structure that no scalar can fully resolve. Shannon himself pointed out that the entropy of a source measures only the amount of information and not the information itself, since two sources may produce information at the same rate without producing the same information \cite{Shannon1953}. He therefore proposed to identify the information of a source with everything that can be computed from it, ordering sources by whether one can be obtained from the other by reversible encoding, which yields a lattice in which the join of two sources is the pair, their total information, and the meet is the most that can be computed from each of them separately, their common information. This order is only partial, and the same is true of the orders by which uncertainty and information have been compared since. For instance, majorization formalizes uncertainty as a non-total preorder on probability distributions \cite{Marshall2011,Arnold2018}, with applications in economics \cite{Veinott1971}, physics \cite{Ruch1976}, quantum information \cite{Nielsen2010}, and decision theory \cite{Gottwald2019}. Closely related, Blackwell's ordering of experiments \cite{Blackwell1953} compares statistical experiments by their usefulness for arbitrary decision problems, again resulting in only a partial order. Similarly, when Kolmogorov \cite{Kolmogorov1958} and Sinai \cite{Sinai1959} carried entropy into the theory of dynamical systems, they defined it through measurable partitions of the underlying probability space, taking a supremum over the partition refinement order, which is itself only partial (see also \cite{Cornfeld1982}).

In classical probability theory it is common to identify the partition $\Pi(X)=\{X^{-1}(x)\}_x$ of the sample space $\Omega$, or alternatively the associated $\sigma$-algebra $\sigma(X)$, itself as the \textit{information} carried by a random variable $X$ \cite{Aumann1976, Geanakoplos1992,Billingsley2012}. Partitions represent the knowledge gainable through measurements, say $X=x$, by specifying \textit{what} is learnable about the actual outcome $\omega\in \Omega$ from the value $x$, namely, that $\omega$ belongs to the block $X^{-1}(x)$ corresponding to $x$. The partition $\Pi(X)$ thus determines the granularity at which subsets of $\Omega$ can be distinguished by observing $X$, i.e., observations with finer partitions are more informative than those with coarser partitions. Since not all partitions of $\Omega$ can be compared with respect to coarseness, this notion of information is a partial order. An actual information \textit{measure}, however, requires an additional objective with respect to which arbitrary partitions can be evaluated, effectively turning the partial order into a total preorder.

In this work, we focus on one family of such measures that unifies the coding-theoretic and the partition-based notions of information. Specifically, partitions are evaluated by the extent to which they enable optimization of a given objective function $a \mapsto l(X,a)$, when decisions $a$ are restricted to the coarse-graining level induced by the partition in terms of its knowledge about $X$. We identify information as the reduction in risk when passing from one partition to another, with entropy appearing as the special case corresponding to risk reduction from the trivial partition ${\bot\,\coloneqq\,\{\Omega\}}$ (no knowledge) to the partition $\Pi(X)$ generated by $X$ itself (full knowledge). This construction subsumes a variety of known uncertainty and information measures, where Shannon entropy and mutual information, in particular, are recovered when the objective function corresponds to minimal expected codeword length. Nevertheless, many characteristic properties usually associated with Shannon's measures, such as the chain rule and connections to divergence measures, are retained in the general case. The chain rule reveals how total information naturally decomposes into non-negative subterms, each representing risk reduction between successive levels of partial knowledge. This perspective links classical concepts from probability theory, such as Aumann’s notion of common knowledge \cite{Aumann1976}, to partial information decomposition (PID) approaches \cite{Williams2010, Griffith2014, Bertschinger2014}, consequently grounding them within the established framework of probability theory.

To this end, we distinguish knowledge about outcomes $\omega\in \Omega$ and knowledge about variables of interest $X$. As an example, consider the classical $\mathsf{Xor}$ gate: Two fair coins $Y$ and $Z$ are tossed independently, and $X$ records whether they show different faces. Seeing $Y$ alone teaches us something about the world, namely which face the first coin shows, but nothing about $X$. Whatever $Y$ shows, $X$ is still equally likely to be either value, and the same holds for $Z$. However, seeing both coins determines $X$ exactly. Knowledge about the world behaves differently than knowledge about $X$: The two observations together tell us exactly what a single joint observation of the pair would, but each observation separately contributes nothing to our knowledge about $X$, so combining what they contribute yields nothing. What an observation tells us about the world and what it tells us about $X$ are thus two different things, and only the latter matters for information about $X$.

This means that instead of working with the observation partitions $\Pi(Y)$ directly, we move to the lattice of distribution partitions $\pi_Y \coloneqq \Pi(\mathbb P_{X|Y})$ associated with the conditional laws $\mathbb P_{X|Y}$ of $X$ given random variables $Y$. This does not affect the standard information measures, such as $I(X;Y)$ or $I(X;Y|Z)$, because the risk associated with an observation $Y$ only depends on $Y$ through $\mathbb P_{X|Y}$. While the partition $\Pi(Y)$ represents the knowledge obtainable about $\omega$, the partition induced by $\mathbb P_{X|Y}$ represents the knowledge obtainable about $X$. For $\mathsf{Xor}$, $\Pi(Y)$ and $\Pi(Z)$ each split the sample space in two different halves, whereas $\pi_Y$ and $\pi_Z$ are trivial. For observation partitions, the combined knowledge $\Pi(Y)\vee\Pi(Z)$ (coarsest common refinement) is always the same as the knowledge $\Pi(Y,Z)$ of the joint observation $(Y,Z)$, just like for $\mathsf{Xor}$, where the coarsest common refinement of the two orthogonal splittings $\Pi(Y) = \{Y{=}0, Y{=}1\}$ and $\Pi(Z)=\{Z{=}0,Z{=}1\}$ is the partition into all four combinations of $Y,Z\in \{0,1\}$. For the distribution partitions this is not generally true, i.e., $\pi_Y\vee\pi_Z\neq\pi_{Y,Z}$, as in $\mathsf{Xor}$, where $\pi_Y\vee\pi_Z$ is trivial while the distinctions in $\pi_{Y,Z}=\Pi(X)$ fully determine $X$. The difference between the corresponding risks is what we identify as synergy, which is non-negative for every loss (cf.~Lemma \ref{lma:synergygap}). In $\mathsf{Xor}$, we also see explicitly that $\Pi(Y,Z)$ contains more distinctions than are relevant for $X$, whereas $\pi_{Y,Z}$ correctly only discriminates $\{(1,1),(0,0)\}$ from $\{(0,1),(1,0)\}$ (see also Fig.\,\ref{fig:examples1}).

Information measures in our approach are thus risk differences between states on the lattice of distribution partitions, whose elements encode the knowledge about $X$ gainable through observations and combine by meet and join. Existing partial information decompositions rest instead on an assumption that is rarely questioned: the unique information of a source is taken to be the same quantity whether it is measured on its own or after the other sources have been observed. This assumption is based on a view of informational quantities as primitive state functions, defined independently of the knowledge context in which they are measured and therefore combinable like parts of a whole. Redundant, unique, and synergistic information are then treated as quantities attached to the sources themselves, so that specifying any one of them, typically redundancy or synergy, determines the others by subtraction.


Instead, we identify each term of an information decomposition as risk difference of one transition between knowledge states. Redundant information is the transition to common knowledge, complementary information is the transition from common knowledge to combined knowledge, and synergy is the transition from combined to joint knowledge, i.e., the risk gap between $\pi_{Y}\vee \pi_{Z}$ and $\pi_{Y,Z}$ noted above. Complementary information decomposes further into unique contributions of the individual sources, whose values depend on the order in which the sources are observed. For example, in the case of two sources $Y$ and $Z$, complementary information decomposes into unique and conditional unique contributions, just as $I(X;(Y,Z))$ splits into the information $I(X;Y)$ of $Y$ about $X$ and the conditional information $I(X;Z|Y)$ of $Z$ about $X$ given $Y$. In particular, complementary information decomposes differently depending on whether the route passes through $\pi_Y$ or $\pi_Z$ first. On a range of discrete and continuous examples, for which the literature reports diverging and occasionally negative values, the existing PID proposals often share terms with one of these two routings, which we trace to the single unique term being forced to serve in both roles, before and after the other source is observed.

\section{From knowledge to information} \label{sec:knowledge-to-information}

\subsection{The knowledge lattice of partitions}

In probability theory \cite{Kallenberg2002}, \textit{observations}, or \textit{measurements}, are modeled as \textit{random variables}, or \textit{signals}, $X:\Omega\to \mathcal X$. That is, they are modeled as functions from the underlying sample space $\Omega$ to some well-behaved space $\mathcal X$ such as the real numbers (see Section~\ref{sec:sigmaAlgebraOfKnowableSets} for a precise definition). The idea is that the set $\Omega$ is rich enough to jointly model the uncertainty of all measurements we are interested in. Once $\omega \in \Omega$ is sampled, the measurement $X(\omega)$ is uniquely specified.

In all but the simplest cases, the reverse is not true: observing a particular measurement $X=x$ does not generally allow us to determine a unique elementary event $\omega\in\Omega$. Instead, it only narrows down the possible $\omega$ that are compatible with this observation. Specifically, when observing $X=x$, we know $\omega$ is somewhere in the preimage
\[ \Pi(x) \coloneqq X^{-1}(x) =  \{\omega\in\Omega \mid X(\omega) = x\}, \]
which is also simply denoted by $\{X\,{=}\,x\}$. Since $X$ is a (deterministic) function on $\Omega$, the collection $\Pi(X)\coloneqq \{\Pi(x)\}_x$ of non-empty preimages is a \textit{partition} of $\Omega$, i.e., $\Omega= \bigcup_{x} \Pi(x)$ and $\Pi(x) \cap \Pi(x') = \emptyset$ if $x\neq x'$.

\begin{figure}
\centering
\includegraphics[width=.46\textwidth]{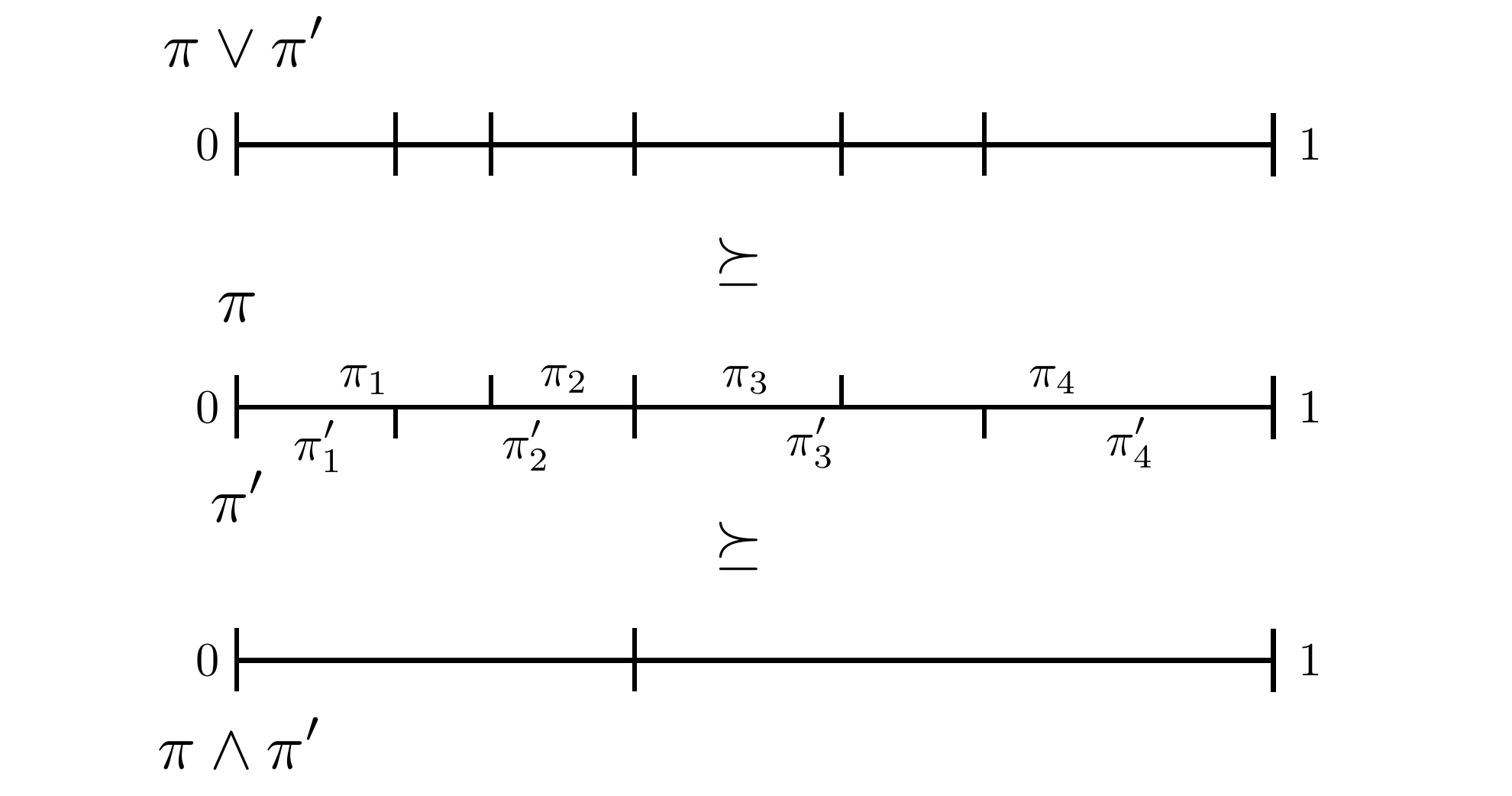}
\caption{Examples of partitions $\pi$ and $\pi'$ of $\Omega = [0,1]$, including their coarsest common refinement (join $\pi\vee \pi'$) and finest common coarsening (meet $\pi\wedge\pi'$). The blocks of $\pi\vee \pi'$ are knowable from the combination of $\pi$ and $\pi'$, whereas the blocks of $\pi\wedge \pi'$ are sets of common knowledge, because these sets are knowable by both $\pi$ and $\pi'$. In the terminology of Aumann \cite{Aumann1976}, a decision-maker who performs measurement $\pi$ knows the block of $\pi\wedge \pi'$ that contains $\omega$ and \textit{also} knows that a decision-maker with measurement $\pi'$ knows that $\omega$ is in that block. They also know that the other decision-maker knows that they know, and so on.}
\label{fig:joinAndMeet}
\end{figure}

Partitions encode a notion of \textit{knowledge}: coarse partitions correspond to \textit{less informative} measurements, finer partitions to \textit{more informative} measurements. Formally, this idea is captured by the partial order (a reflexive, transitive, and antisymmetric relation) of partitions, given by $\pi\preceq \pi'$ if and only if $\pi'$ is \textit{finer} than $\pi$, that is, if any element of $\pi$ can be written as a union of elements of $\pi'$. Note that in the set theory literature, this relation is sometimes defined in reverse order. We adopt the usual order-theoretic language, e.g., $\pi$ and $\pi'$ are called \textit{comparable} if $\pi\preceq \pi'$ or $\pi'\preceq \pi$. Even if two partitions $\pi$ and $\pi'$ are not comparable, we can form their least upper bound $\pi \vee \pi'$ (their \textit{join}), which is their \textit{coarsest common refinement}, whereas their greatest lower bound $\pi\wedge \pi'$ (their \textit{meet}) is their \textit{finest common coarsening} (cf.~Fig.~\ref{fig:joinAndMeet}).

Given a sample space $\Omega$ and a measurement with partition $\pi$ of $\Omega$, we call a set $B\subseteq\Omega$ \textit{$\pi$-knowable} if it has the property that each block of $\pi$ is either contained in $B$ or in $B^c$ (sets that are knowable by observing $\pi$). In the measure theory literature, such sets are also known as $\pi$-sets \cite{Bogachev2007}, \textit{events} with respect to $\pi$, or \textit{informed sets} \cite{HervesBeloso2013}. Intuitively, knowable sets $B$ do not overlap the blocks of $\pi$, they are aligned with the partition. Sets $B\subseteq\Omega$ that are knowable by each of $\pi$ and $\pi'$ are called sets of \textit{common knowledge} \cite{Aumann1976}. We call a set of common knowledge \textit{minimal} if it contains no proper nonempty subset that is also a set of common knowledge.

This terminology allows for a concise description of the partial order $\preceq$ and its meet and join, which we summarize in the following definition.

\begin{figure}
\centering
\includegraphics[width=.5\textwidth]{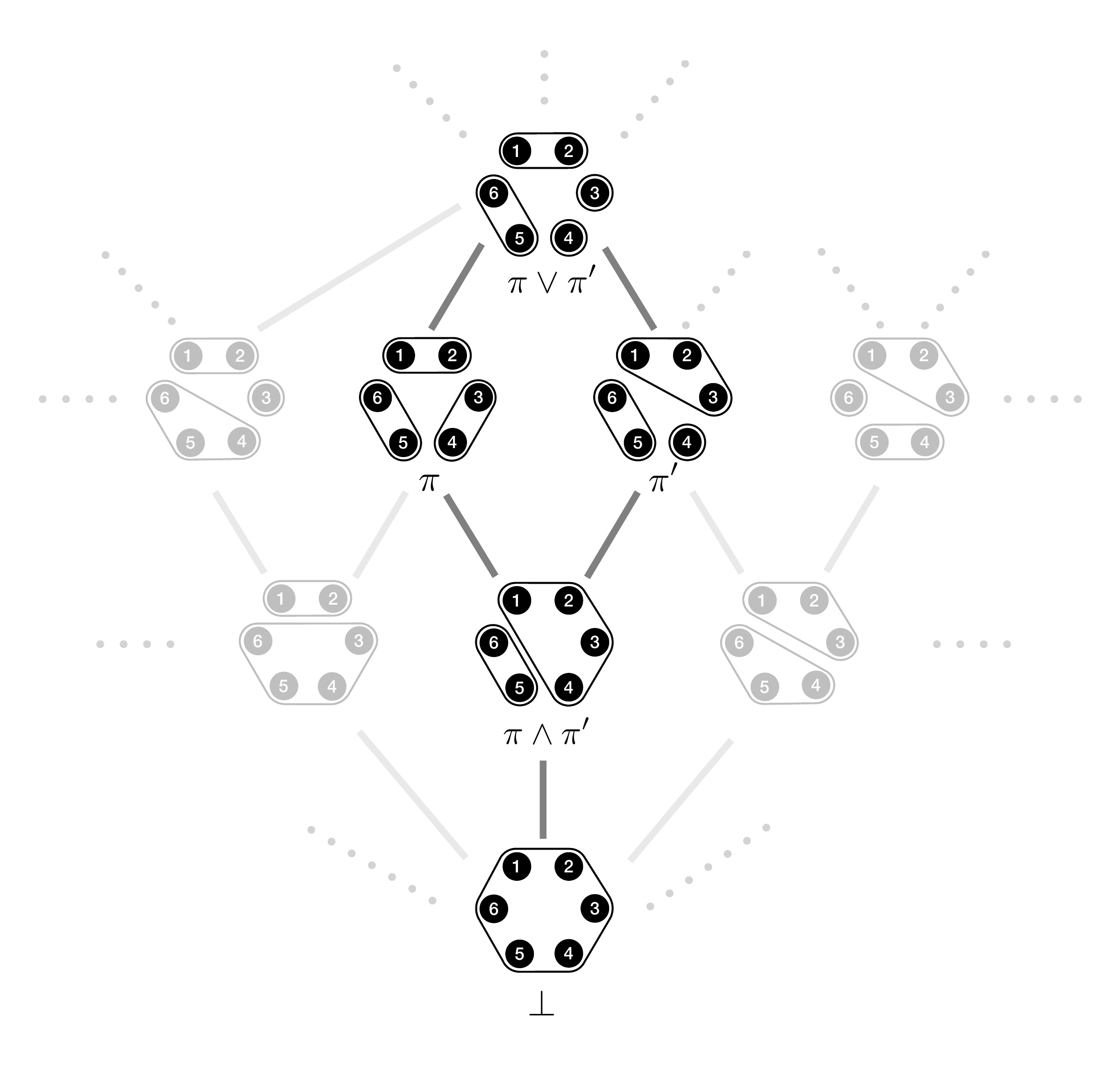}
\caption{A small section of the knowledge lattice of a $6$-element sample space $\Omega=\{1,\dots,6\}$. Partitions connected through a path are comparable with respect to $\preceq$, with upper ones being finer than lower ones. Note that $\pi$ and $\pi'$ are not comparable with respect to $\preceq$, whereas both are bound by their meet and join. $\pi$ could for example be the partition corresponding to the random variable $X$ given by $X(\omega)=\lfloor (\omega -1)/2 \rfloor$ (integer division by 2 of $\omega - 1$).}
\label{fig:knowledgeLattice}
\end{figure}

\begin{definition}[Knowledge lattice I] \label{def:knowledgelattice1}
We call a partition $\pi$ of a probability space $\Omega$ a \textit{measurement} or \textit{knowledge state}. The set of all partitions $\pi$ of $\Omega$ is partially ordered through refinement:
\begin{enumerate}[$(i)$]
\item $\pi\preceq \pi'$ if and only if all observations $A\in\pi$ are $\pi'$-knowable.
\item The coarsest common refinement $\pi\vee\pi'$ contains all non-empty intersections of elements from $\pi$ and $\pi'$. This partition corresponds to the obtainable knowledge when measurements of both $\pi$ and $\pi'$ are observed.
\item The finest common coarsening $\pi\wedge\pi'$ is the partition formed by all minimal sets of common knowledge. It corresponds to the shared knowledge obtainable by either measurement, $\pi$ or $\pi'$.
\end{enumerate}
We refer to the set of partitions equipped with this partial order as the \textit{knowledge lattice of partitions} of $\Omega$ (cf.~Fig.~\ref{fig:knowledgeLattice} for a finite example).
\end{definition}

Partitions enable a comparison of signals based on their informational content: each signal induces a partition of the state space and finer partitions represent more informative signals. This representation has a long tradition in probability \cite{Blackwell1950}, decision theory \cite{Aumann1976}, and economics \cite{Milgrom1994}. Comparing signals by their induced partitions underlies Blackwell's classical ordering of experiments \cite{Blackwell1953}, which considers one experiment more informative than another if it yields lower Bayes risk for \emph{every} loss function. We return to the relation between this order and our construction at the end of Section~\ref{sec:information}.

Continuous signals $X$, however, induce partitions with uncountably many blocks $X^{-1}(x)$ which are typically measure-zero sets under the underlying probability measure. To describe conditioning on such events, modern probability theory is built on $\sigma$-algebras instead, which provide the minimal structure needed to describe events and their probabilities in general spaces. To connect the intuitive picture offered by partitions to the framework of probability theory, it is therefore necessary to pass from partitions to suitable $\sigma$-algebras.

\subsection{The knowledge lattice of $\sigma$-algebras} \label{sec:sigmaAlgebraOfKnowableSets}

A \textit{$\sigma$-algebra} on $\Omega$ is a collection of subsets $A\subseteq\Omega$ that contains $\Omega$ and is closed under complements and countable unions. $\sigma$-algebras form the basic building blocks of \textit{measure theory}: a set $\Omega$ together with a $\sigma$-algebra $\Sigma$ is known as a \textit{measurable space} $(\Omega,\Sigma)$, elements of $\sigma$-algebras are called \textit{measurable sets}, and \textit{measures} $\mu$ are special mappings defined on $\sigma$-algebras that are additive under disjoint unions. Functions between measurable spaces are called $\mathcal F$-measurable if the preimages of measurable sets lie in the $\sigma$-algebra $\mathcal F$. A measure $\mathbb P$ on $(\Omega,\Sigma)$ is a \textit{probability measure} if $\mathbb P(\Omega)=1$, in which case $(\Omega,\Sigma,\mathbb P)$ (or just $\Omega$) is called a \textit{probability space} and the elements $B\in \Sigma$ are called \textit{events}. A random variable (or \textit{random element}) $X:(\Omega,\Sigma)\to (\mathcal X,\mathcal B)$ is any $\Sigma$-measurable mapping from a probability space $\Omega$ to some other measurable space $(\mathcal X,\mathcal B)$.

Random variables induce a sub-$\sigma$-algebra $\sigma(X)\subseteq\Sigma$, which is the smallest $\sigma$-algebra on $\Omega$ with respect to which $X$ is measurable, i.e., $\sigma(X)$ is the collection of all preimages $X^{-1}(B)$ of measurable sets $B\in\mathcal B$. Unless specified otherwise, random variables are usually considered to be mappings to some Euclidean space $\mathcal X$, e.g., $\mathcal X=\mathbb R$, equipped with the \textit{Borel $\sigma$-algebra} $\mathcal B(\mathcal X)$, which is the smallest $\sigma$-algebra containing all open sets. For example, $\mathcal B(\mathbb R)$ is the smallest $\sigma$-algebra that contains all open intervals $(a,b)$. The collection of sub-$\sigma$-algebras on $\Omega$ is partially ordered through inclusion and obeys a lattice structure with meet $\mathcal F\wedge \mathcal F'$ given by the intersection of $\mathcal F$ and $\mathcal F'$, and join $\mathcal F\vee \mathcal F'$ given by the smallest $\sigma$-algebra containing the union of $\mathcal F$ and $\mathcal F'$.

An important probabilistic tool is the \textit{conditional expectation} $\mathbb E[X|\mathcal F]$ of a random variable $X$ with respect to the $\sigma$-algebra $\mathcal F$. In particular, it enables conditioning on other (even continuous) random variables $Y$ by choosing $\mathcal F = \sigma(Y)$. Intuitively, $\mathbb E[X|\mathcal F]$ corresponds to a coarse-graining of $X$ compatible with the knowledge encoded by $\mathcal F$. In fact, one of the many equivalent ways to define conditional expectation \cite{Kallenberg2002} is as the \textit{best} such approximation in terms of mean squared error,
\begin{equation}\label{eq:Banerjee}
\mathbb E[X|\mathcal F] = \argmin_{Z\ \mathcal F\text{-measurable}} \mathbb E[(X-Z)^2].
\end{equation}
That is, $\mathbb E[X|\mathcal F]$ is the $L^2$-projection of $X$ onto the subspace of $\mathcal F$-measurable functions. It can even be shown that it is the unique $\mathcal F$-measurable predictor that minimizes the discrepancy to $X$ with respect to any \textit{Bregman} loss, and that this property characterizes Bregman divergences \cite{Banerjee2005a}.

Hence, the conditional expectation $\mathbb E[X|\mathcal F]$ interpolates between the single number $\mathbb E[X]=\mathbb E[X|\{\emptyset,\Omega\}]$  representing $X$ under no knowledge, and the full random variable $X=\mathbb E[X|\sigma(X)]$, suggesting modeling knowledge through $\sigma$-algebras. The \textit{tower property} further supports this interpretation, since $\mathbb E[\mathbb E[X|\mathcal F']|\mathcal F] = \mathbb E[X|\mathcal F]$ whenever $\mathcal F\subseteq \mathcal F'$, i.e., under the smaller $\sigma$-algebra $\mathcal F$, $X$ cannot be distinguished from its coarse-grained version $\mathbb E[X|\mathcal F']$.

Since knowledge states and measurements were represented in the previous
section as partitions $\pi$ of the sample space, these properties motivate
passing to the associated $\sigma$-algebras $\sigma(\pi)$,
which capture precisely the events that are distinguishable under a given
state of knowledge. In particular, there are further properties of $\sigma$-algebras that align well with the intuition of informational
content. For example, if $Z$ is a $\sigma(Y)$-measurable random variable, that is, if $\sigma(Z)\subseteq\sigma(Y)$, then by the Doob-Dynkin Lemma \cite{Doob1962}, $Z = f(Y)$ for some measurable $f$, i.e., if $Y$ is known then $Z$ is known.

In fact, in the discrete case, partitions $\pi$ and $\sigma$-algebras have a very simple relationship: if $\pi=\{\pi_i\}_i$ is a countable partition, for example generated by a discrete random variable $X$, where $\pi_i = X^{-1}(x_i)$, then the smallest $\sigma$-algebra that contains $\pi$ is given by the collection of all $\pi$-knowable sets, i.e., all unions of blocks from $\pi$, $\sigma(\pi) = \{\bigcup_{i\in I}\pi_i \, | \, I\subseteq\mathbb N\}$. It then follows immediately that
\begin{equation}\label{eq:sigmalatticeprop}
\pi\preceq \pi' \quad  \Leftrightarrow \quad  \sigma(\pi) \subseteq \sigma(\pi'),
\end{equation}
because the property that the blocks of $\pi$ are countable unions of blocks of $\pi'$ translates directly to the corresponding $\sigma$-algebras.

However, this does not directly transfer to the continuous case, because $\sigma$-algebras are only closed with respect to countable unions, and so the smallest $\sigma$-algebra that contains an uncountable partition $\pi = \{\Pi(x)\}_x$ is the collection of all subsets $B\subseteq\Omega$ such that $B$ or its complement $B^c$ is the union of a countable number of blocks of $\pi$. This generally results in $\sigma$-algebras that are too small to satisfy \eqref{eq:sigmalatticeprop}, as can be seen from the following example from \cite{Dubra2004}: Consider the Borel space $([0,1],\mathcal B)$ and the two partitions of $[0,1]$ given by
\[\pi = \{[0,1/2),[1/2,1]\}, \quad \pi' = \{\{x\}|x\in[0,1]\}.
\]
Then $\pi\preceq \pi'$, and the corresponding smallest $\sigma$-algebra that contains $\pi$ is $\mathcal F = \{\emptyset,[0,1/2), [1/2,1], [0,1]\}$, whereas the smallest $\sigma$-algebra that contains $\pi'$ is the collection of subsets of $[0,1]$ that are either countable or have countable complement. In particular, this $\sigma$-algebra does not contain $\mathcal F$ (they are not comparable with respect to inclusion).

Rather than defining $\sigma(\pi)$ as the smallest $\sigma$-algebra containing the blocks of $\pi$, one chooses $\sigma(\pi)$ to be the $\sigma$-algebra induced by the corresponding random variable, which exists whenever $\pi$ is a \textit{measurable} partition \cite{Rohlin1952,Bogachev2007}.  Such partitions have a \textit{basis} $\{B_i\}_{i\in I}$, that is, a countable collection of $\pi$-knowable sets that separates the blocks of $\pi$ in the sense that for any two blocks $C$ and $D$ of $\pi$, there exists $i\in I$ such that either $C\subseteq B_i$ and $D\not \subseteq B_i$, or $C\not\subseteq B_i$ and $D\subseteq B_i$. In our example above, a basis for $\pi'$ would be the countable collection of all open sub-intervals of $[0,1]$ with rational endpoints. Another standard assumption is that the relations $\pi\preceq \pi'$ and $\pi=\pi'$ are considered only up to $\mathbb P$-null sets, i.e., sets of measure $0$ with respect to the underlying measure $\mathbb P$. More precisely, given two collections $\mathcal C$, $\mathcal D$ of subsets of $\Omega$, we write \textit{$\mathcal C\subseteq \mathcal D$ modulo $\mathbb P$-null sets}, if every $C\in\mathcal C$ agrees with some $D\in\mathcal D$ up to a $\mathbb P$-null set (i.e., $\mathbb P(C\Delta D)=0$), and equality means this holds in both directions \cite{Bogachev2007}. It then follows that the meet $\pi\wedge\pi'$ and join $\pi\vee\pi'$ of two measurable partitions $\pi$ and $\pi'$ are again measurable \cite{Rohlin1967}. The $\sigma$-algebra $\sigma(\pi)$ of a measurable partition $\pi$ is the collection of all measurable $\pi$-knowable sets. For our above example, $\sigma(\pi')$ is therefore the full Borel algebra $\mathcal B([0,1])$, and so $\sigma(\pi)\subset\sigma(\pi')$. Notice, it is this construction of $\sigma(\pi)$ through $\pi$-knowable sets (rather than as the smallest $\sigma$-algebra containing $\pi$), that translates the comparability to the corresponding $\sigma$-algebras, whereas the measurability of $\pi$ makes this construction possible.

Shannon's information lattice \cite{Shannon1953} is the analogue of this structure, with sources ordered by whether one can be computed from the other in place of $\sigma$-algebras ordered by refinement, and he already observes that it is at least as general as the lattice of finite partitions. In fact, the $\sigma$-algebras induced by measurable partitions are in one-to-one correspondence with the underlying partitions, and, importantly, \eqref{eq:sigmalatticeprop} is satisfied up to null sets \cite{Rohlin1967}. Moreover, the knowledge lattice of measurable partitions directly translates to their $\sigma$-algebras \cite{Rohlin1967}: the common knowledge partition $\pi\,{\wedge}\,\pi'$ is represented by the meet of their induced $\sigma$-algebras, whereas $\pi\,{\vee}\,\pi'$ is represented by their join. These observations motivate the following definition.

\begin{definition}[Knowledge lattice II] \label{def:knowledgelattice2}
Given a measurable partition $\pi$, let $\sigma(\pi)$ denote the collection of all $\pi$-knowable measurable sets. Then \cite{Rohlin1967}:
\begin{enumerate}[$(i)$]
\item $\sigma(\pi)$ represents the knowledge in $\pi$ in the set of $\sigma$-algebras on $\Omega$, i.e.,
\begin{equation}
\pi\preceq \pi' \ \Leftrightarrow \ \sigma(\pi) \subseteq \sigma(\pi')\, .
\end{equation}
\item The knowable sets with respect to the finest common coarsening of two partitions $\pi$ and $\pi'$ are exactly those sets that are knowable by both $\pi$ and $\pi'$, i.e., the sets that are shared by $\sigma(\pi)$ and $\sigma(\pi')$,
\begin{equation}\label{def:meet2}
\sigma(\pi\wedge \pi') = \sigma(\pi)\wedge \sigma(\pi')\, .
\end{equation}
Similarly, the knowable sets with respect to the coarsest common refinement of $\pi$ and $\pi'$ form the smallest $\sigma$-algebra containing $\sigma(\pi)\cup \sigma(\pi')$,
\begin{equation}\label{def:join2}
\sigma(\pi \vee \pi') = \sigma(\pi)\vee \sigma(\pi').
\end{equation}
\end{enumerate}
We refer to the set of $\sigma$-algebras induced by measurable partitions equipped with the partial order of inclusion as the \textit{knowledge lattice of $\sigma$-algebras} of $\Omega$, where meet and join operations are given by \eqref{def:meet2} and \eqref{def:join2}, respectively. The \textit{knowledge lattice of $\Omega$} can refer to either of the equivalent knowledge lattices of partitions (Definition~\ref{def:knowledgelattice1}) or $\sigma$-algebras.
\end{definition}
We note that measurable partitions and their induced $\sigma$-algebras are exactly those partitions and respective $\sigma$-algebras that are generated by random variables. In fact, a classical result states that a partition is measurable if and only if its blocks can be represented as level sets $f^{-1}(\{x\})$ of some measurable function $f:\Omega\to[0,1]$ \cite{Bogachev2007}. From now on, all partitions are assumed to be measurable.

\subsection{Risk minimization} \label{sec:riskminimization}

As we have seen in the previous sections, knowledge about $\omega\in\Omega$ obtainable through measurements, represented by the underlying partitions, or equivalently by their $\sigma$-algebras, is only partially ordered. However, by explicitly specifying how useful an observation is, we can compare any partition with respect to that objective, projecting the knowledge lattice onto one dimension of scalar values, as visualized in Fig. \ref{fig:knowledgeToInformation}. For example, we could use the surprise generated by a particular observation about some unknown quantity, in terms of the reduction in codeword length ($\log$ loss), or the reduction in variance of an unknown signal (square error), etc. In general, information measures can be seen as risk reduction with respect to a chosen \textit{risk function} $R$ defined on measurements $\pi$ that quantifies the usefulness of the knowledge in $\pi$ for a particular task.

We focus on the class of risk functions induced by \textit{loss functions} $l:\mathcal X\times\mathcal A\to \mathbb R$, defined on two variables $x$ and $a$, where $x$ denotes an \textit{outcome} or \textit{state} and $a$ denotes an \textit{action}. Such losses allow us to quantify how useful the knowledge encoded by a partition $\pi$ is by optimizing $l(x,a)$ with respect to $a$ under the knowledge in $\pi$. We scope our work to the following common requirements.

\smallskip

\textbf{Assumptions.} Throughout, $(\Omega,\Sigma,\mathbb P)$ is a probability space, $\mathcal{X}$ and $\mathcal{A}$ are standard Borel spaces, and $l$ is a jointly measurable real-valued function on $\mathcal X\times\mathcal A$. Also, $X$ always denotes a random variable on $\Omega$ with target space $\mathcal X$. Moreover, all measurability statements are understood modulo $\mathbb P$-null sets.

\smallskip
Let us first consider the two extreme cases of full knowledge about $X$, encoded by the partition $\Pi(X)=\{X^{-1}(x)\}_x$, and no knowledge about $X$, encoded by the partition $\bot= \{\Omega\}$. Since knowledge of the realization $x$ of $X$ enables pointwise optimization of the loss $a\mapsto l(x,a)$, whereas without any knowledge only the expected loss $a\mapsto \mathbb E[l(X,a)]$ can be optimized, we have
\begin{equation} \label{ineq:extremes}
\mathbb E\Big[\inf_{a\in\mathcal A} l(X,a)\Big] \leq \inf_{a\in\mathcal A} \mathbb E[l(X,a)].
\end{equation}
\textit{Some} knowledge about $X$, realized by an arbitrary partition $\pi$ of $\Omega$, can lead to an expected loss between these two extremes. The key observation is that a decision-maker with state of knowledge $\pi$ can only choose actions that are compatible with the distinctions available under $\pi$, i.e., actions that are $\sigma(\pi)$-measurable. The two extremes in \eqref{ineq:extremes} correspond to the coarsest and finest possible knowledge: under no knowledge only constant actions ($\sigma(\bot)$-measurable actions) are possible, whereas under full knowledge, the action can be a function of $X$, i.e., it is $\sigma(X)$-measurable. This motivates the following definition for arbitrary partitions $\pi$ of $\Omega$.

\begin{definition}[Risk]\label{def:risk}

Under our standing assumptions, given a measurable partition $\pi$ of $\Omega$, let the \textit{risk of $\pi$} be defined by
\begin{equation}\label{eq:risk}
R(\pi)\, \coloneqq  \inf_{A\ \sigma(\pi)\text{-meas.}} \mathbb E[l(X,A)],
\end{equation}
where the minimization is over $\sigma(\pi)$-measurable random variables $A:\Omega\to\mathcal A$.
For random variables $Y$, we write $R(Y)$ for the risk of the partition $\pi=\Pi(Y)$ induced by $Y$.
\end{definition}

Expectations of losses are understood in the extended sense, with $\mathbb E[l(X,A)]  = \mathbb E[l(X,A)^+] - \mathbb E[l(X,A)^-]\in [-\infty,+\infty]$, with the convention that the value is $+\infty$ if both parts are infinite. Differences of risks are understood whenever they are defined, that is, unless both risks are infinite with the same sign.

As we show in the proof of Proposition \ref{prop:properties} in Appendix \ref{prf:properties}, for the special case $\pi = \Pi(X)$, we obtain
\[
R(X) = \mathbb E\Big[ \inf_{a\in\mathcal A} l(X,a) \Big],
\]
from which follows that for any partition $\pi$ of $\Omega$ we have
\begin{equation}\label{ineq:risks}
R(\bot) \geq R(\pi) \geq R(X).
\end{equation}
In particular, once a loss function $l$ is given, we can compare knowledge states $\pi$ based on how well their blocks align with the blocks of $\Pi(X)$ in terms of optimizing $l(X,\cdot)$.

\subsection{Information as risk reduction} \label{sec:information}
First, consider the case of \textit{total risk reduction}, from no knowledge to full knowledge about $X$, i.e., the difference between the two sides of \eqref{ineq:extremes}, which we identify as the \textit{entropy} of $X$,
\begin{equation}\label{def:entropy}
H(X) \coloneqq \inf_{a\in\mathcal A} \mathbb E\big[l(X,a)\big] - \mathbb E\Big[\inf_{a\in\mathcal A} l(X,a)\Big].
\end{equation}
Hence, $H(X)$ is the average difference in loss between not knowing and knowing $X$, representing the total \textit{uncertainty} in $X$ with respect to the loss $l$. Different loss functions may capture different aspects of uncertainty. For example, if the loss function is taken to be the square error, then $H$ is the variance, measuring uncertainty in the sense of the average square distance to the expected value, whereas for the log loss we recover Shannon entropy (see Appendix \ref{app:examples}). Note that \eqref{def:entropy} differs from the decision-theoretic notion of entropy in \cite{Gruenwald2004} by subtracting the average loss incurred even if $x$ is known perfectly. In many practical examples, $l$ is chosen such that this term vanishes, by ensuring that $l(x,a) \geq 0$ with equality at the minimum. However, in principle, $l$ can have arbitrary minimal values, which do not affect the value of $H(X)$ in \eqref{def:entropy}. Including this term in the definition makes it possible to treat entropy and information simply as two cases of risk reduction between different knowledge states, as we will see in the following.

\begin{figure}
\centering
\includegraphics[width=.5\textwidth]{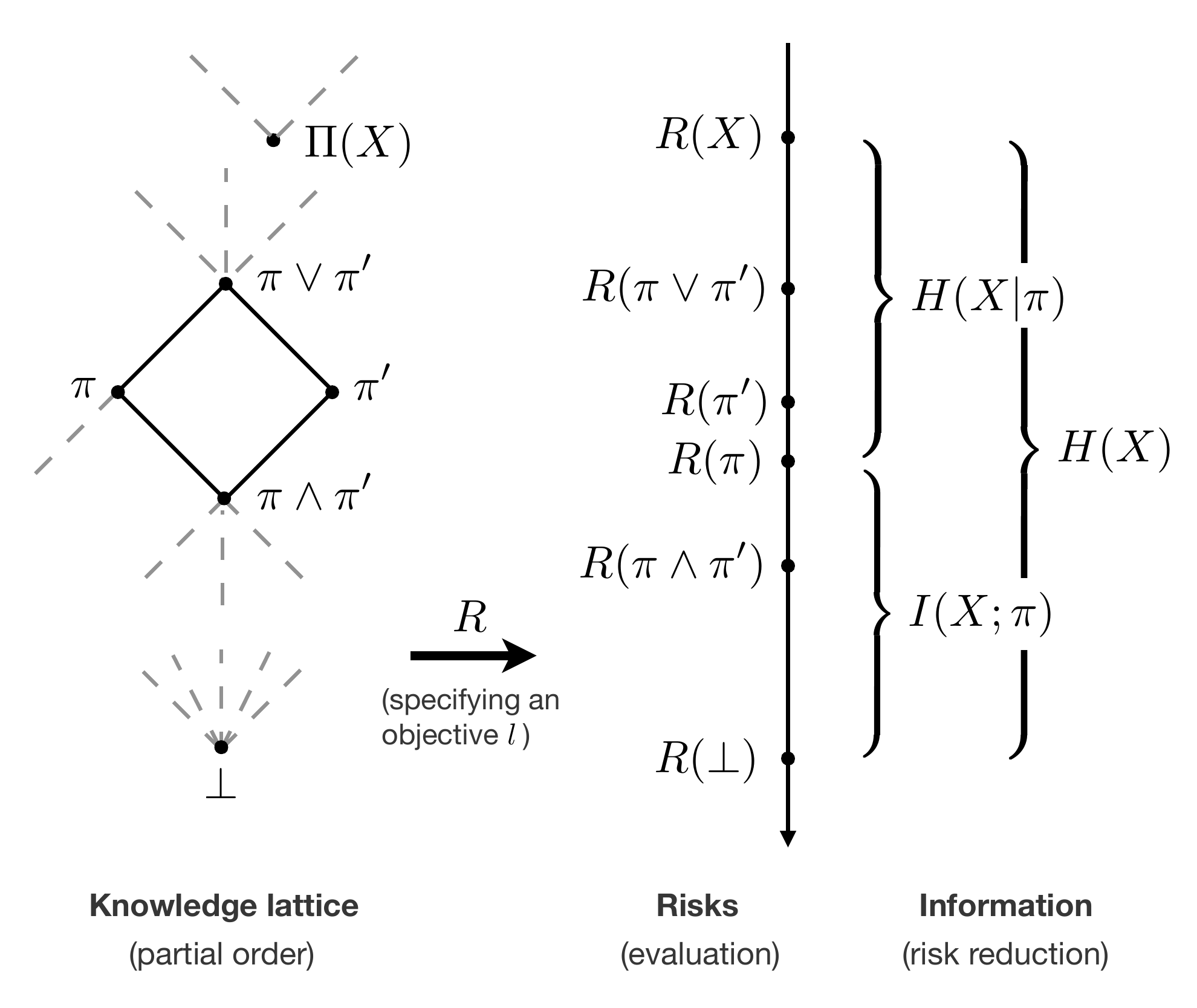}
\caption{Schematic illustration of how specifying a loss projects the knowledge lattice on $\Omega$ to the real line of risks, inducing informational quantities, such as entropy $H(X)$, conditional entropy $H(X|\pi)$, and information $I(X;\pi)$, through risk reduction from one knowledge state to another.}
\label{fig:knowledgeToInformation}
\end{figure}

Since entropy \eqref{def:entropy} is the difference of the risks $R(\bot)$ and $R(X)$, inequality \eqref{ineq:risks} suggests decomposing $H(X)$ into two non-negative terms,
\[
H(X) = \underbrace{R(\bot) - R(\pi)}_{\text{Information } I(X;\pi)} + \underbrace{R(\pi) - R(\Pi(X)),}_{\text{Conditional entropy } H(X|\pi)}
\]
where we identify the risk reduction from some knowledge $\pi$ to full knowledge $\Pi(X)$ as the \textit{conditional entropy of $X$ given $\pi$}, and the risk reduction from no knowledge $\bot$ to some knowledge $\pi$ as \textit{information in $\pi$ about $X$} (cf.~Fig.~\ref{fig:knowledgeToInformation}). In particular, the usual decomposition of entropy into mutual information and conditional Shannon entropy, $H(X)=I(X;Y) + H(X|Y)$, is a basic feature of our general notion of information for arbitrary loss functions, which we specify in the following.

\begin{definition}[Uncertainty, entropy, and information]\label{def:uncertainty-entropy-inf}
Under our standing assumptions, let $\pi$ and $\pi'$ be two not necessarily comparable partitions of $\Omega$. The \textit{uncertainty reduction about $X$ from $\pi$ to $\pi'$} (with respect to $l$) is
\begin{align} \nonumber
& U_{\pi\to\pi'}(X)  \coloneqq R(\pi) - R(\pi') \\ \label{def:information}
&\quad \ \  = \inf_{A\,\sigma(\pi)\text{-meas.}} \mathbb E[l(X,A)] - \inf_{A\,\sigma(\pi')\text{-meas.}} \mathbb E[l(X,A)]\, .
\end{align}
Furthermore, we distinguish the following quantities.
\begin{enumerate}[$(i)$]
\item The \textit{entropy of $X$ conditioned on $\pi$} is given by
\begin{equation}\label{eq:conditionalentropy}
H(X|\pi) \coloneqq U_{\pi \to \Pi(X)}(X) \, ,
\end{equation}
i.e., conditional entropy is the risk reduction from some knowledge state $\pi$ to \textit{full knowledge} $\Pi(X)$. The case $\pi{=}\bot$ corresponds to the maximum risk reduction associated with $X$, which is the \textit{entropy} \eqref{def:entropy} of $X$,
 \begin{equation}
H(X) = H(X|\bot) = U_{\bot\to\Pi(X)}(X)\,.
\end{equation}
Moreover, if $Y$ denotes a random variable with partition $\pi$, then we write $H(X|Y)\coloneqq H(X|\Pi(Y))$.
\item The \textit{information about $X$ in $\pi'$ conditioned on $\pi$} is
\begin{equation}
I(X;\pi'|\pi) \coloneqq U_{\pi\to\pi\vee\pi'}(X).
\end{equation}
For $\pi=\bot$ we obtain the \textit{information about $X$ in $\pi'$},
\begin{equation}
I(X;\pi) \coloneqq I(X;\pi|\bot) = U_{\bot\to\pi}(X),
\end{equation}
since $\bot \vee \pi = \pi$. We write $I(X;Y) \coloneqq U_{\bot \to\Pi(Y)}(X)$ and, moreover, since $\Pi(Y,Z) = \Pi(Y)\vee \Pi(Z)$, we also write
\[
I(X;Y|Z)\coloneqq U_{\Pi(Z)\to\Pi(Y,Z)}(X)
\]
for random variables $Y,Z$.
\end{enumerate}
\end{definition}

Without further assumptions on $\pi$ and $\pi'$, we cannot say whether $U_{\pi\to\pi'}$ is positive or negative, corresponding to a decrease or increase in uncertainty, respectively. The special cases $(i)$ and $(ii)$, however, are non-negative, as we show in Proposition \ref{prop:properties} below. Concrete examples for common loss functions, including square error, log loss, general Bregman losses, and proper scoring rules, are discussed in Appendix~\ref{app:examples}. There we also discuss that in the continuous case, entropy takes values in the extended reals $[0,\infty]$ and can diverge for certain losses (e.g., the log loss and the Hyvärinen loss), because full knowledge permits driving the pointwise loss unboundedly low. Information, by contrast, remains finite if the corresponding conditional law of $X$ at the given knowledge state has a density with finite risk. Only knowledge that fixes $X$ (or a coordinate of $X$) exactly, drives the corresponding risk to be arbitrarily negative (cf.~Example~\ref{ex:gaussrdn}). Since information is defined directly as a risk difference rather than as a difference of two entropies, finite quantities require no cancellation of infinities.

Note also that $I(X;Y)$ is inherently asymmetric with respect to interchanging $X$ and $Y$, since $X$ directly determines the loss $l(X,\cdot)$, whereas $Y$ generally contributes only through the partition $\Pi(Y)$. This asymmetry reflects the fact that $I(X;Y)$ quantifies the knowledge gain about $X$ when moving from no knowledge to observing $Y$, whereas $I(Y;X)$ quantifies the knowledge gain about $Y$ when moving from no knowledge to observing $X$. These two quantities are therefore conceptually distinct and coincide only for specific choices of the objective $l$ with respect to which the knowledge gain is measured, such as the log loss.

 The following properties of entropy and information are shown in Appendix \ref{prf:properties}.

\begin{Proposition}\label{prop:properties}
Under our standing assumptions, for any random variable $X:\Omega\to\mathcal X$, we have:
\begin{enumerate}[$(i)$]
\item Maximality of $\Pi(X)$: $U_{\pi\to\Pi(X)}(X)\geq 0$ for any $\pi$, in particular $H(X|\pi)$ is non-negative.
\item Non-negativity: If $\pi\preceq\pi'$, then $R(\pi')\leq R(\pi)$ (i.e., $-R$ is a monotone with respect to $\preceq$), and therefore $U_{\pi\to\pi'}\geq 0$. In particular, $I(X;\pi'|\pi)$ is non-negative.
\item No uncertainty: If $\mathbb P_X =\delta_x$ (the Dirac measure at $x$) for some $x\in\mathcal X$ then $H(X) = 0$.
\item Independence: If $X$ and $Y$ are stochastically independent then
\[I(X;Y)=0 \, , \quad H(X|Y) = H(X)\,.\]
\item Chain rule: $I(X;(Y,Z)) = I(X;Y) + I(X;Z|Y)$, and in general, for any sequence $(\pi_n)_{n=0}^N$ with $n\mapsto R(\pi_n)$ being non-increasing,
\begin{equation}\label{eq:generalchainrule}
U_{\pi_0\to\pi_N}(X) = \sum_{n=1}^N U_{\pi_{n-1}\to\pi_{n}}(X)
\end{equation}
is a non-negative decomposition.
\end{enumerate}
\end{Proposition}

Proposition \ref{prop:properties} shows that many well-known properties of Shannon entropy and information are retained when allowing the loss function to be arbitrary. Not only that, but in Section \ref{sec:partialinformationdecomp}, we show that the underlying structure of the knowledge lattice enables a principled treatment of partial information decompositions (PIDs) using \eqref{eq:generalchainrule}.

Using Bayes risk to study entropy and information goes back to early statistical decision theory \cite{DeGroot1962}, was made precise as a notion of generalized entropy by Gr\"unwald and Dawid \cite{Gruenwald2004}, and plays the key role in Blackwell's ordering of experiments \cite{Blackwell1953}, where Bayes risks are compared to measure when one experiment is more informative than another. Our treatment, however, differs from the concepts studied in these classical approaches in two fundamental ways. The first concerns the difference between entropy and information: our entropy \eqref{def:entropy} matches the generalized entropy of \cite{Gruenwald2004} whenever the residual risk $\mathbb E[\inf_a l(X,a)]$ vanishes, as it does for most used losses in practice, but retaining this term lets us treat entropy and information as a single concept rather than as two separately defined quantities. Both are instances of risk reduction $U_{\pi\to\pi'}$ differing only in the choice of endpoints. The second and more structural difference concerns the object on which the comparison acts: Blackwell \cite{Blackwell1953} orders experiments by Bayes-risk dominance simultaneously for all losses, a partial order that retains the achievable risks of a given experiment. Fixing a loss instead gives a total preorder, but rather than reducing knowledge states to scalars, we retain the knowledge states as the primitive objects, with their meets and joins.

Importantly, we use the loss only to identify informational quantities as evaluations of transitions between the lattice nodes. This identification relates the concepts of information and entropy to previously unrelated properties of partitions and enables the general chain rule \eqref{eq:generalchainrule} from which the partial information decompositions of Section~\ref{sec:partialinformationdecomp} follow. In summary, the novelty of our approach lies not in Bayes risk reduction itself, but in using the underlying knowledge lattice to understand the general structure of classical information measures and in using that structure to discover new informational quantities, e.g., for decomposing those classical measures.

\subsection{Knowledge about $X$}\label{sec:knowledge-about-X}
Recall that the risk $R(Y)$ is the minimal expected loss $\mathbb E[l(X,A)]$ achievable by actions $A$ measurable with respect to $\sigma(Y)$. Since the expected loss depends on $Y$ only through the expectation of $\mathbb E[l(X,A)|Y]$ due to the tower property of conditional expectations, $R(Y)$ only depends on the random conditional probability measure $\mathbb P_{X\mid Y}$ and not on the full observation $Y$ itself. We note that $\mathbb P_{X|Y}$ is a regular conditional probability because $\mathcal X$ is assumed to be standard Borel. Therefore, the infimum over $\sigma(Y)$-measurable actions may be restricted to the smaller $\sigma$-algebra $\sigma(\mathbb P_{X\mid Y})$. In terms of partitions, this corresponds to replacing the observation partition $\Pi(Y)$ by the coarser partition
\[
\pi_{Y} \coloneqq \Pi(\mathbb P_{X\mid Y}) \preceq \Pi(Y),
\]
which identifies outcomes of $Y$ that induce the same conditional law of
$X$. We summarize this in the following proposition.

\begin{Lemma}\label{prop:problattice} For any random variable $Y$ on $\Omega$, we have
\begin{align*}
R(Y) = R(\pi_{Y}).
\end{align*}
In particular, while $\Pi(Y)$ encodes the knowledge in $Y$ about $\omega$, for the risk $R$ only the knowledge in $Y$ about $X$ is important, which is encoded by the partition $\pi_{Y}\preceq \Pi(Y)$.
\end{Lemma}
Here, the target space of $Y$ is arbitrary, since $R$ only depends on $Y$ through the $\sigma$-algebra of $\mathbb P_{X|Y}$, and the conditionals are regular for arbitrary sub-$\sigma$-algebras because $\mathcal X$ is standard Borel by our standing assumptions.

By Lemma \ref{prop:problattice}, we may replace $\Pi(Y)$ by $\pi_{Y}$ in the definition of entropy and information, i.e.
\begin{align*}
H(X|Y) & = R(\pi_{Y}) - R(X),  \\
I(X;Y) & = R(\bot) - R(\pi_{Y}), \\
I(X;Z|Y) & = R(\pi_{Y}) - R(\pi_{Y,Z})\,.
\end{align*}

Note that, if $\mathbb P_X$ is viewed as the (constant) random variable $\omega\mapsto \mathbb P_X$, then $\Pi(\mathbb P_X) = \bot$, and moreover, for $Y=X$, we have $\Pi(\delta_X) = \Pi(X)$, where $\delta_x$ denotes the Dirac measure centered at $x\in\mathcal X$. We can therefore interpret a path of risk reduction from no knowledge to partial knowledge obtained through an observation $Y$ to full knowledge about $X$ as a sequence of uncertainty reductions from $\mathbb P_X$ to $\mathbb P_{X\mid Y}$ and finally to $\delta_X$ (see also Example \ref{sec:transformationcosts} in Appendix \ref{app:examples}). As we show in the following, when passing to the corresponding knowledge lattice, such a path can be refined further via the meet and join of the associated partitions. These refinements result in partial information decompositions grounded directly in the structure of the knowledge lattice.

\section{Partial information decomposition} \label{sec:partialinformationdecomp}

Since its publication, the seminal paper \cite{Williams2010} on \textit{partial information decomposition} (PID) by Williams and Beer has attracted considerable attention in the literature. The practical relevance of PIDs lies in the ability to distinguish qualitatively different types of multivariate dependence. Whereas Shannon information aggregates all dependence into a single scalar, PIDs separate redundant contributions (shared among the sources), unique contributions (source-specific), and synergistic contributions (obtainable only from joint observations). This distinction is particularly useful in fields such as neuroscience \cite{Wibral2017,Cramer2020}, computational biology \cite{Sherrill2021,Zhang2025}, machine learning \cite{Dewan2024,Schneider2025}, and complex systems analysis \cite{Luppi2024,Mediano2025}, where understanding the shared, unique, and synergistic contributions of signals is important. For example, in neural coding, failing to separate synergistic from redundant contributions can lead to fundamentally different conclusions about whether a neural population encodes information collectively or individually \cite{Schneidman2003,Latham2005,Sherrill2021}. Due to this sustained interest, there have been many foundational contributions, \cite{Harder2013,Griffith2014,Griffith2014a,Bertschinger2014,Barrett2015,Ince2017,James2018,Rosas2020,Ay2020,Makkeh2021,Kolchinsky2022,Gomes2024}, to name a few.

However, recently it has been shown that most current PID formulations are incompatible with core information-theoretic properties such as the chain rule and non-negativity \cite{Matthias2025}. In the following, we offer a possible solution to this problem. Our approach yields information decompositions that, unlike existing proposals, define no term by subtraction from another. Instead, redundancy, synergy, and the unique contributions are each read off a transition on the lattice of distribution partitions. Moreover, whereas existing PIDs are specific to Shannon information, analogous decomposition questions exist particularly for variance decompositions \cite{Cramer1946,Spurrell1967,Mood1971}, while facing similar conceptual difficulties \cite{Leps2003}. Our approach results in partial information decompositions for any suitable loss function, unifying these settings. We also include an extensive set of discrete and continuous examples (cf.~Tables~\ref{tab:pid_comparison_main} and \ref{tab:pid_comparison_continuous} in the following sections, and Tables~\ref{tab:pid_comparison_appendix1}-\ref{tab:pid_comparison_appendix3} in the appendix) supporting these points by comparing our decomposition against many established PID measures. We attribute the general disagreement demonstrated by those examples to the path-independence assumption, which cannot be satisfied exactly, and so the resulting discrepancies leak into redundant or synergistic terms, for each approach in a slightly different way.

\subsection{Current PID assumptions} \label{sec:currentPID}

First, consider the case of three random variables $X$, $Y_1$, and $Y_2$, where $X$ is considered to be the quantity of interest, and $Y_i$ are examined with respect to their information about $X$. Using the chain rule of mutual information, we can decompose total Shannon information $I_S(X;(Y_1,Y_2))$ between $X$ and $(Y_1,Y_2)$ into the Shannon information in $Y_1$ about $X$, $I_S(X;Y_1)$, plus the Shannon information in $Y_2$ about $X$ given $Y_1$, $I_S(X;Y_2|Y_1)$. PIDs attempt to further decompose these terms into \textit{redundant}, \textit{unique}, and \textit{synergistic} information,
\begin{equation}\label{def:pid}
I_S(X;(Y_1,Y_2)) = R_{1,2} + U_{1} + U_{2} + S_{1,2},
\end{equation}
where the redundant and synergistic terms, $R_{1,2}$ and $S_{1,2}$, are symmetric with respect to interchanging $Y_1$ and $Y_2$, whereas the unique terms are not, and all terms are assumed to be non-negative. Note that $U_1$, the unique information of $Y_1$ about $X$, also depends on the random variable $Y_2$ with respect to which it has to be unique, and vice versa. The decomposition \eqref{def:pid} results from the following assumptions on the single-variable information terms:
\begin{align}\label{def:pid1}
    I_S(X;Y_i) & = R_{1,2} + U_i \\ \label{def:pid2}
    I_S(X;Y_i|Y_j) & = U_i + S_{1,2} \, ,
\end{align}
where $i,j\in\{1,2\}$ with $i\neq j$.

Notably, the same unique information $U_i$ appears in both quantities, no matter whether $Y_i$ is observed on its own or after $Y_j$. As noted in the introduction, this is based on a view of information measures as state functions, rather than process quantities, insensitive to the knowledge contexts in which they are measured. As we discuss below in more detail, such path independence is appropriate at the level of information \textit{content} (i.e., knowledge states), not at the level of \textit{measuring} information. This assumption turns \eqref{def:pid1}--\eqref{def:pid2} into a linear system of rank three in the four unknowns $R_{1,2},U_1,U_2,S_{1,2}$, so that a single additional condition fixes the remaining terms by subtraction.

Most proposals in the literature impose one such additional assumption in order to specify how either redundant, unique, or synergistic information should be identified, so-called \textit{PID-inducing conditions} \cite{Gutknecht2025}, from which the full decomposition then follows through \eqref{def:pid1} and \eqref{def:pid2}. For example, Williams and Beer \cite{Williams2010} define redundant information as the average of the minimum information about outcomes $X=x$ provided by each of the variables $Y_i$. Bertschinger \textit{et al.} \cite{Bertschinger2014} define unique information $U_i$ as the minimum of $I_S(X;Y_i|Y_j)$ for $j\neq i$ over all joint distributions preserving the marginals $\mathbb P_{X,Y_i}$. Similarly, Griffith and Koch \cite{Griffith2014} quantify synergistic information, motivated by \cite{Maurer1999}, as the difference between the total Shannon information $I_S(X;(Y_1,Y_2))$ when evaluated using the actual joint $\mathbb P_{X,Y_1,Y_2}$ and when evaluated using the joint $q_{X,Y_1,Y_2}$ that minimizes the total information under the constraint that the marginals $q_{X,Y_i}$ are the same as $\mathbb P_{X,Y_i}$. For two sources, this construction coincides with the decomposition of Bertschinger \textit{et al.} \cite{Bertschinger2014}. From a more decision-theoretic standpoint, Banerjee \textit{et al.}\ \cite{Banerjee2018} define unique information through (weighted) Le~Cam deficiencies, quantifying how well one channel can be approximated from the other by randomization at its input or output. Kolchinsky \cite{Kolchinsky2022}, on the other hand, relies on the Blackwell ordering of experiments \cite{Blackwell1953} to define redundancy as the maximal Shannon information $I_S(X;Q)$ over auxiliary variables $Q$ that are Blackwell-dominated by every source, a maximization that is needed because the Blackwell order, unlike the knowledge lattice, admits no meets \cite{Bertschinger2014a}. It seems to be the only current appraoch that does not rely on the assumptions \eqref{def:pid1}-\eqref{def:pid2}, with the downside of overshooting the full information by a small amount on some of the examples we have tested (see, e.g., Table \ref{tab:pid_comparison_main}).

Note that, for ease of presentation, we have discussed the case of two sources here, but PIDs are often extended to the case of arbitrarily many sources $Y_1,\dots, Y_n$, with increasingly many terms in the decomposition, e.g., 18 terms for $n=3$ \cite{Griffith2014} (cf. Appendix \ref{app:threesources}, where we show one path through the knowledge lattice resulting in a finer PID of 32 terms).

\begin{figure}
\centering
\includegraphics[width=.5\textwidth]{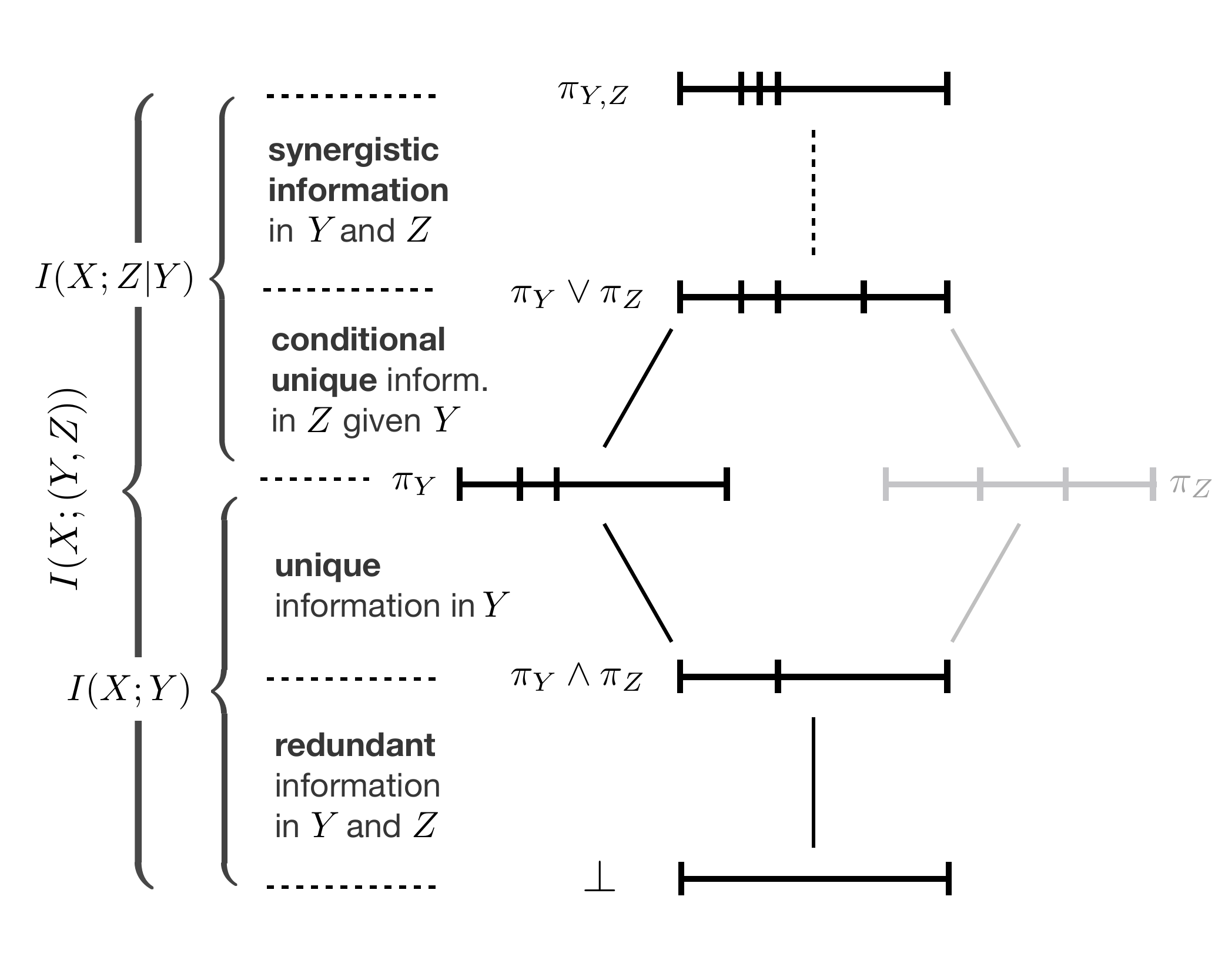}
\vspace{-20pt}
\caption{A 4-term partial information decomposition for two source variables. When traversing the knowledge lattice along the shown path the total information $I(X;(Y,Z))$  decomposes into four terms, corresponding to redundant, unique, conditional unique, and synergistic information. As an example we show interval partitions as in Fig.~\ref{fig:joinAndMeet}. The dashed line between $\pi_{Y,Z}$ and the join of $\pi_{Y}$ and $\pi_{Z}$ indicates that the partitions are not necessarily comparable, whereas the corresponding risks are always non-increasing (cf.~Lemma \ref{lma:synergygap}).}
\label{fig:simplePID}
\end{figure}

The assumption of the two unique terms to be identical is a result of treating informational quantities as state functions attached to sources, which can then be added, compared, and said to be ``contained'' in one another. On the knowledge lattice, however, the only state function is the risk $R(\pi)$. Information between two knowledge states is a difference of this state function, and hence independent of the route taken between them. The information terms of a decomposition, by contrast, are increments along a chosen chain and depend on the chain wherever the lattice allows for a choice (cf.~Corollary~\ref{cor:PID}). Additivity of informational quantities beyond the chain rule therefore has no justification, and the idea of one term being contained in another has limited formal meaning. Besides of clamping the unique terms in \eqref{def:pid1} and \eqref{def:pid2} to the same value $U_i$, other examples of this point of view include the definition of redundancy in \cite{SchickPoland2021} through a collection of terms required to be contained in each $I_S(X;Y_i)$, and the part-whole lattices of \cite{Gutknecht2021,Gutknecht2025}, in which information terms are considered \textit{part} of other terms. The most visible instance is the use of Venn diagrams in the PID literature, known as PI-diagrams \cite{Griffith2014}: The interpretation of standard Venn diagrams representing relationships between informational quantities through overlapping two-dimensional regions is conceptually difficult, both because they may contain regions corresponding to negative terms and because they suggest inappropriate set-theoretic analogies \cite{MacKay2002}. While PI-diagrams are defined to contain only non-negative regions, they still give a false set-theoretic intuition for quantities that are purely algebraic. The same applies to variance decompositions \cite{Leps2003}, which also use Venn diagrams to represent overlap between contributions. It is the knowledge states $\pi$ that admit a compositional structure, through meets and joins. Informational quantities are what those compositions cost or gain in terms of risk measured with respect to a given loss function.

\subsection{Redundancy, uniqueness, and synergy} \label{sec:PIDours}

Due to the chain rule \eqref{eq:generalchainrule}, the number of possible decompositions of informational quantities of the form $R(\pi)-R(\pi')$ discussed in Section \ref{sec:information} is limited only by the number of partitions of $\Omega$ and the possible risk-reducing paths through the knowledge lattice connecting $\pi$ with $\pi'$.

However, depending on the particular informational measure to be decomposed, certain paths play a special role. For instance, for $H(X) = R(\bot)-R(X)$, we might consider another random variable $Y$, which we want to study in terms of how much information it contains with respect to $X$. Then the path $\gamma = (\bot, \pi_{Y},\Pi(X))$ results in the standard decomposition
\[
H(X)  =  I(X;Y) + H(X|Y) \, .
\]
Similarly, for $I(X;(Y,Z)) = R(\bot)-R(\pi_{Y,Z})$, we might consider two alternative routings from $\bot$ to $\pi_{Y,Z}$, either through $\pi_{Y}$ or through $\pi_{Z}$, each resulting in a different decomposition. For the $\pi_{Y}$-routing $\gamma = (\bot,\pi_{Y},\pi_{Y,Z})$ (e.g., chosen because $Y$ is observed before $Z$), we obtain
\begin{equation}\label{eq:decompJointInfo0}
I(X;(Y,Z)) = I(X;Y) + I(X;Z|Y) \, .
\end{equation}
In the first example, the waypoint $\pi_{Y}$ is forced by the question being asked, where we want to know how much of the uncertainty in $X$ is resolved by $Y$, such that the only natural intermediate state is $\pi_{Y}$. In the second example, by contrast, the two waypoints $\pi_{Y}$ and $\pi_{Z}$ are equally valid, and the choice between them is an extra degree of freedom. We call a decomposition \textit{path-dependent} when such a choice arises, that is, when the path includes a routing through one of several incomparable knowledge states.

We now turn to a natural class of such decompositions motivated by the partial information decompositions discussed in the previous section. To this end, note that in the second example above, leading to \eqref{eq:decompJointInfo0}, irrespective of the chosen routing there are two special waypoints that require no ordering choice: the common knowledge $\pi_{Y} \wedge \pi_{Z}$ and the combined knowledge $\pi_{Y} \vee \pi_{Z}$.

The common knowledge $\pi_{Y} \wedge \pi_{Z}$ of the distribution partitions must be distinguished from the common knowledge $\Pi(Y)\wedge\Pi(Z)$ of the observation partitions. Since $\pi_{Y} \preceq \Pi(Y)$ and $\pi_{Z} \preceq \Pi(Z)$, the latter is in general finer: the observations $Y$ and $Z$ may share knowledge about $\omega$ that is irrelevant to $X$. Since the distribution partitions $\pi_{Y} = \Pi(\mathbb P_{X|Y})$ retain only those distinctions in $Y$ that affect the conditional laws of $X$, $\pi_{Y} \wedge \pi_{Z}$ captures common knowledge \textit{about $X$}, which motivates identifying the corresponding risk reduction from $\bot$ as \textit{redundant information}.

Moreover, in contrast to the observation partitions, for which $\Pi(Y,Z) = \Pi(Y)\vee \Pi(Z)$, the analogous identity for $\pi_{Y,Z}$ generally fails.

\begin{Lemma}[synergy gap]\label{lma:synergygap}
Under our standing assumptions, let $Y,Z$, and $X$ be random variables. Then, the join $\pi_Y\vee \pi_Z$ of the single variable distribution partitions $\pi_Y = \Pi(\mathbb P_{X|Y})$ and $\pi_Z = \Pi(\mathbb P_{X|Z})$ needs not to be comparable to the partition $\pi_{Y,Z} = \Pi(\mathbb P_{X|Y,Z})$ with respect to the refinement order. However, we always have
\begin{equation}\label{eq:synergygap}
R(\pi_{Y,Z})\leq R(\pi_Y\vee\pi_Z)
\end{equation}
for every measurable loss $l$.
\end{Lemma}

The difference of the risks in \eqref{eq:synergygap} is what we call synergistic information, because \textit{synergy} describes what is gained when the observations $Y$ and $Z$ are available together in comparison to having observed $Y$ and $Z$ separately. The proof of Lemma \ref{lma:synergygap}, which can be found in Appendix \ref{prf:properties}, extends to arbitrarily many sources, so that we obtain the following general 3-term information decomposition.

\begin{Proposition}[Path-independent 3-term PID]\label{prop:redundancy-synergy}
Under our standing assumptions, let $Y_1,\dots,Y_n$ be
random variables considered with respect to what their observations tell
about $X$. Recall that $\pi_{Y_i} = \Pi(\mathbb P_{X|Y_i})$ and define:
\begin{enumerate}[$(i)$]
\item The
\textit{redundant information} in $Y_1,\dots,Y_n$ about $X$ is the risk
reduction from no knowledge to common knowledge,
\begin{equation}\nonumber
I_r(X;(Y_1,\dots,Y_n)) \coloneqq R(\bot) -
  R \Big(\bigwedge\nolimits_{i=1}^n \pi_{Y_i}\Big) .
\end{equation}
\item The \textit{synergistic information} of $Y_1,\dots,Y_n$ about $X$ is the
risk reduction from combined knowledge to joint knowledge,
\begin{equation}\nonumber
I_s(X;(Y_1,\dots,Y_n)) \coloneqq
  R \Big(\bigvee\nolimits_{i=1}^n \pi_{Y_i}\Big) - R(\pi_{Y_1,\dots,Y_n}) .
\end{equation}
\item The \textit{complementary information}
\begin{equation}\nonumber
I_c(X;(Y_1,\dots,Y_n)) \coloneqq
  R\Big(\bigwedge\nolimits_{i=1}^n \pi_{Y_i}\Big) -
  R\Big(\bigvee\nolimits_{i=1}^n \pi_{Y_i}\Big)
\end{equation}
is the residual that accounts for all source-specific contributions.
\end{enumerate}
Then
\begin{align} \nonumber I(X;(Y_1,\dots,Y_n)) \ & = \  I_r(X;(Y_1,\dots,Y_n)) \\  \label{3-termgeneral}& \quad + I_c(X;(Y_1,\dots,Y_n)) \\ & \quad + I_s(X;(Y_1,\dots,Y_n)) \nonumber
\end{align}
is a non-negative path-independent decomposition. Any further decomposition of $I_c$ requires choosing a route through intermediate states such as individual $\pi_{Y_i}$ or their partial meets and joins, and is therefore path-dependent in the sense defined above.
\end{Proposition}

Even though the final step, i.e., from combined knowledge $\bigvee_i \pi_{Y_i}$ to $\pi_{Y_1,\dots,Y_n}$, is a transition between two generally incomparable states, it is still non-increasing for any Bayes risk (displayed as a dashed edge in the figures). We could, however, create a $\preceq$-ordered step by replacing $\pi_{Y_1,\dots,Y_n}$ by its join with $\bigvee_i \pi_{Y_i}$ without changing the risk, because $\pi_{Y_1,\dots,Y_n}$ carries all knowledge about $X$ obtainable through the joint observation $(Y_1,\dots,Y_n)$.

\begin{figure*}
\centering
\includegraphics[width=\textwidth]{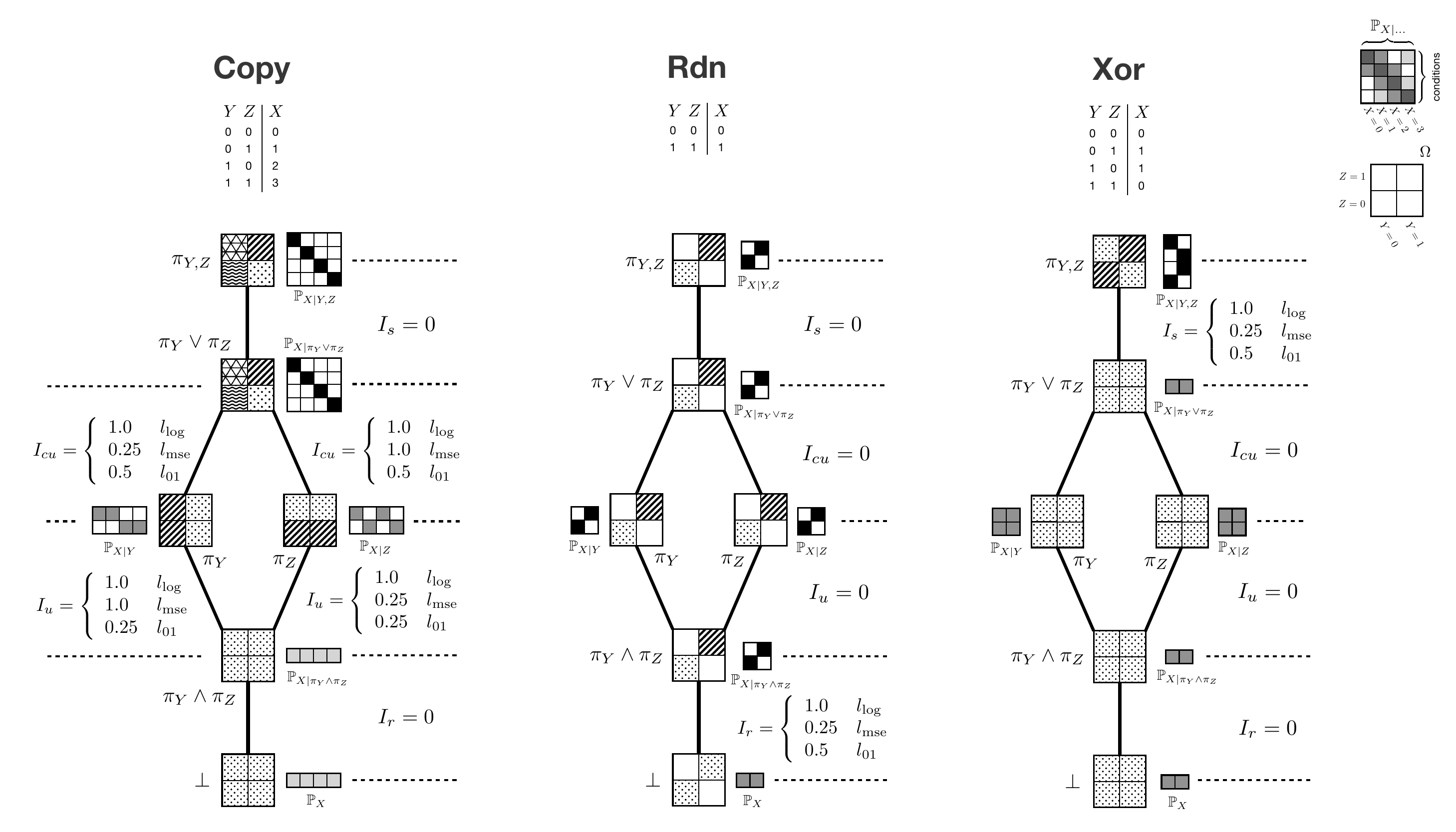}
\caption{Knowledge lattices for the three most basic examples of three variables $X,Y$ and $Z$, each example focusing on a different type of information: $\mathsf{Copy}$ only contains complementary information (unique and conditional unique terms), $\mathsf{Rdn}$ only contains redundant information, and $\mathsf{Xor}$ only contains synergistic information. The nodes, each representing a partition of $\Omega = \{(0,0),(0,1),(1,0),(1,1)\}$, are visualized using different textures for different blocks of the corresponding partition. Next to the partitions, we display the associated conditional laws $\mathbb P_{X|\pi}\coloneqq \mathbb P_{X|\sigma(\pi)}$. The non-zero risk reduction terms are calculated for the log loss, mean squared error and zero-one loss, discussed in Appendix \ref{app:examples}.}
\label{fig:examples1}
\end{figure*}

For two sources $Y$ and $Z$, the simplest routing that decomposes complementary information $I_c$ further, passes through either $\pi_Y$ or $\pi_Z$, splitting $I_c$ into a unique and a conditional unique term (cf.~Fig.~\ref{fig:simplePID}).

\begin{Corollary}[Path-dependent 4-term PID for two sources] \label{cor:PID}
Decomposing $I(X;(Y,Z))$ for two arbitrary random variables $Y,Z$ along the path
\[
\gamma = (\bot, \pi_Y\wedge \pi_Z, \pi_Y,\pi_Y\vee \pi_Z, \pi_{Y,Z})
\]
results in a path-dependent partial information decomposition with redundant, synergistic, and two unique contributions,
\begin{align}
I(X;Y) & = I_r(X;(Y,Z)) + I_u(X;Y{\setminus} Z) \\
I(X;Z|Y) & = I_{cu}(X;Z|Y) + I_s(X;(Y,Z)) \, ,
\end{align}
and thus
\begin{align}
I(X;(Y,Z)) \ = & \ I_r(X;(Y,Z)) + I_u(X;Y{\setminus}Z) \nonumber \\
& + I_{cu}(X;Z|Y) + I_s(X;(Y,Z)) \, , \label{eq:pidfromknowledge}
\end{align}
where unique information $I_u$ and conditional unique information $I_{cu}$ of a variable $Y$ about $X$ are distinguished depending on the knowledge context available when $Y$ is measured,
\begin{align}
I_u(X;Y{\setminus} Z) &  \coloneqq R(\pi_Y\wedge \pi_Z) - R(\pi_Y), \\
I_{cu}(X;Y|Z) & \coloneqq R(\pi_Z) - R(\pi_Y\vee \pi_Z) \, . \label{def:condunique}
\end{align}
\end{Corollary}

The notation $Y{\setminus}Z$ (\textit{$Y$ without $Z$}) in the expression for unique information $I_u$ is not a set difference, but is used to distinguish it from $Y|Z$ (\textit{$Y$ given $Z$}) in conditional unique information, because $I_u(X;Y{\setminus}Z)$ measures the transition from knowledge common to $Y$ and $Z$ to knowledge just in $Y$.

In Appendix \ref{prf:properties}, we give a simple proof of Proposition \ref{prop:redundancy-synergy}. As can be seen from the examples in the following section, the partial information terms introduced above are generally non-trivial.

The four-term decomposition \eqref{eq:pidfromknowledge} resembles existing partial information decompositions in the literature, as it consists of redundant, synergistic, and two unique terms. Unlike existing formulations, however, the two unique contributions are of different type: moving from common knowledge to $\pi_Y$ and moving from $\pi_Z$ to $\pi_Y\vee\pi_Z$ are two separate transitions on the lattice, with different risk reductions in general, corresponding to unique information $I_u(X;Y{\setminus}Z)$ and conditional unique information $I_{cu}(X;Y|Z)$, respectively. Moreover, the decomposition records whether the route passes through $\pi_Y$ or $\pi_Z$ before reaching $\pi_Y\vee\pi_Z$. Existing approaches based on \cite{Williams2010} identify unique and conditional unique information by assumption \eqref{def:pid1}--\eqref{def:pid2}, with the consequences described in Section \ref{sec:currentPID}, as illustrated by the examples in the following section.

In most of the PID literature a set-based notation of the form $\{i\cdots j,\dots,k\cdots l\}$ is used to label terms in the PID of $I(X;(Y_1,\dots,Y_n))$ studied there for $n>2$ \cite{Williams2010,Griffith2014}, where commas indicate redundant contributions, joint indices without commas denote synergistic terms, and unique information terms do not include the other variables. For example, for $n=2$, the four PID components in \eqref{def:pid} are denoted by $\{1,2\}$ for the redundant information, $\{1\}$ and $\{2\}$ for the unique information terms, and $\{12\}$ for the synergistic information. If we write $\{2|1\}$ for conditional unique information \eqref{def:condunique} of $Y_2$ about $X$ given $Y_1$ then the 2-source 4-term decomposition \eqref{eq:pidfromknowledge} consists of the four terms  $\{1,2\},\{1\},\{2|1\},\{12\}$.

For $n=3$, the PID literature considers 18 distinct informational terms \cite{Griffith2014}. As is clear from the above exposition, there are many possible paths from $\bot$ to $\pi_{Y_1,Y_2,Y_3}$ through the knowledge lattice that each induce a non-negative decomposition of $I(X;(Y_1,Y_2,Y_3))$. For example, Fig.~\ref{fig:pid3vars} in the appendix illustrates one such path consisting of 32 non-trivial contributions.

Moreover, we note that \eqref{3-termgeneral} is not the only symmetric non-trivial decomposition of joint information. For example, we can split up redundant information further by taking the meet with other knowledge states $\pi \not= \pi_{Y_i}$ before reaching  common knowledge $\bigwedge_{i} \pi_{Y_i}$. In the case of two source variables, this means that a path of the shape $\gamma = (\bot, \pi_Y\wedge \pi_Z,\dots,\pi_{Y,Z})$ is refined to
\[
\gamma = (\bot, \pi_Y\wedge \pi_Z \wedge \pi, \pi_Y\wedge \pi_Z, \dots, \pi_{Y,Z}).
\]
Similarly, we could further decompose conditional unique information, e.g., by including the meet with another partition $\pi$ as a middle step between $\pi_Y$ and $\pi_Y\vee \pi_Z$, i.e., $\gamma = (\bot,\dots,\pi_Y,\pi_Y\vee \pi_Z,\pi_{Y,Z})$ could be refined to
\[
\gamma = (\bot,\dots,\pi_Y,\pi_Y\vee (\pi_Z \wedge \pi), \pi_Y\vee \pi_Z,\pi_{Y,Z}).
\]
As an example, in Appendix \ref{sec:finerdecomp}, we create such non-trivial refinements of $I_r(X;(Y,Z))$ and $I_{cu}(X;Y|Z)$ using $\pi=\pi_{Y,Z}$, enabling to qualitatively distinguish the examples $\mathsf{And}$ and $\mathsf{Copy}$ discussed in the following section, offering a possible solution to the ongoing debate in the literature about the PID terms of $\mathsf{And}$ in particular \cite{Griffith2014,Ay2020}.

\subsection{Discrete examples} \label{sec:pid_examples}

In this section, we provide two different kinds of comparisons regarding decompositions of $I(X;(Y,Z))$ for discrete random variables $X, Y$, and $Z$. First, in $(i)$ below we discuss our decomposition on three basic example gates for different loss functions: the log loss, the mean squared error loss and the $01$-loss. Whereas most of the PID proposals in the literature agree with these decompositions for the log loss, in $(ii)$, we discuss three example gates on which the PID literature disagrees (cf.~Table \ref{tab:pid_comparison_main}). Throughout we visualize the corresponding knowledge lattices including the explicit partitions for each reduction step.

$(i)$ Consider the three examples shown in Fig.~\ref{fig:examples1}. $\mathsf{Copy}$ maps the sources $Y$ and $Z$ to $X = 2Y + Z$, which encodes copying $Y$ and $Z$ to $X\cong (Y,Z)$. The partitions $\pi_Y$ and $\pi_Z$ each divide the sample space in two orthogonal halves (cf.~Fig.~\ref{fig:examples1}), so that common knowledge is trivial, i.e., no redundancy, and, moreover, the combined knowledge $\pi_Y \vee \pi_Z$ already captures all knowledge, so that there is no synergy either. In contrast, the example $X=\mathsf{Rdn}(Y,Z)$ only requires knowing one of $Y$ or $Z$ to determine $X$, so that the information is fully redundant because $\pi_Y\wedge \pi_Z = \pi_{Y,Z}$. $\mathsf{Xor}$, on the other hand, requires full knowledge about both $Y$ and $Z$ in order to know anything about $X$, and so there is only synergistic information. As can be seen in Fig.~\ref{fig:examples1} on the right, this is realized by all partitions $\pi_Y\wedge \pi_Z$, $\pi_Y$, $\pi_Z$, and $\pi_{Y}\vee \pi_Z$ being trivial, so that only the final step to $\pi_{Y,Z} = \Pi(\mathbb P_{X|Y,Z})$ results in non-zero information $I_s(X;(Y,Z)) = R(\bot)-R(\pi_{Y,Z})$. These basic examples can be combined to create gates where any combination of information terms are non-zero, such as the example $\mathsf{RdnXor}$ shown in Table \ref{tab:pid_comparison_appendix1} in the appendix, which is a continuation of Table \ref{tab:pid_comparison_main} that we will discuss next.

\begin{table*}[p]
\centering
\caption{Partial information decomposition of $I(X;(Y,Z))$ (in bits) for example gates. See also Tables \ref{tab:pid_comparison_appendix1}, \ref{tab:pid_comparison_appendix2}, and \ref{tab:pid_comparison_appendix3}\label{tab:pid_comparison_main}}
\renewcommand{\arraystretch}{1.7}
\setlength{\tabcolsep}{5pt}
\begin{tabular}{@{}l@{\hskip 3pt}l@{\hskip 3pt}rrrrrr@{}}
\toprule
Example & & Measure & $R$ & $U_Y$ & $U_Z$ & $S$ & $I$ \\
\midrule
\multirow{12}{*}{\begin{tabular}{@{}c@{}}\textbf{ImperfectRdn}\\[8pt]{\scriptsize\parbox[t][1.2cm][t]{3.3cm}{\centering $X=Y$, with $Z$ being an imperfect copy of $Y$}}\\[6pt]{\scriptsize\setlength{\tabcolsep}{3pt}\begin{tabular}{@{}cc|c|c@{}}$Y$ & $Z$ & $X$ & $\mathbb P$ \\0 & 0 & 0 & 0.400 \\0 & 1 & 0 & 0.100 \\1 & 1 & 1 & 0.500 \\\end{tabular}}\end{tabular}} & \multirow{12}{*}{\raisebox{0.4cm}{\includegraphics[height=6.3cm]{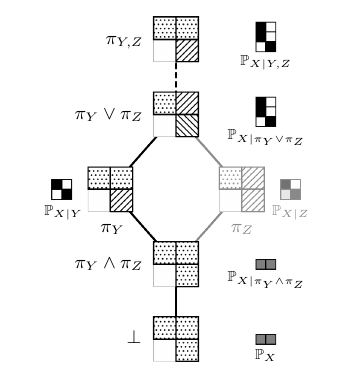}}} & \textbf{Ours ($\pi_Y$-routing)} & \textbf{0.000} & \textbf{1.000} & \textbf{0.000} & \textbf{0.000} & \multirow{12}{*}{1.000} \\
 &  & \textcolor[gray]{0.5}{\textbf{Ours ($\pi_Z$-routing)}} & \textcolor[gray]{0.5}{\textbf{0.000}} & \textcolor[gray]{0.5}{\textbf{0.390}} & \textcolor[gray]{0.5}{\textbf{0.610}} & \textcolor[gray]{0.5}{\textbf{0.000}} &  \\
 &  & Williams-Beer ($I_{\min}$) & 0.610 & \textcolor[gray]{0.5}{\textbf{0.390}} & \textbf{0.000} & \textbf{0.000} &  \\
 &  & Harder \textit{et al.}\ ($I_{\mathrm{proj}}$) & 0.610 & \textcolor[gray]{0.5}{\textbf{0.390}} & \textbf{0.000} & \textbf{0.000} &  \\
 &  & Griffith \textit{et al.} ($I_{\mathrm{GK}}$) & \textbf{0.000} & \textbf{1.000} & \textcolor[gray]{0.5}{\textbf{0.610}} & -0.610 &  \\
 &  & Bertschinger \textit{et al.}\ ($I_{\mathrm{BROJA}}$) & 0.610 & \textcolor[gray]{0.5}{\textbf{0.390}} & \textbf{0.000} & \textbf{0.000} &  \\
 &  & Barrett ($I_{\mathrm{MMI}}$) & 0.610 & \textcolor[gray]{0.5}{\textbf{0.390}} & \textbf{0.000} & \textbf{0.000} &  \\
 &  & Ince ($I_{\mathrm{ccs}}$) & 0.768 & 0.232 & -0.158 & 0.158 &  \\
 &  & James \textit{et al.}\ ($I_{\mathrm{dep}}$) & 0.610 & \textcolor[gray]{0.5}{\textbf{0.390}} & \textbf{0.000} & \textbf{0.000} &  \\
 &  & Ay-Polani-Virgo & $-$ & 0.695 & 0.305 & \textbf{0.000} &  \\
 &  & Kolchinsky & 0.610 & \textcolor[gray]{0.5}{\textbf{0.390}} & \textbf{0.000} & \textbf{0.000} &  \\
 &  & Gomes-Figueiredo & 0.610 & \textcolor[gray]{0.5}{\textbf{0.390}} & \textbf{0.000} & \textbf{0.000} &  \\
\addlinespace[7pt]\midrule\addlinespace[7pt]
\multirow{12}{*}{\begin{tabular}{@{}c@{}}\textbf{ReducedCopy}\\[8pt]{\scriptsize\parbox[t][1.2cm][t]{3.3cm}{\centering Copy with the $(Y,Z)=(1,1)$ state removed, uniformly distributed}}\\[6pt]{\scriptsize\setlength{\tabcolsep}{3pt}\begin{tabular}{@{}cc|c|c@{}}$Y$ & $Z$ & $X$ & $\mathbb P$ \\0 & 0 & 0 & 0.333 \\0 & 1 & 1 & 0.333 \\1 & 0 & 2 & 0.333 \\\end{tabular}}\end{tabular}} & \multirow{12}{*}{\raisebox{0.4cm}{\includegraphics[height=6.3cm]{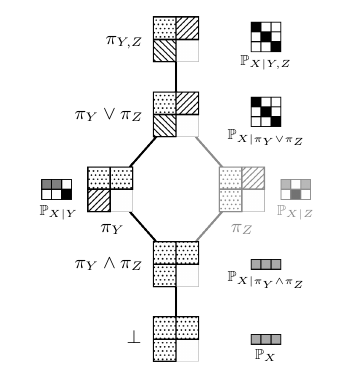}}} & \textbf{Ours ($\pi_Y$-routing)} & \textbf{0.000} & \textbf{0.918} & \textbf{0.667} & \textbf{0.000} & \textbf{1.585} \\
 &  & \textcolor[gray]{0.5}{\textbf{Ours ($\pi_Z$-routing)}} & \textcolor[gray]{0.5}{\textbf{0.000}} & \textcolor[gray]{0.5}{\textbf{0.667}} & \textcolor[gray]{0.5}{\textbf{0.918}} & \textcolor[gray]{0.5}{\textbf{0.000}} & \textcolor[gray]{0.5}{\textbf{1.585}} \\
 &  & Williams-Beer ($I_{\min}$) & 0.585 & 0.333 & 0.333 & 0.333 & 1.585 \\
 &  & Harder \textit{et al.}\ ($I_{\mathrm{proj}}$) & 0.252 & \textcolor[gray]{0.5}{\textbf{0.667}} & \textbf{0.667} & \textbf{0.000} & 1.585 \\
 &  & Griffith \textit{et al.} ($I_{\mathrm{GK}}$) & \textbf{0.000} & \textbf{0.918} & \textcolor[gray]{0.5}{\textbf{0.918}} & -0.252 & 1.585 \\
 &  & Bertschinger \textit{et al.}\ ($I_{\mathrm{BROJA}}$) & 0.252 & \textcolor[gray]{0.5}{\textbf{0.667}} & \textbf{0.667} & \textbf{0.000} & 1.585 \\
 &  & Barrett ($I_{\mathrm{MMI}}$) & 0.918 & 0.000 & 0.000 & 0.667 & 1.585 \\
 &  & Ince ($I_{\mathrm{ccs}}$) & 0.390 & 0.528 & 0.528 & 0.138 & 1.585 \\
 &  & James \textit{et al.}\ ($I_{\mathrm{dep}}$) & 0.252 & \textcolor[gray]{0.5}{\textbf{0.667}} & \textbf{0.667} & \textbf{0.000} & 1.585 \\
 &  & Ay-Polani-Virgo & $-$ & 0.792 & 0.792 & \textbf{0.000} & 1.585 \\
 &  & Kolchinsky & \textbf{0.000} & \textbf{0.918} & \textcolor[gray]{0.5}{\textbf{0.918}} & \textbf{0.000} & 1.837 \\
 &  & Gomes-Figueiredo & 0.252 & \textcolor[gray]{0.5}{\textbf{0.667}} & \textbf{0.667} & \textbf{0.000} & 1.585 \\
\addlinespace[7pt]\midrule\addlinespace[7pt]
\multirow{12}{*}{\begin{tabular}{@{}c@{}}\textbf{ExtendedXor}\\[8pt]{\scriptsize\parbox[t][1.2cm][t]{3.3cm}{\centering $\mathsf{Xor}$ extended by a third $Y$ value on which $X$ is fixed only together with $Z$, and a state $(3,2)$ recognized by both}}\\[6pt]{\scriptsize\setlength{\tabcolsep}{3pt}\begin{tabular}{@{}cc|c@{}}$Y$ & $Z$ & $X$ \\0 & 0 & 0 \\0 & 1 & 1 \\1 & 0 & 1 \\1 & 1 & 0 \\2 & 0 & 2 \\2 & 1 & 0 \\3 & 2 & 3 \\\end{tabular}}\end{tabular}} & \multirow{12}{*}{\raisebox{0.4cm}{\includegraphics[height=6.3cm]{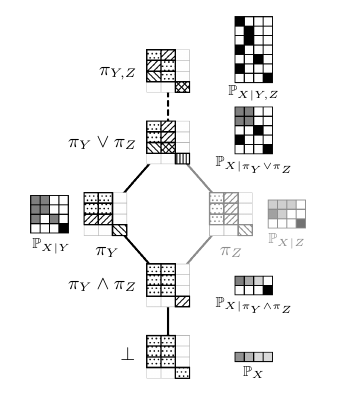}}} & \textbf{Ours ($\pi_Y$--routing)} & \textbf{0.592} & \textbf{0.394} & \textbf{0.286} & \textbf{0.571} & \textbf{1.842} \\
 &  & \textcolor[gray]{0.5}{\textbf{Ours ($\pi_Z$--routing)}} & \textcolor[gray]{0.5}{\textbf{0.592}} & \textcolor[gray]{0.5}{\textbf{0.501}} & \textcolor[gray]{0.5}{\textbf{0.178}} & \textcolor[gray]{0.5}{\textbf{0.571}} & \textcolor[gray]{0.5}{\textbf{1.842}} \\
 &  & Williams--Beer ($I_{\min}$) & 0.735 & 0.251 & 0.035 & 0.822 & 1.842 \\
 &  & Harder \textit{et al.}\ ($I_{\mathrm{proj}}$) & 0.700 & 0.286 & 0.070 & 0.787 & 1.842 \\
 &  & Griffith \textit{et al.} ($I_{\mathrm{GK}}$) & \textbf{0.592} & \textbf{0.394} & \textcolor[gray]{0.5}{\textbf{0.178}} & 0.679 & 1.842 \\
 &  & Bertschinger \textit{et al.}\ ($I_{\mathrm{BROJA}}$) & 0.627 & 0.359 & 0.143 & 0.714 & 1.842 \\
 &  & Barrett ($I_{\mathrm{MMI}}$) & 0.770 & 0.216 & 0.000 & 0.857 & 1.842 \\
 &  & Ince ($I_{\mathrm{ccs}}$) & 0.587 & 0.398 & 0.182 & 0.675 & 1.842 \\
 &  & James \textit{et al.}\ ($I_{\mathrm{dep}}$) & 0.627 & 0.359 & 0.143 & 0.714 & 1.842 \\
 &  & Ay--Polani--Virgo & $-$ & 0.743 & 0.528 & \textbf{0.571} & 1.842 \\
 &  & Kolchinsky & \textbf{0.592} & \textbf{0.394} & \textcolor[gray]{0.5}{\textbf{0.178}} & 0.714 & 1.877 \\
 &  & Gomes--Figueiredo & 0.627 & 0.359 & 0.143 & 0.714 & 1.842 \\
\addlinespace[7pt]
\bottomrule
\end{tabular}
\end{table*}

$(ii)$ Table \ref{tab:pid_comparison_main} compares our knowledge lattice-based decomposition with ten PID proposals from the literature on three finite examples for which no consensus exists. More specifically, we list values for the minimum-information redundancy $I_{\min}$ of Williams and Beer \cite{Williams2010}, the information-projection measure $I_{\mathrm{proj}}$ of Harder \textit{et al.}~\cite{Harder2013}, the Gács-Körner common-information measure $I_{\mathrm{GK}}$ of Griffith \textit{et al.} \cite{Griffith2014a}, the unique-information measure $I_{\mathrm{BROJA}}$ of Bertschinger \textit{et al.}~\cite{Bertschinger2014}, the minimum-mutual-information decomposition $I_{\mathrm{MMI}}$ of Barrett \cite{Barrett2015}, the common-change-in-surprisal measure $I_{\mathrm{ccs}}$ of Ince \cite{Ince2017}, the dependency decomposition $I_{\mathrm{dep}}$ of James \textit{et al.}~\cite{James2018}, the cooperative-game decomposition of Ay, Polani and Virgo \cite{Ay2020}, the Blackwell-order redundancy of Kolchinsky \cite{Kolchinsky2022}, and the union-information decomposition of Gomes and Figueiredo \cite{Gomes2024}. Cells of the other measures that coincide with ours are set in bold. The middle columns $U_Y, U_Z$ hold what the PID literature refers to as unique information of each source, where for ours they carry the unique and conditional-unique terms, whose assignment to $Y$ and $Z$ is path-dependent. The two bold rows contain our two routings, so, depending on the row, either $U_Y = I_u(X;Y{\setminus} Z)$ ($\pi_Y$-routing) or $U_Y=I_{cu}(X;Y|Z)$ ($\pi_Z$-routing).

Some remarks about the other decompositions are necessary. The Ay-Polani-Virgo decomposition has no redundancy term. The Griffith \textit{et al.} and Ince approaches are known to not always result in non-negative terms \cite{Griffith2014a, Ince2017}, whereas for Kolchinsky we report the ``native synergy'' $I-I_\cup\ge0$, whose four terms need not sum to $I$. The last column gives $I$, or each row's term sum where these differ. All values have been computed either using the dit library \cite{Dit2018} or implemented from scratch for the newer approaches.

Whereas existing formulations mostly disagree on the examples in Table \ref{tab:pid_comparison_main}, assigning widely different values to the redundant, unique, and synergistic terms of the same gates, the values produced by our two routings also appear in several other decompositions (highlighted in bold). This observation may be explained as follows: current approaches are built on the assumption that unique information is path-independent, i.e., that the same term $U_Y$ expresses unique information irrespective of whether $Y$ was observed before or after $Z$. From Corollary \ref{cor:PID},
\begin{align*} & I_u(X;Y{\setminus}Z)-I_{cu}(X;Y|Z) \\
& \qquad = \big(I(X;Y)-I(X;Y|Z)\big)-\big(I_r - I_s\big), \end{align*}
so under \eqref{def:pid1}--\eqref{def:pid2}, which force the left side to vanish, the term $I(X;Y)-I(X;Y|Z)$ (also known as \textit{interaction} or \textit{co-information}, even though it can be negative) must equal the difference of redundant and synergistic information, $I_r-I_s$. On the knowledge lattice this identity does not hold, and the difference is what PIDs based on \eqref{def:pid1}--\eqref{def:pid2} distribute into their redundant or synergistic terms. The gates in Table~\ref{tab:pid_comparison_main} (and Tables \ref{tab:pid_comparison_appendix1}-\ref{tab:pid_comparison_appendix3} in the appendix) make this leakage visible, for example by negative terms or non-zero contributions at edges whose knowledge transitions are trivial.

The example $\mathsf{ImperfectRdn}$, taken from \cite{Griffith2015}, demonstrates that moving from trivial common knowledge to $\pi_Y$ resolves full uncertainty about $X$, since $\pi_Y$ already discriminates the same sets as $\pi_{Y,Z}$, whereas the $\pi_Z$-routing reduces uncertainty in two steps, because observing $Z$ does not fully fix $X$ ($Z=1$ reduces the probability of the outcome $X=0$ but still allows it to happen).

Table \ref{tab:pid_comparison_main} also contains $\mathsf{ReducedCopy}$, which we created as a simple example showing path dependence, by removing the last instance from $\mathsf{Copy}$. As can be seen from the distributions and corresponding partitions, here the two paths from common knowledge $\pi_Y\wedge \pi_Z$ to combined knowledge $\pi_Y\vee \pi_Z$ via $\pi_Y$ and $\pi_Z$ are symmetric, so that $I_u(X;Y{\setminus}Z)=I_u(X;Z{\setminus}Y)$ and $I_{cu}(X;Y|Z)=I_{cu}(X;Z|Y)$. However, the unique and conditional unique contributions are not the same,
\[
I_u(X;Y{\setminus}Z) \neq I_{cu}(X;Y|Z).
\]
This example illustrates that while the knowledge \textit{content} of $Y$, represented by $\pi_Y$, is independent of getting to know $Z$ before or after observing $Y$, the order does make a difference when \textit{measuring} this content through a specified loss function.

We created the third example in Table \ref{tab:pid_comparison_main}, $\mathsf{ExtendedXor}$, as an instance of all four terms being non-zero and the unique contributions all being different. As a side note, since $I_\mathrm{dep}$ and the Gomes-Figueiredo decomposition happen to agree in all examples we found in the literature, even though their constructions are very different, we created the example $\mathsf{NoisyRdn}$, shown in Table \ref{tab:pid_comparison_appendix3}, in the appendix by a random search producing an actual difference.

\subsection{Continuous examples} \label{sec:continuousexamples}

The gates of Section~\ref{sec:pid_examples} are finite, but nothing in Proposition~\ref{prop:redundancy-synergy} or Corollary~\ref{cor:PID} relies on this. The same decompositions hold for continuous variables once the lattice operations are read on the $\sigma$-algebras induced by measurable partitions (Definition~\ref{def:knowledgelattice2}), modulo $\mathbb P$-null sets. We describe the three examples of Table~\ref{tab:pid_comparison_continuous} below and defer the derivations to Appendix~\ref{app:continuous}. In every case the partial information terms are finite, whereas the remaining uncertainty $H(X|Y,Z)$ is finite for the square error and infinite for the log loss (cf.~Example~\ref{ex:logloss}).

For a joint Gaussian ($\mathsf{JointGaussian}$ in Table \ref{tab:pid_comparison_continuous}), we show in Appendix \ref{app:continuous} that $\sigma(\pi_Y)=\sigma(Y)$, $\sigma(\pi_Z)=\sigma(Z)$, the meet is $\{\emptyset, \Omega\}$, and the join is
$\sigma(Y,Z)$. Hence all information
is complementary, i.e., $I_r(X;(Y,Z))=I_s(X;(Y,Z))=0$, regardless of the chosen loss. In Table \ref{tab:pid_comparison_continuous} we compare our decomposition for the square error with the classical commonality analysis \cite{Spurrell1967,Mood1971}, where, similarly to some of the finite examples, we can see that our $\pi_Y$- and $\pi_Z$-routings each share one term of the decomposition given by commonality analysis. For the log loss, we compare with the minimal-mutual-information redundancy $\min\{I_S(X;Y),I_S(X;Z)\}$ of Barrett \cite{Barrett2015}, which has non-zero redundancy and synergy for this example, and importantly, there is no unique contribution by $Z$, even though the example is symmetric in $Y$ and $Z$ except for the different correlation values. In contrast, our decomposition shows a relatively weaker unique contribution of $Z$ compared to $Y$ in both routings, properly reflecting the correlation differences.

In the example $\mathsf{RdnGaussian}$, the Gaussian sources $Y$ and $Z$ additionally share a discrete bit $W$, and $X$ depends on their sum as well as on Gaussian noise $N$ seen by neither source. We show in Appendix \ref{app:continuous} (Example \ref{ex:gaussrdn}) that $\sigma(\pi_Y) = \sigma(W,V_1)$, $\sigma(\pi_Z)=\sigma(W,V_2)$, $\sigma(\pi_{Y,Z}) = \sigma(W,\,V_1{+}V_2)$, the meet is $\sigma(\pi_Y)\wedge\sigma(\pi_Z) = \sigma(W)$
modulo null sets, and the join is $\sigma(W,V_1)\vee\sigma(W,V_2) = \sigma(W,V_1,V_2)$. In particular, there is no synergy, redundancy is given by the shared bit ($\mathbb V(W)$ for squared error, $1$ bit for the log loss), and the unique contributions for the square error are the variances of the Gaussian parts of the sources, agreeing with commonality analysis. The PID based on MMI \cite{Barrett2015} overcounts redundancy ($1.292$ bits) and contains synergy, but has no unique terms.  Note that replacing $W$ by a continuous Gaussian leaves the lattice unchanged and keeps redundancy $\mathbb V(W)$ finite under the
square error, but sends $I_r(X;(Y,Z))$ to infinity under the log loss.

The example $\mathsf{SignXor}$, $X=Y\,\mathrm{sign}(Z)+N$, is a continuous
analogue of $\mathsf{Xor}$. As we show in Example \ref{ex:signxor} in Appendix \ref{app:continuous}, $\sigma(\pi_Z)$ is trivial, $\sigma(\pi_Y)=\sigma(|Y|)=\sigma(\pi_Y)\vee \sigma(\pi_Z)$, and $\sigma(\pi_{Y,Z})=\sigma(Y\,\mathrm{sign}(Z))$.
Under the square error it is purely synergistic, whereas under the log loss the unique term carries the continuous magnitude $|Y|$ and the synergy carries the sign. Here MMI coincides with our values, whereas commonality analysis has two unique contributions but negative redundancy.

\begin{table*}[t]
\centering
\caption{Partial information decomposition for the continuous examples of Section~\ref{sec:continuousexamples}.\label{tab:pid_comparison_continuous}}

\begin{tabular}{llrrrrrr}
\toprule
Example & Measure & $I_r$ & $U_Y$ & $U_Z$ & $I_s$ & $I$ & $H(X|Y,Z)$ \\
\midrule
\multirow{6}{*}{\begin{tabular}{@{}c@{}}\textbf{JointGaussian}\\[2pt]{\footnotesize\parbox{4.1cm}{\centering $(X,Y,Z)$ jointly Gaussian, $\rho_Y=0.8$, $\rho_Z=0.6$, $\rho_{YZ}=0.4$}}\end{tabular}} & \textbf{Ours, sq.\ error ($\pi_Y$)} & \textbf{0.000} & \textbf{0.640} & \textbf{0.093} & \textbf{0.000} & \multirow{3}{*}{0.733} & \multirow{3}{*}{0.267} \\
 & \textcolor[gray]{0.5}{\textbf{Ours, sq.\ error ($\pi_Z$)}} & \textcolor[gray]{0.5}{\textbf{0.000}} & \textcolor[gray]{0.5}{\textbf{0.373}} & \textcolor[gray]{0.5}{\textbf{0.360}} & \textcolor[gray]{0.5}{\textbf{0.000}} &  &  \\
 & Commonality analysis & 0.267 & \textcolor[gray]{0.5}{\textbf{0.373}} & \textbf{0.093} & $-$ &  &  \\
\addlinespace[3pt]
 & \textbf{Ours, log loss ($\pi_Y$)} & \textbf{0.000} & \textbf{0.737} & \textbf{0.216} & \textbf{0.000} & \multirow{3}{*}{0.953} & \multirow{3}{*}{$\infty$} \\
 & \textcolor[gray]{0.5}{\textbf{Ours, log loss ($\pi_Z$)}} & \textcolor[gray]{0.5}{\textbf{0.000}} & \textcolor[gray]{0.5}{\textbf{0.632}} & \textcolor[gray]{0.5}{\textbf{0.322}} & \textcolor[gray]{0.5}{\textbf{0.000}} &  &  \\
 & Barrett (MMI) & 0.322 & 0.415 & 0.000 & 0.216 &  &  \\
\midrule
\multirow{6}{*}{\begin{tabular}{@{}c@{}}\textbf{RdnGaussian}\\[2pt]{\footnotesize\parbox{4.1cm}{\centering $Y{=}(W,V_1)$, $Z{=}(W,V_2)$, $X{=}(W,V_1{+}V_2{+}N)$; $W\sim\mathrm{Bern}(\tfrac12)$, $V_i, N \sim \mathcal N(0,1)$}}\end{tabular}} & \textbf{Ours, sq.\ error ($\pi_Y$)} & \textbf{0.250} & \textbf{1.000} & \textbf{1.000} & \textbf{0.000} & \multirow{3}{*}{2.250} & \multirow{3}{*}{1.000} \\
 & \textcolor[gray]{0.5}{\textbf{Ours, sq.\ error ($\pi_Z$)}} & \textcolor[gray]{0.5}{\textbf{0.250}} & \textcolor[gray]{0.5}{\textbf{1.000}} & \textcolor[gray]{0.5}{\textbf{1.000}} & \textcolor[gray]{0.5}{\textbf{0.000}} &  &  \\
 & Commonality analysis & \textbf{0.250} & \textbf{1.000} & \textbf{1.000} & $-$ &  &  \\
\addlinespace[3pt]
 & \textbf{Ours, log loss ($\pi_Y$)} & \textbf{1.000} & \textbf{0.292} & \textbf{0.500} & \textbf{0.000} & \multirow{3}{*}{1.792} & \multirow{3}{*}{$\infty$} \\
 & \textcolor[gray]{0.5}{\textbf{Ours, log loss ($\pi_Z$)}} & \textcolor[gray]{0.5}{\textbf{1.000}} & \textcolor[gray]{0.5}{\textbf{0.500}} & \textcolor[gray]{0.5}{\textbf{0.292}} & \textcolor[gray]{0.5}{\textbf{0.000}} &  &  \\
 & Barrett (MMI) & 1.292 & 0.000 & 0.000 & 0.500 &  &  \\
\midrule
\multirow{6}{*}{\begin{tabular}{@{}c@{}}\textbf{SignXor}\\[2pt]{\footnotesize\parbox{4.1cm}{\centering $X = Y\,\mathrm{sign}(Z) + N$; $Y,Z\sim\mathcal N(0,1)$, $\sigma_N^2{=}0.5$}}\end{tabular}} & \textbf{Ours, sq.\ error ($\pi_Y$)} & \textbf{0.000} & \textbf{0.000} & \textbf{0.000} & \textbf{1.000} & \multirow{3}{*}{1.000} & \multirow{3}{*}{0.500} \\
 & \textcolor[gray]{0.5}{\textbf{Ours, sq.\ error ($\pi_Z$)}} & \textcolor[gray]{0.5}{\textbf{0.000}} & \textcolor[gray]{0.5}{\textbf{0.000}} & \textcolor[gray]{0.5}{\textbf{0.000}} & \textcolor[gray]{0.5}{\textbf{1.000}} &  &  \\
 & Commonality analysis & -1.000 & 1.000 & 1.000 & $-$ &  &  \\
\addlinespace[3pt]
 & \textbf{Ours, log loss ($\pi_Y$)} & \textbf{0.000} & \textbf{0.318} & \textbf{0.000} & \textbf{0.474} & \multirow{3}{*}{0.792} & \multirow{3}{*}{$\infty$} \\
 & \textcolor[gray]{0.5}{\textbf{Ours, log loss ($\pi_Z$)}} & \textcolor[gray]{0.5}{\textbf{0.000}} & \textcolor[gray]{0.5}{\textbf{0.318}} & \textcolor[gray]{0.5}{\textbf{0.000}} & \textcolor[gray]{0.5}{\textbf{0.474}} &  &  \\
 & Barrett (MMI) & \textbf{0.000} & \textbf{0.318} & \textbf{0.000} & \textbf{0.474} &  &  \\
\bottomrule
\end{tabular}
\end{table*}

\section{Conclusion}
\label{sec:conclusion}

The concept of information plays an important role throughout science and engineering, yet it is often implicitly identified with Shannon entropy or mutual information. Whereas these quantities originate from optimal coding with respect to message length, their broader use as generic uncertainty measures can obscure the distinction between informational content, understood as \textit{what} is known, and information as a scalar quantity, understood as how knowledge is \textit{evaluated}. In this work, we made this distinction explicit and provided a principled separation between informational content and its numerical assessment. In particular, our goal is not to argue against the use of Shannon information, but rather to expose the general knowledge lattice structure that underlies informational quantities, from which, for example, principled decompositions of information follow naturally.

More precisely, we identified a general structural description of information and entropy in terms of transitions between knowledge states and their quantification through loss-based optimization. Knowledge is modeled through partitions of the underlying probability space, or equivalently by the corresponding $\sigma$-algebras of knowable sets, forming a knowledge lattice ordered by refinement. Scalar informational quantities are obtained once a loss function is specified, which enables us to measure the value of knowledge states via their achievable risks. Accordingly, entropy and information are unified as instances of risk reduction relative to different states of knowledge. The idea of quantifying knowledge through entropy reduction also appears in applied settings, albeit with the conceptual order reversed, taking entropy as primitive rather than the knowledge states \cite{Wang2023a}.

While many well-known information and uncertainty measures fit into the risk-based structure adopted here, such as those induced by Bregman losses (including Shannon information and variance) and proper scoring rules (e.g., Tsallis and Brier scores), it does not include all generalized information measures. For example, Fisher information quantifies parametric sensitivity rather than uncertainty reduction between knowledge states. Moreover, for orders $\alpha>1$, R\'enyi entropy generally fails to be a risk difference (cf. Example \ref{sec:transformationcosts} in Appendix~\ref{app:examples}). A natural way to generalize our approach  would be to replace risk differences by divergences between the conditional laws directly, such as $f$-divergences \cite{Csiszar1972}. The knowledge lattice would persist, but the chain rule would no longer come for free.

In continuous settings, the possibility of unbounded uncertainty reduction is due to the use of pointwise loss functions defined on continuous action spaces. This is in alignment with the divergence of Shannon entropy under increasingly fine discretizations. However, since information is not defined as a difference of entropies in our approach, cancellations of infinities are not required to obtain finite informational quantities.

A key application of the knowledge lattice-based formulation is the decomposition of informational quantities. Taking knowledge states rather than information values as the objects, a partial information decomposition is no longer a matter of choosing one term and solving for the rest. The three familiar terms are three transitions on the knowledge lattice of the distribution partitions $\pi_{Y_i}=\Pi(\mathbb P_{X|Y_i})$: moving from $\bot$ to their meet, i.e., Aumann's common knowledge about $X$, is redundancy, the gap between their join and $\pi_{Y_1,\dots,Y_n}$ (which is absent for observation partitions) is synergy, and passing through the individual $\pi_{Y_i}$ results in unique contributions, which therefore depend on the chosen route, i.e., on which observations $Y_i$ are considered first. Like entropy and information themselves, each term is a risk difference between lattice nodes.

Treating unique information as a state function, as most current PID approaches do through \eqref{def:pid1}–\eqref{def:pid2} (with \cite{Kolchinsky2022} as an exception) leads to widely disagreeing decompositions, especially for non-textbook examples. The reported values scatter and several proposals assign negative contributions, whereas our routings remain non-negative throughout and often reproduce, term by term, single values of those measures that happen to align with one of our routing orders.

Moreover, the continuous examples of Section~\ref{sec:continuousexamples} illustrate that the decomposition terms are finite as soon as the risks at the nodes of the chosen path are finite, whereas any uncertainty that can only be resolved by a continuous quantity is isolated in the residual term $H(X|Y,Z)$, which can diverge without affecting the decomposition itself.

Several lattice-based approaches in the PID literature, most prominently the redundancy lattice \cite{Williams2010} and more recent parthood lattices \cite{Gutknecht2021}, organize decomposition terms under the assumption that informational quantities themselves admit part-whole relationships. These constructions are therefore combinatorial rather than epistemic in nature, serving primarily to identify and index terms rather than to capture relationships between knowledge states. A notable exception is the constraint lattice of \cite{Ay2020}, whose elements represent probabilistic structures rather than scalar quantities, making it potentially closer in spirit to the knowledge lattice introduced here, but without distinguishing redundant and unique information. Closest to our construction is the intersection information of Griffith \textit{et al.}\ \cite{Griffith2014a}, which is defined on the information lattice of Shannon \cite{Shannon1953} and of Li and Chong \cite{Li2011}, where random variables are ordered by almost sure functional dependence, and takes redundancy to be the Shannon information carried by the meet of the sources, their Gács-Körner common random variable. The remaining terms are obtained in \cite{Griffith2014a} by \eqref{def:pid1}--\eqref{def:pid2}, which is where the negative synergies in the tables originate. Closely related is also the decision-theoretic construction of Banerjee \textit{et al.}\ \cite{Banerjee2018}, but the difference lies again in the primitive object: their Le~Cam deficiencies compare channels pairwise, whereas we compare knowledge states on a shared lattice.

Investigating formal connections between these frameworks more closely is an interesting direction for future work. More broadly, our approach suggests that the foundations of information decomposition should be sought in the structure of probability spaces themselves, in the meets, joins, and further refinements of knowledge states, rather than in algebraic relationships among the scalar quantities they induce. Moreover, our approach is flexible, covers general loss functions, applies to discrete and continuous variables, and the non-negativity of the resulting information terms is guaranteed by the lattice structure, for every loss simultaneously. These properties may make it useful in fields such as deep learning, quantum information, and neuroscience, where information decompositions are in active demand.

\appendix

\subsection{Deferred proofs}
\label{prf:properties}
\noindent \textbf{Proof of Proposition \ref{prop:properties}.}
\noindent $(i)$ First of all, we show that
\begin{equation}\label{eq:Rx}
R(X) = \mathbb E\Big[\inf_{a\in\mathcal A}l(X,a)\Big].
\end{equation}

If we write $m(x)\coloneqq \inf_{a\in\mathcal A} l(x,a)\in[-\infty,\infty)$, then for $M>0$, the truncated loss $l_M(x,a)\coloneqq \max\{l(x,a),-M\}$ is jointly measurable and bounded below, and
\[
\inf_{a\in\mathcal A} l_M(x,a) = \max\{m(x),-M\}
\]
Since $\mathcal A$ and $\mathcal X$ are standard Borel, by the Jankov-von Neumann theorem \cite{Bertsekas1978}, the map $x\mapsto \max\{m(x),-M\}$ is universally measurable, and for every $\varepsilon>0$ there exists a universally measurable selection $f_{M,\varepsilon}:\mathcal X\to\mathcal A$ with
\[
l(x,f_{M,\varepsilon}(x)) \leq l_M(x,f_{M,\varepsilon}(x)) \leq \max\{m(x),-M\} + \varepsilon
\]
for all $x\in\mathcal X$. In particular, $m$ is universally measurable under the pointwise limit $M\to\infty$, and $f_{M,\varepsilon}(X)$ is $\sigma(X)$-measurable up to $\mathbb P$-null sets, so that
\begin{align}\nonumber
\inf_{A\,\sigma(X)\text{-meas.}}\mathbb E[l(X,A)] & \leq \mathbb E[l(X,f_{M,\varepsilon}(X))]\\
& \leq \mathbb E\big[\max\{m(X),-M\}\big] + \varepsilon .\label{ineq:truncated}
\end{align}
If $\mathbb E[m(X)^+]=\infty$, then $\mathbb E[l(X,A)^+]=\infty$ for every $A$, because $l(x,a)\geq m(x)$ for all $a$, and by our convention both sides of \eqref{eq:Rx} equal $+\infty$. Otherwise, $\mathbb E[m(X)^+]\,{<}\,\infty$, and since $\max\{m(X),-M\}$ decreases pointwise to $m(X)$ as $M$ goes to $\infty$, monotone convergence shows that $\mathbb E[\max\{m(X),-M\}]$ converges to $\mathbb E[m(X)]\in[-\infty,\infty)$. Letting first $M\to\infty$ and then $\varepsilon\to 0$ in \eqref{ineq:truncated} yields $\inf_{A\,\sigma(X)\text{-meas.}}\mathbb E[l(X,A)]\leq \mathbb E[m(X)]$. The opposite inequality holds because $l(X,A)\geq m(X)$ pointwise for any $A$, and therefore $\mathbb E[l(X,A)]\geq \mathbb E[m(X)]$ in the extended sense. This proves \eqref{eq:Rx}.

Finally, we have
\[
\inf_{A\,\sigma(\pi)\text{-meas.}}\mathbb E[l(X,A)] \geq \mathbb E\Big[\inf_{a\in\mathcal A}l(X,a)\Big] \stackrel{\eqref{eq:Rx}}{=} R(X).
\]

\noindent $(ii)$ If $\pi\preceq \pi'$, then $\sigma(\pi)\subseteq \sigma(\pi')$, and so if $A$ is $\sigma(\pi)$-measurable, it is also $\sigma(\pi')$-measurable, i.e., $R(\pi')\leq R(\pi)$ and thus $U_{\pi\to\pi'}(X)\geq 0$.

\noindent $(iii)$ If $\mathbb P_X = \delta_x$ for some $x\in \mathcal X$, then $\mathbb E[l(X,a)] = l(x,a)$ and therefore $R(\bot) = R(X)$, i.e., $H(X)=0$.

\noindent $(iv)$ If $X$ and $Y$ are independent and $A$ is $\sigma(Y)$-measurable, then, by the Doob-Dynkin Lemma,
\[\mathbb E[l(X,A)]=\mathbb E_{\mathbb P_Y}[L(\mathbb P_X,f(Y))]
\]
for some measurable function $f$ and $L(p,a)\coloneqq \mathbb E_{p}[l(X,a)]$. It follows that
\[
\inf_{A\,\sigma(Y)\text{-meas.}} \mathbb E[l(X,A)] = \inf_a L(\mathbb P_X,a) = \inf_{a\in\mathcal A} \mathbb E[l(X,a)]
\]
and therefore $I(X;Y) = 0$.

\noindent $(v)$ By definition, $I(X;(Y,Z)) = U_{\bot \to \Pi(Y,Z)}$ and since $\pi_0=\bot$, $\pi_1=\Pi(Y)$, $\pi_2=\Pi(Y,Z)$ defines a sequence with non-increasing risks, $R(\bot)\geq R(Y) \geq R(Y,Z)$, the chain rule for $I(X;(Y,Z))$ follows from the general chain rule \eqref{eq:generalchainrule}. The latter is just a telescope sum of terms of the form $R(\pi_{i})-R(\pi_{i+1})$, and each term is non-negative by assumption of $i\mapsto R(\pi_i)$ being monotonically decreasing.

\bigskip
\noindent\textbf{Proof of Lemma \ref{lma:synergygap}.} It follows from the example section that the two partitions $\pi_Y\vee \pi_Z$ and $\pi_{Y,Z}$ are generally not comparable (see, e.g., $\mathsf{ExtendedXor}$ in Table \ref{tab:pid_comparison_main}). In order to see \eqref{eq:synergygap}, we note that for arbitrary $n\in\mathbb N$, by Lemma \ref{prop:problattice}, $R(\pi_{Y_1,\dots,Y_n}) = R(Y_1,\dots,Y_n)$, so it is enough to prove that
\[
R(Y_1,\dots,Y_n) \leq R\Big(\bigvee\nolimits_{i=1}^n \pi_{Y_i}\Big).
\]

First, since $\mathcal X$ is standard Borel, for each $i$ the mapping $\omega\mapsto\mathbb P_{X|Y_i=Y_i(\omega)}$ can be chosen $\sigma(Y_i)$-measurable and such that $B\mapsto\mathbb P_{X| Y_i=Y_i(\omega)}(B)$ is a probability measure for a.e. $\omega$, giving a regular conditional distribution. In particular, by the $\sigma(Y_i)$-measurability, $\sigma(\mathbb P_{X|Y_1},\dots,\mathbb P_{X|Y_n})\subseteq \sigma(Y_1,\dots,Y_n)$, and so the inequality follows from
\[\sigma(\mathbb P_{X|Y_1},\dots,\mathbb P_{X|Y_n}) = \sigma(\bigvee\nolimits_i\pi_{Y_i}).\]
This, however, follows by definition, since for any random variables $S$ and $T$, the smallest $\sigma$-algebra $\sigma(S,T)$ such that both $S$ and $T$ are measurable is the same as the smallest $\sigma$-algebra containing $\sigma(S)\cup\sigma(T)$, which by definition (cf.~Section \ref{sec:sigmaAlgebraOfKnowableSets}) is $\sigma(S)\vee \sigma(T)=\sigma(\Pi(S)\vee \Pi(T))$. Applied to the regular conditional probabilities $\mathbb P_{X|Y_i}$ this finishes the proof.

We note that, by the lattice equivalence discussed in Section \ref{sec:sigmaAlgebraOfKnowableSets}, this also shows that
\[
\bigvee\nolimits_i \pi_{Y_i} \preceq \bigvee\nolimits_i \Pi(Y_i) = \Pi(Y_1,\dots Y_n).
\]

\bigskip
\label{app:nonnegPID}
\noindent\textbf{Proof of Proposition \ref{prop:redundancy-synergy}.} The decomposition is a telescope sum and thus follows from the chain rule, once we have seen that the sequence of risks is non-increasing. Redundant and complementary information are trivially non-negative since the corresponding partitions are an ordered chain with respect to refinement,
\[
\bot \ \preceq \ \bigwedge\nolimits_{i=1}^n \pi_{Y_i} \ \preceq \ \bigvee\nolimits_{i=1}^n \pi_{Y_i} \, ,
\]
and by the proof of Lemma \ref{lma:synergygap} synergistic information is non-negative in the general case as well.

\subsection{Information for specific losses}\label{app:examples}

In the following, we work out the resulting notions of uncertainty and information for a range of commonly used loss functions. For square error loss, uncertainty reduces to variance and information to explained variance, recovering classical $L^2$ decompositions. For log loss, our definitions recover Shannon entropy and mutual information in the discrete case, whereas in the continuous case entropy becomes infinite but information remains well-defined in terms of expected KL divergences. More generally, for Bregman losses, information coincides with Bregman information (or Jensen gaps), linking our framework to bias-variance decompositions and prediction uncertainty. Finally, for arbitrary losses admitting Bayes acts, information admits an interpretation as the cost of transforming distributions, generalizing the familiar relationship between Shannon information and KL divergence. These examples illustrate that the proposed theory both subsumes classical information measures and extends naturally to objectives beyond coding.

\begin{example}[Square error]\label{ex:squareerror} The \textit{square error} on the real line $\mathcal X = \mathbb R$ is the loss function $l$ defined on $\mathbb R^2$ (here $\mathcal A = \mathcal X$) given by
$$
l(x,a) = (x-a)^2.
$$
The conditional expectation $\mathbb E[X|\mathcal F]$ is the best $\mathcal F$-measurable predictor of $X$ with respect to the square error $l$ \cite{Rockafellar1970} (see also \eqref{eq:Banerjee}).
In particular, we obtain
\begin{align*}
H(X)&= \mathbb E[(X-\mathbb E[X])^2] = \mathbb V(X), \\
H(X|Y)&= \mathbb E[(X-\mathbb E[X|Y])^2] = \mathbb E[\mathbb V(X|Y)],\\
I(X;Y)&= \mathbb V(X)-\mathbb E[\mathbb V(X|Y)] = \mathbb V(\mathbb E[X|Y]),\\
I(X;Y|Z)&= \mathbb E_Z[\mathbb V_Y(\mathbb E[X|Y,Z])].
\end{align*}
The variance $\mathbb V(X)$ of a random variable $X$ thus corresponds to its total uncertainty measured with respect to the square error. Conditional variance $\mathbb V(X|Y)$ is the total uncertainty left when $Y$ is known, whereas $I(X;Y)$ is the variance of the random variable $\mathbb E[X|Y]$ resulting from compressing $X$ according to the knowledge in $\sigma(Y)$. Conditional information is a combination of these two when another source of knowledge is taken into account. The analogous results for the square error on $\mathbb R^d$ follow along the same lines, replacing $(\cdot)^2$ by $\|\cdot\|^2$.

Similar to how Shannon entropy is the minimal expected codeword length, variance is the optimal square Euclidean distance between $X$ and a single number representing $X$ under no knowledge, capturing the variability of $X$. Conditional variance is an important tool in econometrics, especially in autoregressive conditional heteroskedasticity (ARCH) models \cite{Bollerslev1986}.

\end{example}

\begin{example}[Log loss]\label{ex:logloss} For the discrete case, let $\mathcal A=\Delta$ be the $d$-simplex of discrete probability distributions \[q=(q(x_1),\dots,q(x_d))\]
on a finite space $\mathcal X$ with $d$ elements. The \textit{log loss} defined on $\mathcal X\times\mathcal A$ is given by
\[l(x,q) = -\log q(x).\]
It is well-known that the log loss is a so-called \textit{proper} loss, justifying its use in probability estimation, since on average $-\log q(X)$ is minimal at the distribution $q$ over which the average is taken. In particular, for any two probability distributions $p,q\in\mathcal A$, we have
\[
\mathbb E_{X\sim p}[-\log q(X)]\geq \mathbb E_{X\sim p}[-\log p(X)].
\] Moreover, there is no residual risk for the log loss, as $\argmin_{Q\,\sigma(X)\text{-meas.}}\mathbb E[-\log Q(X)] = \delta_X$, where $\delta_x$ denotes the Dirac measure centered at $x\in\mathcal X$. We note that here and in the optimization problems over $\sigma(Y)$-measurable random variables $A$ appearing in the expressions for $H(X|Y)$ and $I(X;Y)$, the optimization is over random probability measures $Q$. In total, we obtain
\begin{align*}
H(X) &= \mathbb E[-\log \mathbb P_X] = H_S(X),  \\
H(X|Y) &= \mathbb E[-\log \mathbb P_{X|Y}] = H_S(X|Y),\\
I(X;Y) & = \mathbb E[\log \mathbb P_{X|Y} - \log \mathbb P_X] = I_S(X;Y),\\
I(X;Y|Z) & = \mathbb E[\log \mathbb P_{X|Y,Z} - \log \mathbb P_{X|Z}] = I_S(X;Y|Z)\, ,
\end{align*}
where we suppress the arguments of the conditional distributions inside the expectations for brevity. This recovers non-conditional and conditional Shannon entropy and Shannon information.

For the continuous case, let $\lambda$ be a fixed $\sigma$-finite measure on a measurable space $\mathcal X$, e.g., the Lebesgue measure on the real line $\mathbb R$. If we let $\mathcal A$ represent all probability densities $f$ on $\mathcal X$ with respect to $\lambda$, then the log loss on $\mathcal X\times \mathcal A$ is defined by
\[l(x,f) = - \log f(x)
\]
for $\lambda$-almost every $x\in\mathcal X$ and $f\in\mathcal A$. Note that each $f\in\mathcal A$ corresponds to a probability measure on $\mathcal X$ that is absolutely continuous with respect to $\lambda$ and uniquely determined by $f$ up to a set of measure zero. Moreover, as required by our standing assumptions, $\mathcal A$ is standard Borel because it is a closed subspace of the Polish space $L^1(\mathbb R, \lambda)$. As can be seen by Jensen's inequality, the log loss is proper also in the continuous case \cite{Gruenwald2004}, in particular, $\inf_{G\,\{\emptyset,\Omega\}\text{-meas.}}\mathbb E_f[-\log G(X)] = \mathbb E_f[-\log f(X)]$, where we write $\mathbb E_f$ for the expectation with respect to the measure $fd\lambda$. This expression is known as the \textit{continuous} or \textit{differential entropy} $h(X)$, which is \textit{not} a direct generalization of Shannon entropy. In particular, when taking the limit of smaller and smaller discretizations of the real line, Shannon entropy does not converge to $h(X)$. It becomes infinitely large, corresponding to the fact that one would have to use infinitely long binary sequences in order to encode each $X\,{=}\,x$ \textit{precisely}. The minimization over $\sigma(X)$-measurable $A$ in our definition of $H(X)$ leads to $H(X) = \infty$ simply because $\inf_{G\,\sigma(X)\text{-meas.}}\mathbb E_f[-\log G(X)] = -\infty$, as we can find a sequence $G_n$ of density-valued functions of $x$ that take larger and larger values at $x$ (so that $G_n \lambda$ are approximations to the Dirac measure on $\mathcal X$ centered at $x$). More precisely, we can choose $\sigma(X)$-measurable $G_n^x$ to be a Gaussian density with mean $x$ whose variance converges to zero as $n\to\infty$, for example \smash{$G_n^x(\xi) = \frac{n}{\sqrt{\pi}} e^{-n^2 (\xi-x)^2}$}. Note that the mapping $\mathbb R\to L^1(\mathbb R), x\mapsto G_n^x$ is $L^1$-continuous (a translation of the fixed density $G_n^0$) and therefore Borel, so that its composition with $X$ is $\sigma(X)$-measurable. We have
\[
\inf_{A\,\sigma(X)\text{-meas.}} \mathbb E_f[l(X,A)] \leq \mathbb E_f[l(X,G_n^X)] = -\log \frac{n}{\sqrt{\pi}}\, ,
\]
in particular, $\inf_{A\,\sigma(X)\text{-meas.}} \mathbb E_f[l(X,A)] = -\infty$.

While entropy is infinite, for (unconditional and conditional) information in $Y$ about $X$, where uncertainty is not reduced fully but only to the knowledge level given by $Y$, we obtain results analogous to the discrete case. In summary, we have
\begin{align*}
H(X) &= \infty,  \\
H(X|Y) &= \infty,\\
I(X;Y) & = \mathbb E[D_\mathrm{KL}(f_{X|Y}\|f_X)],\\
I(X;Y|Z) & = \mathbb E[D_\mathrm{KL}(f_{X|Y,Z}\|f_{X|Z})]\, ,
\end{align*}
given that the corresponding (conditional) densities exist and have finite differential entropy, i.e., $\mathbb E[|\log f_X(X)|]<\infty$ and $\mathbb E[|\log f_{X|Y}(X|Y)|]<\infty$ for $I(X;Y)$, and analogously for $I(X;Y|Z)$, so that the risks involved are finite.

Reflecting the fact that we can store an arbitrary amount of information in a continuous random variable, $H(X)$ and $H(X|Y)$ are infinite. In particular, we did not introduce any constraint on the capability of observing the values of $X$, as one would do in practice, where one might, for example, only allow a maximum length of the floating point representations.
\end{example}

\begin{example}[Hyvärinen loss]\label{ex:hyvaerinen} Utilizing both the square error and the log loss, the Hyvärinen objective
\begin{equation} \label{def:hyvaerinenobj}
J(g) = \tfrac{1}{2}\,\mathbb E_f \big[\|\nabla \log g(X) - \nabla \log f(X)\|^2 \big]
\end{equation}
corresponds to matching the so-called \textit{scores} $\nabla \log g$ and $\nabla \log f$ of densities $g$ and $f$ satisfying appropriate regularity conditions. Note that $J(g)$ should not be confused with \textit{Fisher information} of a parametric family, which has a similar functional form, but takes gradients with respect to density parameters rather than the random variables over which the densities are defined. Introduced originally as a loss function for statistical inference \cite{Hyvarinen2005}, the Hyvärinen loss has influenced various approaches to score matching in the machine learning literature, in particular for training energy-based models (EBMs), for example for image denoising and super-resolution \cite{Kingma2010}, and generative modeling in general \cite{Vincent2011,Wang2023}. One of its appealing properties is the invariance of $\nabla \log f$ under multiplicative rescaling of $f$, enabling the use of unnormalized density approximations in the objective function $J$ without having to determine the normalization constants. Moreover, as can be shown by using integration by parts \cite{Hyvarinen2005}, the objective \eqref{def:hyvaerinenobj} can be rewritten as $\mathbb E_f[l(X,g)]$, with the \textit{Hyvärinen loss}
\begin{equation} \label{def:hyvaerinenloss}
l(X,g) = \tfrac{1}{2}\|\nabla \log g\|^2 + \Delta \log g \, .
\end{equation}
Similarly to the log loss, uncertainty reduction from no or some knowledge to full knowledge about $X$ is infinite, because $l(x,g)$ can be made arbitrarily negative for each $x$ separately, by using test functions $g$ with increasing curvature. More precisely, for the Gaussian $G_n^x$ defined in the previous example for $x\in\mathbb R$ and $n\in\mathbb N$, we have
\[
\Delta_\xi \log G_n^x(\xi)\Big|_{\xi=x} = - 2n^2 \quad \forall x\in\mathbb R,
\]
so that also $\inf_{A\,\sigma(X)\text{-meas.}} \mathbb E_f[l(X,A)] = -\infty$. In summary,
\begin{align*}
H(X) &= \infty,  \\
H(X|Y) &= \infty,\\
I(X;Y) & =  \tfrac{1}{2}\, \mathbb E[ \|\nabla \log f_{X|Y} - \nabla \log f_{X} \|^2 ],\\
I(X;Y|Z) & = \tfrac{1}{2}\, \mathbb E[\|\nabla \log f_{X|Y,Z} - \nabla \log f_{X|Z} \|^2]\, ,
\end{align*}
if the corresponding densities have the required regularity and integrability properties. Hence, under the Hyvärinen loss, information $I(X;Y)$ equals the expected reduction in square error when estimating the score function of $X$. Intuitively, it quantifies how much observing $Y$ improves our knowledge of the local geometry of the log-density of $X$, in the sense that it captures the change in direction of the steepest increase of $\log$-density of $X$ from no knowledge to the knowledge of $Y$.

As we have seen, theoretically there is an arbitrary amount of information with respect to the log loss and the Hyvärinen loss in a continuous random variable. In both cases, the reason is that the loss function depends pointwise on the density or its derivatives \textit{and} the action space consists of arbitrary densities, corresponding to perfectly fine-grained control. For example, if we were to restrict the action space to Gaussian distributions with a positive lower bound on the standard deviations, then we could not exploit this pointwise dependence to blow up the loss when $X$ is known perfectly. Despite the theoretical limits of $H(X)$ and $H(X|Y)$ being infinite, it is also clear from our definitions that $I(X;Y)$ and $I(X;Y|Z)$ can be finite, because they are not \textit{defined} through entropy differences of the form $H(X)-H(X|Y)$ (where infinities would have to `cancel out') but through uncertainty reduction from no or partial information to partial information.
\end{example}

\begin{example}[Bregman losses]\label{ex:bregman} Let $\mathcal X$ be (a convex subset of) a Euclidean space. \textit{Bregman losses} (or \textit{Bregman divergences}) \cite{Bregman1967,Csiszar1991} are a large class of loss functions of the form
\begin{equation}\label{eq:bregmandivergence}
l(x,a) = d_\phi(x,a) \coloneqq \phi(x)-\phi(a)-\langle\nabla \phi(a), x-a\rangle \, ,
\end{equation}
where $\phi:\mathcal X\to \mathbb R$ is a given convex and differentiable function on $\mathcal X$, and $\langle \cdot,\cdot\rangle$ denotes the Euclidean inner product on $\mathcal X$. Bregman losses can also be defined on function spaces where ordinary derivatives are not defined, for example by making use of Fréchet derivatives \cite{Frigyik2008}, or by using so-called subgradients $\phi'$ of the convex functions $\phi$ \cite{Ovcharov2018}, which are (generally non-unique) mappings $\phi'$ from the underlying space to its algebraic dual such that $\phi(x)\geq \langle \phi'(y), x-y\rangle + \phi(y)$, where $\langle \cdot,\cdot\rangle$ denotes the corresponding dual pairing, with the dual element (functional, gradient, or subgradient) in the first argument and the primal element in the second. For example, if the underlying space consists of probability measures $p$, then one can use integrable functions as dual and the dual pairing is given by $\langle f,p\rangle = \int f dp$. The inner product in \eqref{eq:bregmandivergence} is then replaced by the dual pairing and the gradient by a subgradient, see Equation \eqref{def:bregman_divergence} below.

Consider the general shape of a (functional or non-functional) Bregman divergence
\begin{equation}\label{def:bregman_divergence}
d_{\phi}(x,y) = \phi(x) - \phi(y) - \langle \phi'(y), x-y\rangle
\end{equation}
for a convex function $\phi$, a subgradient $\phi'$ (gradient in the non-functional case), and the dual pairing $\langle\cdot,\cdot\rangle$ between the underlying space $\mathcal X$ and its algebraic dual $\mathcal X'$. A simple calculation shows
\begin{equation}\label{eq:simpledecomp}
d_{\phi}(x,y)-d_{\phi}(x,z) = d_{\phi}(z,y) + \langle \phi'(z) {-} \phi'(y),x{-}z\rangle
\end{equation}
for arbitrary $x,y,z\in\mathcal X$, and thus, for $\mathcal F$-measurable random variables $Y$, we find
\begin{align}
\mathbb E[d_{\phi}(X,Y){-}d_{\phi}(X,\mathbb E[X|\mathcal F])|\mathcal F]
= d_{\phi}(\mathbb E[X|\mathcal F],Y),\label{eq:bregman_decomp}
\end{align}
if the dual pairing satisfies $\mathbb E[\langle \alpha,X\rangle] = \langle \alpha, \mathbb E[X]\rangle$ for any constant $\alpha\in\mathcal X'$, which holds under fairly general conditions, see e.g. \cite{Bochner1933}. More precisely, \eqref{eq:bregman_decomp} follows from \eqref{eq:simpledecomp} if
\begin{equation}\label{eq:requirement_bregman}
\mathbb E[\langle \phi'(Z),X\rangle|\mathcal F] = \langle \phi'(Z),\mathbb E[X|\mathcal F]\rangle,
\end{equation}
whenever $Z$ is $\mathcal F$-measurable. For example, for non-functional Bregman divergences, where the dual pairing is a Euclidean inner product, \eqref{eq:requirement_bregman} follows from $\mathbb E[Y X|\mathcal F] = Y \, \mathbb E[X|\mathcal F]$ whenever $Y$ is $\mathcal F$-measurable, which is used in \cite{Banerjee2005a} to derive \eqref{eq:bregmanchar} from \eqref{eq:bregman_decomp}. In \cite{Gruber2022}, the same calculation is used to show \eqref{eq:bregman_decomp} with $\mathcal F=\{\emptyset,\Omega\}$ for functional Bregman divergences on random measures.

Equation \eqref{eq:bregman_decomp} shows that the Bregman divergences from $X$ to any $\mathcal F$-measurable $Y$ can be decomposed into two terms consisting of the divergences from $X$ to $\mathbb E[X|\mathcal F]$ and $\mathbb E[X|\mathcal F]$ to $Y$,
\begin{equation}\label{eq:pyth}
\mathbb E[d_\phi(X,Y)] = \mathbb E[d_\phi(X,\mathbb E[X|\mathcal F])] + \mathbb E[d_\phi(\mathbb E[X|\mathcal F], Y)],
\end{equation}
a general version of the Pythagorean theorem \cite{Csiszar1991}. In particular, Bregman divergences (uniquely) keep the property of the square error, by which the divergence from $X$ to $Y$ can be written as the divergence of $X$ to its orthogonal projection $\mathbb E[X|\mathcal F]$ plus the divergence of the projection to the $\mathcal F$-measurable $Y$.

Many loss functions that appear in the literature can be represented as Bregman divergences, see for example Table 1 in \cite{Banerjee2005b} or the examples in \cite{Ovcharov2018}. The basic example of a Bregman loss is the square error, which is generated by $\phi(x) = \|x\|^2$. Bregman divergences are tied to conditional expectations, as they are the only loss functions \cite{Banerjee2005a} that fulfill
\begin{equation}\label{eq:bregmanchar}
\argmin_{A\,\mathcal F\text{-measurable}} \, \mathbb E[d_\phi(X,A)] = \mathbb E[X|\mathcal F]\, ,
\end{equation}
i.e., $\min_{A\,\mathcal F\text{-meas.}} \, \mathbb E[d_\phi(X,A)] = \mathbb E[d_\phi(X,\mathbb E[X|\mathcal F])]$, whenever $X$ and $\phi(X)$ are $\mathbb P$-integrable. Note that, by using the definition of $d_\phi$, this minimum can be further simplified to
\[
\mathbb E[d_\phi(X,\mathbb E[X|\mathcal F])] = \mathbb E[\phi(X)] - \mathbb E[\phi(\mathbb E[X|\mathcal F])] \,,
\]
in particular, the minimum is independent of the chosen subgradient in the functional case. Hence, if $X$ and $\phi(X)$ are integrable, we obtain
\begin{align*}H(X) & = \mathbb E[\phi(X)] - \phi(\mathbb E[X]) \eqqcolon H_\phi(X),\\
H(X|Y) &= \mathbb E[\phi(X)-\phi(\mathbb E[X|Y])] = \mathbb E[H_\phi(X|Y)],\\
I(X;Y) &= \mathbb E[\phi(\mathbb E[X|Y])] - \phi(\mathbb E[X]) = H_\phi(\mathbb E[X|Y]),\\
I(X;Y|Z) & = \mathbb E_Z[H_\phi(\mathbb E[X|Y,Z])],
\end{align*}
generalizing the corresponding expressions for the square error discussed in Example \ref{ex:squareerror}. In the expression for $I(X;Y|Z)$, the application of $H_\phi$ is only over the variable $Y$, so that the resulting expression still depends on $Z$. In the literature, $H_\phi$ is known as \textit{Bregman information} \cite{Banerjee2004} or \textit{Jensen gap}, since it is the difference between the two sides of Jensen's inequality. Bregman information has been shown recently to represent the variance term in a general version \cite{Gruber2022} of the well-known bias-variance trade-off \cite{Geman1992}. In particular, it has been applied to quantify the prediction uncertainty of neural networks trained with the corresponding scoring rule \cite{Serra2024}.

Besides the square error, another well-known Bregman loss is the KL divergence, $l(p,q)=D_\mathrm{KL}(p\|q)$ defined on $\Delta^2$, where $\Delta$ is the $(d{-}1)$-simplex of probability distributions on a finite set with $d$ elements. In general, the random elements $X$ and $A$ in this case are \textit{random distributions}. For example, if $X=\mathbb P_{Y|Z}$, which is a $\sigma(Z)$-measurable random distribution on the finite target space $\mathcal Y$, and $A=Q_Y$ is $\{\emptyset,\Omega\}$-measurable, i.e., it is a single distribution $q=(q(y_1),\dots,q(y_d))$, then by property \eqref{eq:bregmanchar},
\[\argmin_{Q_Y\,\{\emptyset,\Omega\}\text{-meas.}} \mathbb E[D_\mathrm{KL}(\mathbb P_{Y|Z}\|Q_Y)] = \mathbb E_{Z}[\mathbb P_{Y|Z}] =  \mathbb P_Y\, ,\]
which plays an important role in many areas closely related to information theory \cite{Cover2006}, such as the rate-distortion literature \cite{Rose1994}, the information bottleneck method \cite{Tishby1999ib}, and information-theoretic bounded rationality \cite{Genewein2015}. The minimum is the Shannon information, in particular, we have
\[
H_\mathrm{KL}(\mathbb P_{Y|Z})=\min_{Q_Y\,\{\emptyset,\Omega\}\text{-meas.}} \mathbb E[D_\mathrm{KL}(\mathbb P_{Y|Z}\|Q_Y)] = I_S(Y;Z) ,
\]
which means that the total information in $\sigma(Z)$ about $\mathbb P_{Y|Z}$ in terms of the KL loss equals the information in $\sigma(Z)$ about $Y$ in terms of the log loss. The KL divergence thus can be seen as measuring information in the space of probability distributions in such a way that it matches the corresponding information with respect to the log loss.

As we discuss in the following, this relation between the log loss and KL divergence generalizes to loss functions for which the corresponding infima are attained.
\end{example}

\begin{example}[Losses with Bayes acts] \label{sec:transformationcosts}
As shown in \cite{Harsha2010}, Shannon information $I_S(X;Y)$ quantifies the cost of transforming $\mathbb P_X$ into $\mathbb P_{X|Y}$ in terms of the average length of binary strings transmitted from a sender that knows $Y$ to a receiver that wants to generate samples from $\mathbb P_{X|Y}$. The interpretation of information as the cost of transforming distributions is not limited to Shannon information, as mentioned in Section \ref{sec:knowledge-about-X}, where we have seen that the risks $R(Y)$ depend on the random variables $Y$ only through their conditional probability measures $\mathbb P_{X|Y}$. For loss functions for which so-called \textit{Bayes acts} \cite{Ferguson1967} exist, this view becomes explicit, because entropy and information can be directly expressed as divergences between distributions, as we discuss in the following.

A simple example of such a loss function in the discrete case is the \textit{zero-one loss} \cite{Gruenwald2004}, $l_{01}(x,a) = 1 - \delta_{x,a}$, where $\mathcal A=\mathcal X$ and $\delta_{i,j}$ denotes the Kronecker delta. Since for any $a\in\mathcal A$, we have $\mathbb E[l_{01}(X,a)] = 1 - \mathbb P_X(a)$, it follows for any random variable $Y$ that
\[
\inf_{A\ \sigma(Y)\text{-meas.}} \mathbb E[l_{01}(X,A)] = 1 - \mathbb E[\mathbb P_{X|Y}(a(Y))],
\]
where $a(y) \coloneqq \mathrm{argmax}_a \mathbb P_{X|Y=y}(a)$ denotes the Bayes act with respect to $\mathbb P_{X|Y}$.

In general, for a probability measure $q$ on $\mathcal X$, a Bayes act $a_q$ is an optimal action with respect to the expected value of $l$, i.e., $a_q \in \argmin_{a\in\mathcal A}\mathbb E_q[l(X,a)]$. As we have seen in Examples \ref{ex:logloss} and \ref{ex:hyvaerinen}, not all loss functions and sub-$\sigma$-algebras have corresponding Bayes acts. Thus, in addition to our standing assumptions, we assume that Bayes acts exist for all (conditional) distributions of $X$ appearing below, in particular for $\mathbb P_X$, the Dirac measures $\delta_x$, and the conditional laws $\mathbb P_{X|Y=y}$ for $\mathbb P_Y$-almost every $y$, and that a measurable selection $q \mapsto a_q$ has been fixed, which exists under our standing assumptions whenever pointwise minimizers exist, since $\mathcal X$ and $\mathcal A$ are standard Borel and $l$ is jointly measurable \cite{Bertsekas1978}. If there are multiple minimizers, we follow \cite{Gruenwald2004} and fix one Bayes act $a_q$ for each $q$.

Under these assumptions, we find
\begin{align*}H(X) & = \mathbb E[D(\delta_X\|\mathbb P_X)],\\
H(X|Y) &= \mathbb E[D(\delta_X\|\mathbb P_{X|Y})]\, ,\\
I(X;Y) &= \mathbb E[D(\mathbb P_{X|Y}\| \mathbb P_X)] ,\\
I(X;Y|Z) & = \mathbb E[D(\mathbb P_{X|Y,Z}\| \mathbb P_{X|Z})],
\end{align*}
where
\begin{equation}\label{eq:divergence}
D(p\|q) \coloneqq \mathbb E_p[l(X,a_q)-l(X,a_p)]
\end{equation}
is the \textit{divergence} \cite{Osband1985,Dawid2007} between two measures $q$ and $p$ on $\mathcal X$ induced by the proper scoring rule $S(x,q)\coloneqq l(x,a_q)$. Note that, if $l$ is itself a proper scoring rule, then $a_q = q$ and $D$ reduces to the divergence of $l$. For example, in the case of the log loss for discrete random variables, we have $S(x,q) = -\log q(X)$ and $D(p\|q) = D_\mathrm{KL}(p\|q)$. More examples of proper scoring rules include the \textit{Tsallis score} \cite{Tsallis1988} and \textit{Brier score} \cite{Brier1950}, for which $D$ is the Tsallis divergence and $L^2$-divergence of densities, respectively \cite{Dawid2007}. Applications of Tsallis and Brier scores include weather forecasting, statistical mechanics,
and calibration of probabilistic classifiers.

The above results show that, for our knowledge lattice-based definitions of entropy and information, we find the same relationships to $D$ as Shannon entropy and information have to $D_\mathrm{KL}$. Moreover, from the \textit{Bregman representation} \cite{MacCarthy1956,Hendrickson1971,Gneiting2007} of proper scoring rules, it follows that $D$ is also a Bregman divergence, just like KL divergence. In particular, we have
\begin{align*}
&\min_{Q_Y\,\{\emptyset,\Omega\}\text{-meas.}} \mathbb E[D(\mathbb P_{Y|Z}\|Q_Y)] = \mathbb E[D(\mathbb P_{Y|Z}\|\mathbb P_Y)],\\
&\min_{Q_Y\,\sigma(Z)\text{-meas.}} \mathbb E[D(\mathbb P_{Y|Z}\|Q_Y)] = \mathbb E[D(\mathbb P_{Y|Z}\|\mathbb P_{Y|Z})] = 0\, ,
\end{align*}
which means that fully reducing the uncertainty about $\mathbb P_{Y|Z}$ from $\{\emptyset,\Omega\}$ to $\sigma(Z)$ measured under $D$, i.e., the entropy $H(\mathbb P_{Y|Z})$ with respect to $D$, is the same as $I(Y;Z)$, the reduction in uncertainty about $Y$ from $\{\emptyset,\Omega\}$ to $\sigma(Z)$ measured with respect to $l$, generalizing the corresponding result for the KL divergence (cf.~previous example).

For simplicity and in agreement with the literature on proper scoring rules, we have considered $D(p\|q)$ only for losses where both Bayes acts exist, $a_q$ and $a_p$. However, we could extend the above results to a more general definition of $D(p\|q)$ through $\sup_{a}\mathbb E_p[l(X,a_q)-l(X,a)]$, only requiring $a_q$ to exist.
\end{example}

The examples above also show which classical measures the risk-based construction can cover, and in which sense. Every entropy of the form \eqref{def:entropy} is concave in $\mathbb P_X$, being an infimum of affine functions minus a linear term, and on a finite alphabet the converse holds as well: any concave $G$ on the simplex is the Bayes risk of the proper scoring rule $S(x,q)=G(q)+\partial_x G(q)-\sum_y q_y\,\partial_y G(q)$ \cite{Gneiting2007}, with residual term $\mathbb E[G(\delta_X)]$. For Rényi entropy $H_\alpha(p)=\frac{1}{1-\alpha}\log\sum_x p_x^\alpha$ the answer therefore depends on the order and the alphabet. For $0<\alpha\le1$, $H_\alpha$ is concave on every finite alphabet \cite{BenBassat1978}, and the scoring rule above reads
\begin{equation}\label{eq:renyiscore}
S_\alpha(x,q)=H_\alpha(q)+\frac{\alpha}{1-\alpha}\Big(\frac{q_x^{\alpha-1}}{\sum_y q_y^{\alpha}}-1\Big),
\end{equation}
with $S_\alpha(x,\delta_x)=0$, so that \eqref{def:entropy} with $l=S_\alpha$ is Rényi entropy of order $\alpha$. For $\alpha>1$ concavity holds only on small alphabets (on a binary alphabet exactly for $\alpha\le2$) and it fails once the alphabet is large enough: for $p_n=\frac12\delta_x+\frac12u_n$, with $u_n$ uniform on $n$ letters, the sum $\sum_y p_n(y)^\alpha$ tends to $2^{-\alpha}$, so $H_\alpha(p_n)$ stays bounded, whereas concavity would require $H_\alpha(p_n)\ge\frac12H_\alpha(\delta_x)+\frac12H_\alpha(u_n)=\frac12\log n$.

Moreover, representing $H_\alpha$ for $0<\alpha\le1$ as a Bayes risk reduction with $l=S_\alpha$ does not recover conventional Rényi information: the information \eqref{def:information} is $H_\alpha(\mathbb P_X)-\mathbb E[H_\alpha(\mathbb P_{X|Y})]$, which can be written as $\mathbb E[D(\mathbb P_{X|Y}\|\mathbb P_X)]$ with the Bregman divergence $D$ of $S_\alpha$ from \eqref{eq:divergence}. This is not Rényi divergence, which is not a Bregman divergence for $\alpha\neq1$, and accordingly none of the Rényi informations of Sibson, Arimoto, or Csisz\'ar \cite{Verdu2015} is a risk reduction in our sense.

One could attempt to extend our treatment by replacing risk differences $R(\pi)-R(\pi')$ by divergences between the conditional laws $\mathbb P_{X|\pi}$ attached to each lattice node $\pi$. However, this change would mean that the chain rule \eqref{eq:generalchainrule} no longer comes for free. For Bregman divergences, i.e., if the risk reduction from $\pi$ to $\pi'$ were replaced by $\mathbb E[d_\phi(\mathbb P_{X\mid\pi'},\mathbb P_{X\mid\pi})]$ for a convex function $\phi$ on probability measures, the chain rule still holds by the Pythagorean identity discussed above. This is no longer the case for Csisz\'ar's $f$-divergences, for which only weaker versions are known, such as superadditivity under coarse-graining \cite{Gottwald2019}. The Kullback–Leibler divergence is a special case, as it is the only divergence that is simultaneously an $f$-divergence and a Bregman divergence \cite{Csiszar1991}.

Note, however, that replacing risk reductions by Bregman divergences would not extend our approach. In particular, for two partitions $\pi,\pi'$ with $\mathbb E[\mathbb P_{X|\pi'}|\pi] = \mathbb P_{X|\pi}$, we have
\begin{equation} \label{eq:divequalsrisk}
\mathbb E[d_\phi(\mathbb P_{X\mid\pi'},\mathbb P_{X\mid\pi})] =  \mathbb E[\phi(\mathbb P_{X\mid\pi'}) - \phi(\mathbb P_{X\mid\pi})]\, .
\end{equation}
Moreover, it is a known result \cite{Gruenwald2004} that, under certain regularity assumptions, any convex function on probability measures can be written as a (negative) Bayes risk for some loss $l$, so that \eqref{eq:divequalsrisk} is a risk reduction after all.

\begin{table*}[p]
\centering
\caption{Partial information decomposition of $I(X;(Y,Z))$ (in bits) for the remaining example gates, continued from Table~\ref{tab:pid_comparison_main}; the columns, bold highlighting and all conventions are as described there.}\label{tab:pid_comparison_appendix1}
\renewcommand{\arraystretch}{1.7}
\setlength{\tabcolsep}{5pt}
\begin{tabular}{@{}l@{\hskip 3pt}l@{\hskip 3pt}rrrrrr@{}}
\toprule
Example & & Measure & $R$ & $U_Y$ & $U_Z$ & $S$ & $I$ \\
\midrule
\multirow{12}{*}{\begin{tabular}{@{}c@{}}\textbf{Copy}\\[8pt]{\scriptsize\parbox[t][1.2cm][t]{3.3cm}{\centering $X=(Y,Z)$, encoded $2Y+Z$; both sources jointly determine $X$}}\\[6pt]{\scriptsize\setlength{\tabcolsep}{3pt}\begin{tabular}{@{}cc|c@{}}$Y$ & $Z$ & $X$ \\0 & 0 & 0 \\0 & 1 & 1 \\1 & 0 & 2 \\1 & 1 & 3 \\\end{tabular}}\end{tabular}} & \multirow{12}{*}{\raisebox{0.4cm}{\includegraphics[height=6.3cm]{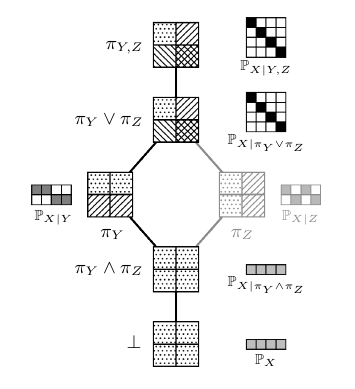}}} & \textbf{Ours ($\pi_Y$-routing)} & \textbf{0.000} & \textbf{1.000} & \textbf{1.000} & \textbf{0.000} & \multirow{12}{*}{2.000} \\
 &  & \textcolor[gray]{0.5}{\textbf{Ours ($\pi_Z$-routing)}} & \textcolor[gray]{0.5}{\textbf{0.000}} & \textcolor[gray]{0.5}{\textbf{1.000}} & \textcolor[gray]{0.5}{\textbf{1.000}} & \textcolor[gray]{0.5}{\textbf{0.000}} &  \\
 &  & Williams-Beer ($I_{\min}$) & 1.000 & 0.000 & 0.000 & 1.000 &  \\
 &  & Harder \textit{et al.}\ ($I_{\mathrm{proj}}$) & \textbf{0.000} & \textbf{1.000} & \textbf{1.000} & \textbf{0.000} &  \\
 &  & Griffith \textit{et al.} ($I_{\mathrm{GK}}$) & \textbf{0.000} & \textbf{1.000} & \textbf{1.000} & \textbf{0.000} &  \\
 &  & Bertschinger \textit{et al.}\ ($I_{\mathrm{BROJA}}$) & \textbf{0.000} & \textbf{1.000} & \textbf{1.000} & \textbf{0.000} &  \\
 &  & Barrett ($I_{\mathrm{MMI}}$) & 1.000 & 0.000 & 0.000 & 1.000 &  \\
 &  & Ince ($I_{\mathrm{ccs}}$) & \textbf{0.000} & \textbf{1.000} & \textbf{1.000} & \textbf{0.000} &  \\
 &  & James \textit{et al.}\ ($I_{\mathrm{dep}}$) & \textbf{0.000} & \textbf{1.000} & \textbf{1.000} & \textbf{0.000} &  \\
 &  & Ay-Polani-Virgo & $-$ & \textbf{1.000} & \textbf{1.000} & \textbf{0.000} &  \\
 &  & Kolchinsky & \textbf{0.000} & \textbf{1.000} & \textbf{1.000} & \textbf{0.000} &  \\
 &  & Gomes-Figueiredo & \textbf{0.000} & \textbf{1.000} & \textbf{1.000} & \textbf{0.000} &  \\
\addlinespace[7pt]\midrule\addlinespace[7pt]
\multirow{12}{*}{\begin{tabular}{@{}c@{}}\textbf{RdnXor}\\[8pt]{\scriptsize\parbox[t][1.2cm][t]{3.3cm}{\centering $Y,Z$ share a redundant block ($\lfloor Y/2\rfloor=\lfloor Z/2\rfloor$), and $\mathsf{Xor}$ within each block}}\\[6pt]{\scriptsize\setlength{\tabcolsep}{3pt}\begin{tabular}{@{}cc|c@{}}$Y$ & $Z$ & $X$ \\0 & 0 & 0 \\0 & 1 & 1 \\1 & 0 & 1 \\1 & 1 & 0 \\2 & 2 & 2 \\2 & 3 & 3 \\3 & 2 & 3 \\3 & 3 & 2 \\\end{tabular}}\end{tabular}} & \multirow{12}{*}{\raisebox{0.4cm}{\includegraphics[height=6.3cm]{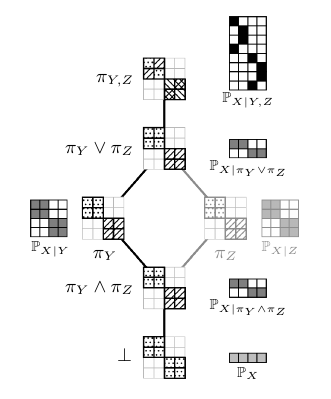}}} & \textbf{Ours ($\pi_Y$-routing)} & \textbf{1.000} & \textbf{0.000} & \textbf{0.000} & \textbf{1.000} & \multirow{12}{*}{2.000} \\
 &  & \textcolor[gray]{0.5}{\textbf{Ours ($\pi_Z$-routing)}} & \textcolor[gray]{0.5}{\textbf{1.000}} & \textcolor[gray]{0.5}{\textbf{0.000}} & \textcolor[gray]{0.5}{\textbf{0.000}} & \textcolor[gray]{0.5}{\textbf{1.000}} &  \\
 &  & Williams-Beer ($I_{\min}$) & \textbf{1.000} & \textbf{0.000} & \textbf{0.000} & \textbf{1.000} &  \\
 &  & Harder \textit{et al.}\ ($I_{\mathrm{proj}}$) & \textbf{1.000} & \textbf{0.000} & \textbf{0.000} & \textbf{1.000} &  \\
 &  & Griffith \textit{et al.} ($I_{\mathrm{GK}}$) & \textbf{1.000} & \textbf{0.000} & \textbf{0.000} & \textbf{1.000} &  \\
 &  & Bertschinger \textit{et al.}\ ($I_{\mathrm{BROJA}}$) & \textbf{1.000} & \textbf{0.000} & \textbf{0.000} & \textbf{1.000} &  \\
 &  & Barrett ($I_{\mathrm{MMI}}$) & \textbf{1.000} & \textbf{0.000} & \textbf{0.000} & \textbf{1.000} &  \\
 &  & Ince ($I_{\mathrm{ccs}}$) & \textbf{1.000} & \textbf{0.000} & \textbf{0.000} & \textbf{1.000} &  \\
 &  & James \textit{et al.}\ ($I_{\mathrm{dep}}$) & \textbf{1.000} & \textbf{0.000} & \textbf{0.000} & \textbf{1.000} &  \\
 &  & Ay-Polani-Virgo & $-$ & 0.500 & 0.500 & \textbf{1.000} &  \\
 &  & Kolchinsky & \textbf{1.000} & \textbf{0.000} & \textbf{0.000} & \textbf{1.000} &  \\
 &  & Gomes-Figueiredo & \textbf{1.000} & \textbf{0.000} & \textbf{0.000} & \textbf{1.000} &  \\
 \addlinespace[7pt]\midrule\addlinespace[7pt]
 \multirow{12}{*}{\begin{tabular}{@{}c@{}}\textbf{PwUnq}\\[8pt]{\scriptsize\parbox[t][1.2cm][t]{3.3cm}{\centering Pointwise-unique: exactly one source is non-zero in each outcome, and $X$ is that non-zero value (shifted by 1)}}\\[6pt]{\scriptsize\setlength{\tabcolsep}{3pt}\begin{tabular}{@{}cc|c@{}}$Y$ & $Z$ & $X$ \\0 & 1 & 0 \\0 & 2 & 1 \\1 & 0 & 0 \\2 & 0 & 1 \\\end{tabular}}\end{tabular}} & \multirow{12}{*}{\raisebox{0.4cm}{\includegraphics[height=6.3cm]{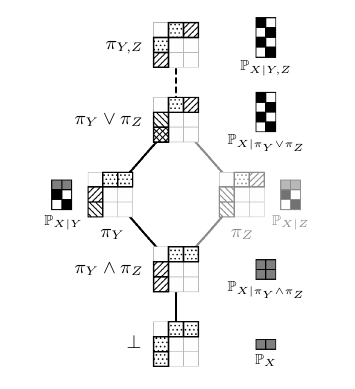}}} & \textbf{Ours ($\pi_Y$-routing)} & \textbf{0.000} & \textbf{0.500} & \textbf{0.500} & \textbf{0.000} & \multirow{12}{*}{1.000} \\
  &  & \textcolor[gray]{0.5}{\textbf{Ours ($\pi_Z$-routing)}} & \textcolor[gray]{0.5}{\textbf{0.000}} & \textcolor[gray]{0.5}{\textbf{0.500}} & \textcolor[gray]{0.5}{\textbf{0.500}} & \textcolor[gray]{0.5}{\textbf{0.000}} &  \\
  &  & Williams-Beer ($I_{\min}$) & 0.500 & 0.000 & 0.000 & 0.500 &  \\
  &  & Harder \textit{et al.}\ ($I_{\mathrm{proj}}$) & 0.500 & 0.000 & 0.000 & 0.500 &  \\
  &  & Griffith \textit{et al.} ($I_{\mathrm{GK}}$) & \textbf{0.000} & \textbf{0.500} & \textbf{0.500} & \textbf{0.000} &  \\
  &  & Bertschinger \textit{et al.}\ ($I_{\mathrm{BROJA}}$) & 0.500 & 0.000 & 0.000 & 0.500 &  \\
  &  & Barrett ($I_{\mathrm{MMI}}$) & 0.500 & 0.000 & 0.000 & 0.500 &  \\
  &  & Ince ($I_{\mathrm{ccs}}$) & \textbf{0.000} & \textbf{0.500} & \textbf{0.500} & \textbf{0.000} &  \\
  &  & James \textit{et al.}\ ($I_{\mathrm{dep}}$) & 0.250 & 0.250 & 0.250 & 0.250 &  \\
  &  & Ay-Polani-Virgo & $-$ & \textbf{0.500} & \textbf{0.500} & \textbf{0.000} &  \\
  &  & Kolchinsky & 0.500 & 0.000 & 0.000 & 0.500 &  \\
  &  & Gomes-Figueiredo & 0.250 & 0.250 & 0.250 & 0.250 &  \\
\addlinespace[7pt]
\bottomrule
\end{tabular}
\end{table*}

\begin{table*}[p]
\centering
\caption{Comparison of PID measures (continued).}\label{tab:pid_comparison_appendix2}
\renewcommand{\arraystretch}{1.7}
\setlength{\tabcolsep}{5pt}
\begin{tabular}{@{}l@{\hskip 3pt}l@{\hskip 3pt}rrrrrr@{}}
\toprule
Example & & Measure & $R$ & $U_Y$ & $U_Z$ & $S$ & $I$ \\
\midrule
\multirow{12}{*}{\begin{tabular}{@{}c@{}}\textbf{And}\\[8pt]{\scriptsize\parbox[t][1.2cm][t]{3.3cm}{\centering $X=Y\wedge Z$}}\\[6pt]{\scriptsize\setlength{\tabcolsep}{3pt}\begin{tabular}{@{}cc|c@{}}$Y$ & $Z$ & $X$ \\0 & 0 & 0 \\0 & 1 & 0 \\1 & 0 & 0 \\1 & 1 & 1 \\\end{tabular}}\end{tabular}} & \multirow{12}{*}{\raisebox{0.4cm}{\includegraphics[height=6.3cm]{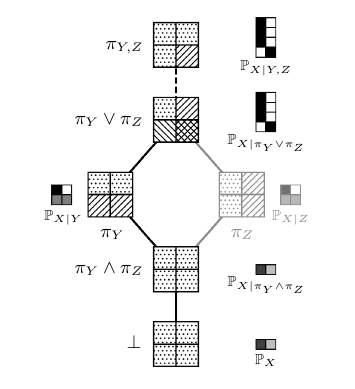}}} & \textbf{Ours ($\pi_Y$-routing)} & \textbf{0.000} & \textbf{0.311} & \textbf{0.500} & \textbf{0.000} & \multirow{12}{*}{0.811} \\
 &  & \textcolor[gray]{0.5}{\textbf{Ours ($\pi_Z$-routing)}} & \textcolor[gray]{0.5}{\textbf{0.000}} & \textcolor[gray]{0.5}{\textbf{0.500}} & \textcolor[gray]{0.5}{\textbf{0.311}} & \textcolor[gray]{0.5}{\textbf{0.000}} &  \\
 &  & Williams-Beer ($I_{\min}$) & 0.311 & 0.000 & 0.000 & 0.500 &  \\
 &  & Harder \textit{et al.}\ ($I_{\mathrm{proj}}$) & 0.311 & 0.000 & 0.000 & 0.500 &  \\
 &  & Griffith \textit{et al.} ($I_{\mathrm{GK}}$) & \textbf{0.000} & \textbf{0.311} & \textcolor[gray]{0.5}{\textbf{0.311}} & 0.189 &  \\
 &  & Bertschinger \textit{et al.}\ ($I_{\mathrm{BROJA}}$) & 0.311 & 0.000 & 0.000 & 0.500 &  \\
 &  & Barrett ($I_{\mathrm{MMI}}$) & 0.311 & 0.000 & 0.000 & 0.500 &  \\
 &  & Ince ($I_{\mathrm{ccs}}$) & 0.104 & 0.208 & 0.208 & 0.292 &  \\
 &  & James \textit{et al.}\ ($I_{\mathrm{dep}}$) & 0.082 & 0.230 & 0.230 & 0.270 &  \\
 &  & Ay-Polani-Virgo & $-$ & 0.406 & 0.406 & \textbf{0.000} &  \\
 &  & Kolchinsky & 0.311 & 0.000 & 0.000 & 0.500 &  \\
 &  & Gomes-Figueiredo & 0.082 & 0.230 & 0.230 & 0.270 &  \\
\addlinespace[7pt]\midrule\addlinespace[7pt]
\multirow{12}{*}{\begin{tabular}{@{}c@{}}\textbf{Sum}\\[8pt]{\scriptsize\parbox[t][1.2cm][t]{3.3cm}{\centering $X=Y+Z$}}\\[6pt]{\scriptsize\setlength{\tabcolsep}{3pt}\begin{tabular}{@{}cc|c@{}}$Y$ & $Z$ & $X$ \\0 & 0 & 0 \\0 & 1 & 1 \\1 & 0 & 1 \\1 & 1 & 2 \\\end{tabular}}\end{tabular}} & \multirow{12}{*}{\raisebox{0.4cm}{\includegraphics[height=6.3cm]{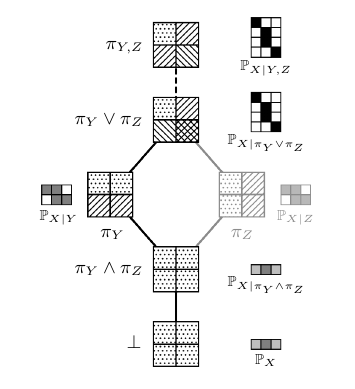}}} & \textbf{Ours ($\pi_Y$-routing)} & \textbf{0.000} & \textbf{0.500} & \textbf{1.000} & \textbf{0.000} & \multirow{12}{*}{1.500} \\
 &  & \textcolor[gray]{0.5}{\textbf{Ours ($\pi_Z$-routing)}} & \textcolor[gray]{0.5}{\textbf{0.000}} & \textcolor[gray]{0.5}{\textbf{1.000}} & \textcolor[gray]{0.5}{\textbf{0.500}} & \textcolor[gray]{0.5}{\textbf{0.000}} &  \\
 &  & Williams-Beer ($I_{\min}$) & 0.500 & 0.000 & 0.000 & 1.000 &  \\
 &  & Harder \textit{et al.}\ ($I_{\mathrm{proj}}$) & 0.500 & 0.000 & 0.000 & 1.000 &  \\
 &  & Griffith \textit{et al.} ($I_{\mathrm{GK}}$) & \textbf{0.000} & \textbf{0.500} & \textcolor[gray]{0.5}{\textbf{0.500}} & 0.500 &  \\
 &  & Bertschinger \textit{et al.}\ ($I_{\mathrm{BROJA}}$) & 0.500 & 0.000 & 0.000 & 1.000 &  \\
 &  & Barrett ($I_{\mathrm{MMI}}$) & 0.500 & 0.000 & 0.000 & 1.000 &  \\
 &  & Ince ($I_{\mathrm{ccs}}$) & \textbf{0.000} & \textbf{0.500} & \textcolor[gray]{0.5}{\textbf{0.500}} & 0.500 &  \\
 &  & James \textit{et al.}\ ($I_{\mathrm{dep}}$) & 0.189 & 0.311 & 0.311 & 0.689 &  \\
 &  & Ay-Polani-Virgo & $-$ & 0.750 & 0.750 & \textbf{0.000} &  \\
 &  & Kolchinsky & 0.500 & 0.000 & 0.000 & 1.000 &  \\
 &  & Gomes-Figueiredo & 0.189 & 0.311 & 0.311 & 0.689 &  \\
\addlinespace[7pt]\midrule\addlinespace[7pt]
\multirow{12}{*}{\begin{tabular}{@{}c@{}}\textbf{ReducedOr}\\[8pt]{\scriptsize\parbox[t][1.2cm][t]{3.3cm}{\centering $X=Y\vee Z$ with the $(1,1)$ state removed and weights $(\tfrac12,\tfrac14,\tfrac14)$}}\\[6pt]{\scriptsize\setlength{\tabcolsep}{3pt}\begin{tabular}{@{}cc|c|c@{}}$Y$ & $Z$ & $X$ & $\mathbb P$ \\0 & 0 & 0 & 0.500 \\0 & 1 & 1 & 0.250 \\1 & 0 & 1 & 0.250 \\\end{tabular}}\end{tabular}} & \multirow{12}{*}{\raisebox{0.4cm}{\includegraphics[height=6.3cm]{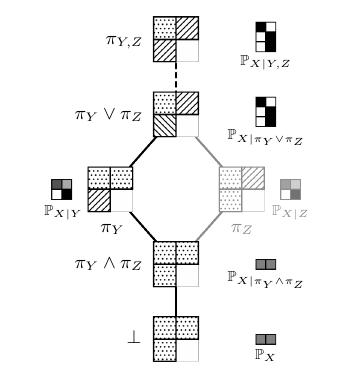}}} & \textbf{Ours ($\pi_Y$-routing)} & \textbf{0.000} & \textbf{0.311} & \textbf{0.689} & \textbf{0.000} & \multirow{12}{*}{1.000} \\
 &  & \textcolor[gray]{0.5}{\textbf{Ours ($\pi_Z$-routing)}} & \textcolor[gray]{0.5}{\textbf{0.000}} & \textcolor[gray]{0.5}{\textbf{0.689}} & \textcolor[gray]{0.5}{\textbf{0.311}} & \textcolor[gray]{0.5}{\textbf{0.000}} &  \\
 &  & Williams-Beer ($I_{\min}$) & 0.311 & 0.000 & 0.000 & 0.689 &  \\
 &  & Harder \textit{et al.}\ ($I_{\mathrm{proj}}$) & 0.311 & 0.000 & 0.000 & 0.689 &  \\
 &  & Griffith \textit{et al.} ($I_{\mathrm{GK}}$) & \textbf{0.000} & \textbf{0.311} & \textcolor[gray]{0.5}{\textbf{0.311}} & 0.377 &  \\
 &  & Bertschinger \textit{et al.}\ ($I_{\mathrm{BROJA}}$) & 0.311 & 0.000 & 0.000 & 0.689 &  \\
 &  & Barrett ($I_{\mathrm{MMI}}$) & 0.311 & 0.000 & 0.000 & 0.689 &  \\
 &  & Ince ($I_{\mathrm{ccs}}$) & \textbf{0.000} & \textbf{0.311} & \textcolor[gray]{0.5}{\textbf{0.311}} & 0.377 &  \\
 &  & James \textit{et al.}\ ($I_{\mathrm{dep}}$) & 0.074 & 0.238 & 0.238 & 0.451 &  \\
 &  & Ay-Polani-Virgo & $-$ & 0.500 & 0.500 & \textbf{0.000} &  \\
 &  & Kolchinsky & 0.311 & 0.000 & 0.000 & 0.689 &  \\
 &  & Gomes-Figueiredo & 0.074 & 0.238 & 0.238 & 0.451 &  \\
\addlinespace[7pt]
\bottomrule
\end{tabular}
\end{table*}

\begin{table*}[p]
\centering
\caption{Comparison of PID measures (continued).}\label{tab:pid_comparison_appendix3}
\renewcommand{\arraystretch}{1.7}
\setlength{\tabcolsep}{5pt}
\begin{tabular}{@{}l@{\hskip 3pt}l@{\hskip 3pt}rrrrrr@{}}
\toprule
Example & & Measure & $R$ & $U_Y$ & $U_Z$ & $S$ & $I$ \\
\midrule
\multirow{12}{*}{\begin{tabular}{@{}c@{}}\textbf{Boom}\\[8pt]{\scriptsize\parbox[t][1.2cm][t]{3.3cm}{\centering Each source pair fixes $X$ except $(Y,Z)=(0,0)$, which gives $X\in\{1,2\}$ at random}}\\[6pt]{\scriptsize\setlength{\tabcolsep}{3pt}\begin{tabular}{@{}cc|c|c@{}}$Y$ & $Z$ & $X$ & $\mathbb P$ \\0 & 0 & 1 & 0.167 \\0 & 0 & 2 & 0.167 \\0 & 2 & 0 & 0.167 \\1 & 2 & 1 & 0.167 \\2 & 0 & 2 & 0.167 \\2 & 1 & 2 & 0.167 \\\end{tabular}}\end{tabular}} & \multirow{12}{*}{\raisebox{0.4cm}{\includegraphics[height=6.3cm]{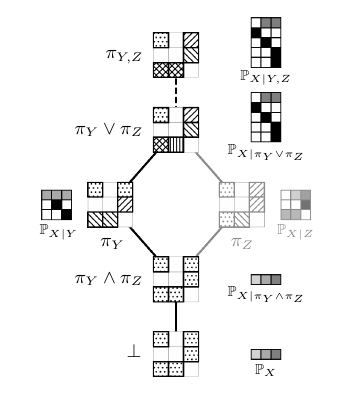}}} & \textbf{Ours ($\pi_Y$-routing)} & \textbf{0.000} & \textbf{0.667} & \textbf{0.459} & \textbf{0.000} & \textbf{1.126} \\
 &  & \textcolor[gray]{0.5}{\textbf{Ours ($\pi_Z$-routing)}} & \textcolor[gray]{0.5}{\textbf{0.000}} & \textcolor[gray]{0.5}{\textbf{0.459}} & \textcolor[gray]{0.5}{\textbf{0.667}} & \textcolor[gray]{0.5}{\textbf{0.000}} & \textcolor[gray]{0.5}{\textbf{1.126}} \\
 &  & Williams-Beer ($I_{\min}$) & 0.500 & 0.167 & 0.167 & 0.292 & 1.126 \\
 &  & Harder \textit{et al.}\ ($I_{\mathrm{proj}}$) & 0.500 & 0.167 & 0.167 & 0.292 & 1.126 \\
 &  & Griffith \textit{et al.} ($I_{\mathrm{GK}}$) & \textbf{0.000} & \textbf{0.667} & \textcolor[gray]{0.5}{\textbf{0.667}} & -0.208 & 1.126 \\
 &  & Bertschinger \textit{et al.}\ ($I_{\mathrm{BROJA}}$) & 0.333 & 0.333 & 0.333 & 0.126 & 1.126 \\
 &  & Barrett ($I_{\mathrm{MMI}}$) & 0.667 & 0.000 & 0.000 & 0.459 & 1.126 \\
 &  & Ince ($I_{\mathrm{ccs}}$) & 0.333 & 0.333 & 0.333 & 0.126 & 1.126 \\
 &  & James \textit{et al.}\ ($I_{\mathrm{dep}}$) & 0.295 & 0.371 & 0.371 & 0.088 & 1.126 \\
 &  & Ay-Polani-Virgo & $-$ & 0.563 & 0.563 & \textbf{0.000} & 1.126 \\
 &  & Kolchinsky & 0.322 & 0.345 & 0.345 & 0.126 & 1.138 \\
 &  & Gomes-Figueiredo & 0.295 & 0.371 & 0.371 & 0.088 & 1.126 \\
\addlinespace[7pt]\midrule\addlinespace[7pt]
\multirow{12}{*}{\begin{tabular}{@{}c@{}}\textbf{NoisyAnd$(0.1)$}\\[8pt]{\scriptsize\parbox[t][1.2cm][t]{3.3cm}{\centering $X=(Y\wedge Z)\oplus N$ with $N\sim\mathrm{Bern}(0.1)$}}\\[6pt]{\scriptsize\setlength{\tabcolsep}{3pt}\begin{tabular}{@{}cc|c|c@{}}$Y$ & $Z$ & $X$ & $\mathbb P$ \\0 & 0 & 0 & 0.225 \\0 & 0 & 1 & 0.025 \\0 & 1 & 0 & 0.225 \\0 & 1 & 1 & 0.025 \\1 & 0 & 0 & 0.225 \\1 & 0 & 1 & 0.025 \\1 & 1 & 0 & 0.025 \\1 & 1 & 1 & 0.225 \\\end{tabular}}\end{tabular}} & \multirow{12}{*}{\raisebox{0.4cm}{\includegraphics[height=6.3cm]{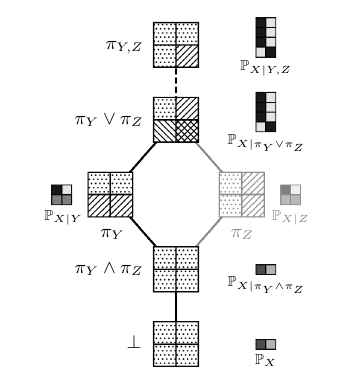}}} & \textbf{Ours ($\pi_Y$-routing)} & \textbf{0.000} & \textbf{0.147} & \textbf{0.266} & \textbf{0.000} & \multirow{12}{*}{0.412} \\
 &  & \textcolor[gray]{0.5}{\textbf{Ours ($\pi_Z$-routing)}} & \textcolor[gray]{0.5}{\textbf{0.000}} & \textcolor[gray]{0.5}{\textbf{0.266}} & \textcolor[gray]{0.5}{\textbf{0.147}} & \textcolor[gray]{0.5}{\textbf{0.000}} &  \\
 &  & Williams-Beer ($I_{\min}$) & 0.147 & 0.000 & 0.000 & 0.266 &  \\
 &  & Harder \textit{et al.}\ ($I_{\mathrm{proj}}$) & 0.147 & 0.000 & 0.000 & 0.266 &  \\
 &  & Griffith \textit{et al.} ($I_{\mathrm{GK}}$) & \textbf{0.000} & \textbf{0.147} & \textcolor[gray]{0.5}{\textbf{0.147}} & 0.119 &  \\
 &  & Bertschinger \textit{et al.}\ ($I_{\mathrm{BROJA}}$) & 0.147 & 0.000 & 0.000 & 0.266 &  \\
 &  & Barrett ($I_{\mathrm{MMI}}$) & 0.147 & 0.000 & 0.000 & 0.266 &  \\
 &  & Ince ($I_{\mathrm{ccs}}$) & 0.064 & 0.083 & 0.083 & 0.183 &  \\
 &  & James \textit{et al.}\ ($I_{\mathrm{dep}}$) & 0.026 & 0.120 & 0.120 & 0.145 &  \\
 &  & Ay-Polani-Virgo & $-$ & 0.167 & 0.167 & 0.077 &  \\
 &  & Kolchinsky & 0.147 & 0.000 & 0.000 & 0.266 &  \\
 &  & Gomes-Figueiredo & 0.026 & 0.120 & 0.120 & 0.145 &  \\
\addlinespace[7pt]\midrule\addlinespace[7pt]
\multirow{12}{*}{\begin{tabular}{@{}c@{}}\textbf{NoisyRdn}\\[8pt]{\scriptsize\parbox[t][1.2cm][t]{3.3cm}{\centering $\mathsf{Rdn}$ corrupted by sparse asymmetric noise}}\\[6pt]{\scriptsize\setlength{\tabcolsep}{3pt}\begin{tabular}{@{}cc|c|c@{}}$Y$ & $Z$ & $X$ & $\mathbb P$ \\0 & 0 & 0 & 0.389 \\0 & 1 & 0 & 0.056 \\0 & 1 & 1 & 0.056 \\1 & 0 & 1 & 0.056 \\1 & 1 & 0 & 0.056 \\1 & 1 & 1 & 0.389 \\\end{tabular}}\end{tabular}} & \multirow{12}{*}{\raisebox{0.4cm}{\includegraphics[height=6.3cm]{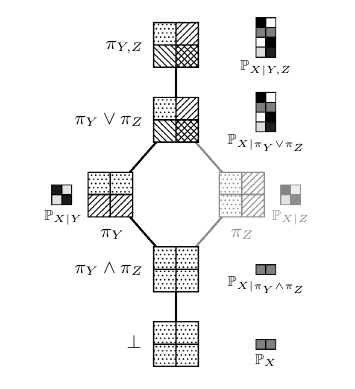}}} & \textbf{Ours ($\pi_Y$-routing)} & \textbf{0.000} & \textbf{0.497} & \textbf{0.151} & \textbf{0.000} & \multirow{12}{*}{0.647} \\
 &  & \textcolor[gray]{0.5}{\textbf{Ours ($\pi_Z$-routing)}} & \textcolor[gray]{0.5}{\textbf{0.000}} & \textcolor[gray]{0.5}{\textbf{0.290}} & \textcolor[gray]{0.5}{\textbf{0.357}} & \textcolor[gray]{0.5}{\textbf{0.000}} &  \\
 &  & Williams-Beer ($I_{\min}$) & 0.357 & 0.139 & 0.000 & 0.151 &  \\
 &  & Harder \textit{et al.}\ ($I_{\mathrm{proj}}$) & 0.357 & 0.139 & 0.000 & 0.151 &  \\
 &  & Griffith \textit{et al.} ($I_{\mathrm{GK}}$) & \textbf{0.000} & \textbf{0.497} & \textcolor[gray]{0.5}{\textbf{0.357}} & -0.207 &  \\
 &  & Bertschinger \textit{et al.}\ ($I_{\mathrm{BROJA}}$) & 0.357 & 0.139 & 0.000 & 0.151 &  \\
 &  & Barrett ($I_{\mathrm{MMI}}$) & 0.357 & 0.139 & 0.000 & 0.151 &  \\
 &  & Ince ($I_{\mathrm{ccs}}$) & 0.472 & 0.025 & -0.115 & 0.265 &  \\
 &  & James \textit{et al.}\ ($I_{\mathrm{dep}}$) & 0.300 & 0.197 & 0.058 & 0.093 &  \\
 &  & Ay-Polani-Virgo & $-$ & 0.347 & 0.207 & 0.093 &  \\
 &  & Kolchinsky & 0.357 & 0.139 & 0.000 & 0.151 &  \\
 &  & Gomes-Figueiredo & 0.207 & \textcolor[gray]{0.5}{\textbf{0.290}} & 0.150 & \textbf{0.000} &  \\
\addlinespace[7pt]
\bottomrule
\end{tabular}
\end{table*}

\subsection{More two-source information decomposition examples}

Table~\ref{tab:pid_comparison_appendix1} and the two tables after it (Tables~\ref{tab:pid_comparison_appendix2} and
\ref{tab:pid_comparison_appendix3}) continue the comparison of Table~\ref{tab:pid_comparison_main} on further discrete gates, using the same measures and the same column conventions. The added gates contain canonical gates from the literature: $\mathsf{Copy}$, $\mathsf{RdnXor}$, $\mathsf{PwUnq}$, $\mathsf{And}$, $\mathsf{Sum}$, $\mathsf{ReducedOr}$, $\mathsf{Boom}$, as well as noisy gates, $\mathsf{NoisyAnd}(0.1)$ and $\mathsf{NoisyRdn}$. We built $\mathsf{NoisyRdn}$ by a random search to separate the dependency decomposition $I_{\mathrm{dep}}$ of James \textit{et al.}\ from the union-information decomposition of Gomes and Figueiredo, which agree on every other gate we tried. Again, notice the leakage between unique, redundant, and synergistic terms of current PID proposals, which results from the clamping of unique and conditional unique terms to be the same and from determining the other terms by subtraction (cf.~Section~\ref{sec:pid_examples}).

\subsection{Details on continuous PID examples} \label{app:continuous}
This appendix collects the derivations for the continuous examples summarized
in Section~\ref{sec:continuousexamples}. All variables live on a common
standard Borel probability space, so the machinery of measurable partitions
\cite{Rohlin1967,Bogachev2007} applies. As in the main text, the lattice
operations are read on the $\sigma$-algebras induced by measurable partitions
(Definition~\ref{def:knowledgelattice2}), modulo $\mathbb P$-null sets.

\begin{example}[Jointly Gaussian sources]\label{ex:gaussianPID}
Let $(X,Y,Z)$ be jointly Gaussian with zero means, unit variances, and correlations $\rho_Y = \operatorname{corr}(X,Y)$,
$\rho_Z = \operatorname{corr}(X,Z)$, $\rho_{YZ} = \operatorname{corr}(Y,Z)$, assumed non-degenerate, i.e., $\rho_Y,\rho_Z\neq 0$ and the covariance matrix of $(X,Y,Z)$ positive definite. Since its determinant equals $(1-\rho_{YZ}^2)(1-R^2)$ with $R^2$ as in \eqref{eq:Rsq} below, positive definiteness is equivalent to $|\rho_{YZ}|<1$ and $R^2<1$, which is what keeps the log-loss terms below finite. The conditional law $\mathbb P_{X|Y=y} = \mathcal N(\rho_Y y,\, 1-\rho_Y^2)$ has a constant variance and a mean that is injective in $y$, so $y\mapsto \mathbb P_{X|Y=y}$ is injective and $\sigma(\pi_Y) = \sigma(Y)$, and also $\sigma(\pi_Z) = \sigma(Z)$. The join is therefore $\sigma(Y,Z)$, and so by Lemma~\ref{prop:problattice}, we have $I_s(X;(Y,Z)) = 0$.

For the meet, suppose $\mathbf 1_B = f(Y) = g(Z)$ almost surely with $0<\mathbb P(B)<1$. Then $f(Y)$ and $g(Z)$ are non-constant and have correlation $\mathrm{corr}(f(Y),g(Z)) = 1$, contradicting the result of Gebelein and Lancaster \cite{Gebelein1941,Lancaster1957} that the correlation of transformed variables must be less or equal in absolute value than $|\rho_{YZ}|$, which is assumed to be smaller than $1$. Hence, modulo $\mathbb P$-null sets, we have $\sigma(\pi_Y)\wedge\sigma(\pi_Z) = \{\emptyset,\Omega\}$ and therefore $I_r(X;(Y,Z)) = 0$.

It follows that all information is complementary, and the path-dependent 4-term decomposition of Corollary~\ref{cor:PID} collapses to the regular chain rule. With the squared multiple correlation
\begin{equation}\label{eq:Rsq}
R^2 \coloneqq \frac{\rho_Y^2 + \rho_Z^2 - 2\rho_Y\rho_Z\rho_{YZ}}{1-\rho_{YZ}^2}\,,
\end{equation}
the $\pi_Y$-routing gives for the log loss
\begin{align*}
I_u(X;Y{\setminus}Z) &= -\tfrac12\log(1-\rho_Y^2)\,,\\
I_{cu}(X;Z|Y) &= -\tfrac12\log\tfrac{1-R^2}{1-\rho_Y^2}\,,
\end{align*}
and $I_u(X;Y{\setminus}Z) = \rho_Y^2$, $I_{cu}(X;Z|Y) = R^2 - \rho_Y^2$,  for the square error. In both cases $I_r = I_s = 0$, and $H(X|Y,Z) = 1-R^2$ for the square error, whereas $H(X|Y,Z)=\infty$ for the log loss (cf.~Example \ref{ex:logloss}). For the square error these are the terms of the classical hierarchical variance decomposition  \cite{Cramer1946}.
\end{example}

\begin{example}[Redundant bit: $\mathsf{RdnGaussian}$]\label{ex:gaussrdn}
Let $W\sim\mathrm{Bern}(\tfrac12)$ and let $V_1,V_2,N$ be independent centered Gaussian variables with variances $\sigma_1^2,\sigma_2^2,\sigma_N^2>0$, independent of $W$. Define
\[
Y = (W,V_1)\,, \quad Z = (W,V_2)\,, \quad X = \big(W,\; V_1+V_2+N\big)\,,
\]
so $X$ takes values in $\{0,1\}\times\mathbb R$ and the noise $N$ is seen by neither source. The conditional laws are
\begin{align*}
\mathbb P_{X|Y} &= \delta_W \otimes \mathcal N(V_1,\,\sigma_2^2+\sigma_N^2),\\
\mathbb P_{X|Y,Z} &= \delta_W \otimes \mathcal N(V_1{+}V_2,\,\sigma_N^2),
\end{align*}
and analogously for $Z$. Each conditional-law map is injective, so
$\sigma(\pi_Y) = \sigma(W,V_1)$, $\sigma(\pi_Z)=\sigma(W,V_2)$, and
$\sigma(\pi_{Y,Z}) = \sigma(W,\,V_1{+}V_2)$.

We claim $\sigma(\pi_Y)\wedge\sigma(\pi_Z) = \sigma(W)$
modulo null sets, i.e., the common knowledge about $X$ is exactly the
shared bit. Indeed, if $\mathbf 1_B = f(W,V_1) = g(W,V_2)$ a.s., then
conditionally on $\{W=w\}$ we have
$f(w,V_1) = g(w,V_2)$ a.s.\ with $V_1$ and $V_2$ independent, so
$f(w,V_1) = \mathbb E[g(w,V_2)\,|\,V_1] = \mathbb E[g(w,V_2)]$ is a.s.\
constant. Hence $B$ agrees with a union of the events $\{W=w\}$ up to null
sets. Conversely, $\sigma(W)$ is contained in both.

The join is
$\sigma(W,V_1)\vee\sigma(W,V_2) = \sigma(W,V_1,V_2)$, i.e., strictly finer than
$\sigma(\pi_{Y,Z})$. In particular, the conditional law of $X$ given $(W,V_1,V_2)$ equals the law given $(W,V_1{+}V_2)$, so the risks agree and thus $I_s(X;(Y,Z)) = 0$ for every
loss.

Writing $\sigma_{12N}^2 \coloneqq \sigma_1^2+\sigma_2^2+\sigma_N^2$ and taking
the reference measure on $\{0,1\}\times\mathbb R$ to be counting measure times
Lebesgue measure, the $\pi_Y$-routing of
Corollary~\ref{cor:PID} for $\mathsf{RdnGaussian}$ is given by
\begin{align*}
I_r(X;(Y,Z)) &= \mathbb V(W) = \tfrac14\,,\\
I_u(X;Y{\setminus}Z) &= \sigma_1^2\,,\\
I_{cu}(X;Z|Y) &= \sigma_2^2\,,\\
I_s(X;(Y,Z)) &= 0\,
\end{align*}
for the square error, and
\begin{align*}
I_r(X;(Y,Z)) &= \log 2\,, \\
I_u(X;Y{\setminus}Z) &= \tfrac12\log\tfrac{\sigma_{12N}^2}{\sigma_2^2+\sigma_N^2}\,,\\
I_{cu}(X;Z|Y) &= \tfrac12\log\tfrac{\sigma_2^2+\sigma_N^2}{\sigma_N^2}\,,\\
I_s(X;(Y,Z)) &= 0\,
\end{align*}
for the log loss. Also, the remaining uncertainty about $X$ is $H(X|Y,Z) = \sigma_N^2$ for the square error and $H(X|Y,Z)=\infty$ for the log loss (as in Example~\ref{ex:logloss}).

Note that, if we replace the bit $W$ by a continuous Gaussian
variable, the lattice is unchanged and $I_r(X;(Y,Z)) = \mathbb V(W)$ under the square
error, but $I_r(X;(Y,Z)) = \infty$ under the log loss, since the common knowledge is now determined by a continuous random variable.
\end{example}

\begin{example}[Continuous $\mathsf{Xor}$: $\mathsf{SignXor}$]\label{ex:signxor}
Let $Y,Z\sim\mathcal N(0,1)$ and $N\sim\mathcal N(0,\sigma_N^2)$ be independent
and set $X = Y\cdot\operatorname{sign}(Z) + N$. The
conditional law $\mathbb P_{X|Z=z} = \mathcal N(0,\,1+\sigma_N^2)$ does not
depend on $z$, so $\sigma(\pi_Z) = \{\emptyset,\Omega\}$. The conditional law
given $Y$ is the symmetric mixture
\[
\mathbb P_{X|Y=y} = \tfrac12 \mathcal N(y,\sigma_N^2)
+ \tfrac12 \mathcal N(-y,\sigma_N^2)\,,
\]
which is independent of the sign of $y$ but injective with respect to $|y|$, so that
$\sigma(\pi_Y) = \sigma(|Y|)$. Finally,
$\mathbb P_{X|Y,Z} = \mathcal N(S,\sigma_N^2)$ with
$S\coloneqq Y\operatorname{sign}(Z)\sim\mathcal N(0,1)$, so
$\sigma(\pi_{Y,Z}) = \sigma(S)$. The $\pi_Y$-routing therefore corresponds to the path
\[
\gamma = (\{\emptyset,\Omega\}, \{\emptyset,\Omega\}, \sigma(|Y|), \sigma(|Y|), \sigma(S))\,,
\]
giving $I_r = I_{cu} = 0$ for every loss. For the square error,
$\mathbb E[X\,|\,|Y|] = 0$, so $R(\pi_Y) = R(\bot) = 1+\sigma_N^2$ and
\[
I_u = 0, \qquad I_s = \mathbb V(\mathbb E[X|S]) = 1\,,
\]
i.e., the gate is purely synergistic, mirroring the discrete $\mathsf{Xor}$ of
Fig.~\ref{fig:examples1}.

For the log loss, note
$h(X|Y{=}y) = h(N) + C(y^2/\sigma_N^2)$, where $h$ is differential entropy and
\[
C(s) = \log 2 - \mathbb E_{G\sim\mathcal N(0,1)}\Big[\log\big(1 +
e^{-2s-2\sqrt{s}\,G}\big)\Big]\,,
\]
a consequence of $h(\varepsilon y + N) = h(N) + I_S(\varepsilon;\varepsilon y + N)$ for a sign random variable $\varepsilon$ independent of $N$. We obtain
\begin{align*}
I_u(X;Y{\setminus}Z) &=
\tfrac12\log\big(1+\sigma_N^{-2}\big)
- \mathbb E_Y\big[C(Y^2/\sigma_N^2)\big]\,,\\
I_s(X;(Y,Z)) &= \mathbb E_Y\big[C(Y^2/\sigma_N^2)\big]\,,
\end{align*}
summing to $I(X;(Y,Z)) = \tfrac12\log(1+\sigma_N^{-2})$.
\end{example}

\begin{figure}
\centering
\includegraphics[width=.5\textwidth]{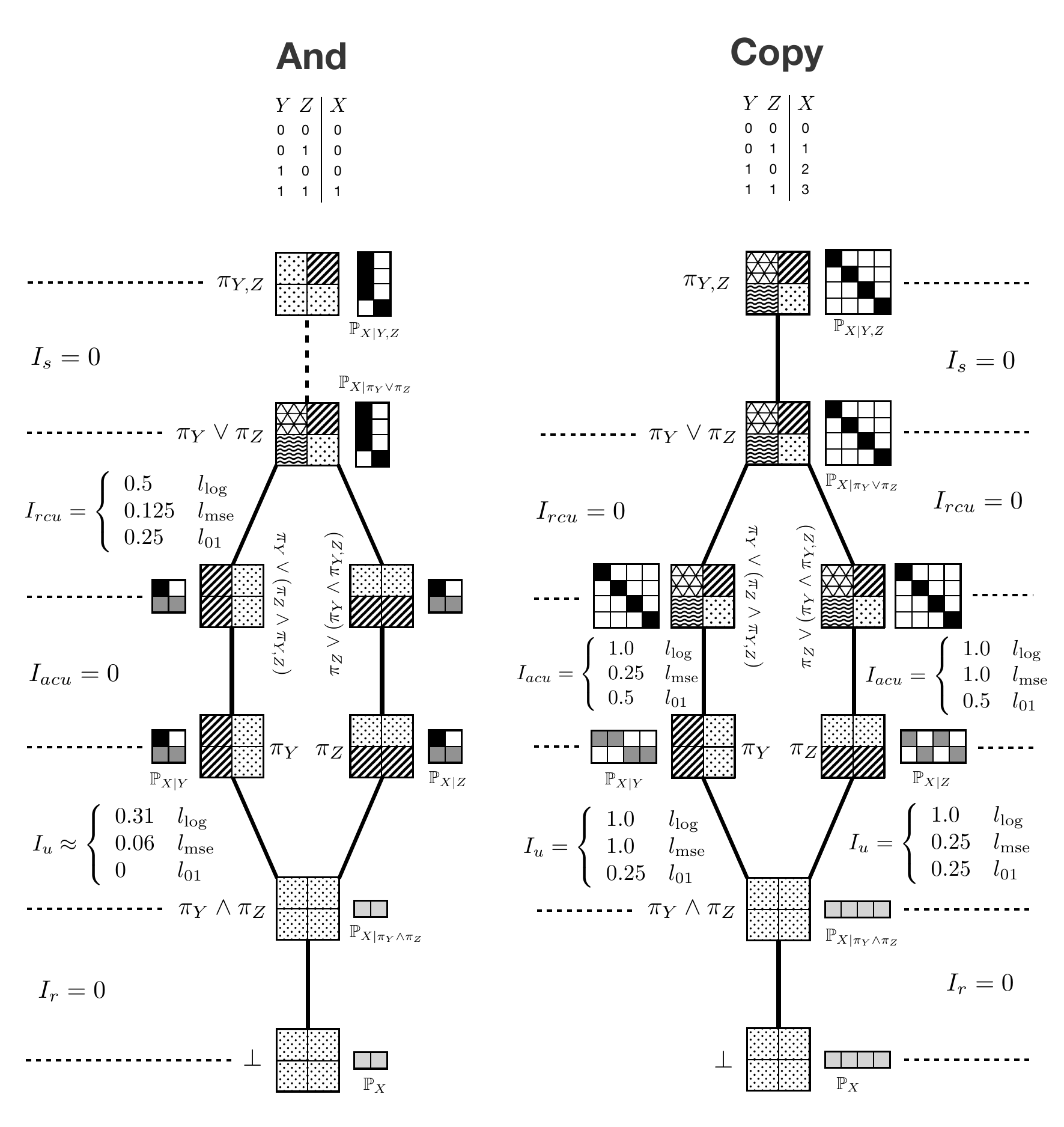}
\caption{Knowledge lattices for $\mathsf{And}$ and $\mathsf{Copy}$ under the finer decomposition of conditional unique information into aligned and residual conditional unique information. For $\mathsf{And}$, we have $I_{acu}=0$ and $I_{rcu} \neq 0$, whereas for $\mathsf{Copy}$, we have $I_{acu}\neq 0$ and $I_{rcu}=0$.}
\label{fig:moreterms}
\end{figure}

\subsection{Finer decompositions} \label{sec:finerdecomp}
Here, we provide two examples of how to refine the path-dependent 4-term PID for two sources from Corollary \ref{cor:PID} further into non-trivial terms. As discussed at the end of Section \ref{sec:PIDours}, including the meet with $\pi_{Y,Z}$ as a separate step results in a decomposition that further disambiguates informational contributions to the total information.

In particular, instead of moving from $\bot$ directly to $\pi_Y\wedge \pi_Z$, we could include the meet with $\pi_{Y,Z}$ as a middle step. More precisely,
\begin{align*}
I_{dr}(X;(Y,Z)) & \coloneqq R(\bot) - R(\pi_Y \wedge \pi_Z \wedge \pi_{Y,Z}), \\
I_{pr}(X;(Y,Z)) & \coloneqq R(\pi_Y \wedge \pi_Z \wedge \pi_{Y,Z}) - R(\pi_Y \wedge \pi_Z),
\end{align*}
where $I_{dr}$ might for example be called \textit{deep redundancy}, capturing the knowledge common in $Y$, $Z$, and $(Y,Z)$, and $I_{pr}$ describes \textit{pairwise redundancy}, capturing the part that is common to $Y$ and $Z$ but not in the joint observation $(Y,Z)$. Then we obtain
\begin{align}\label{eq:refined1}
I_r(X;(Y,Z)) = I_{dr}(X;(Y,Z)) + I_{pr}(X;(Y,Z)),
\end{align}
and both terms are non-negative due to the chain rule.

Similarly, for conditional unique information, instead of moving from, say, $\pi_Y$, directly to $\pi_Y\vee \pi_Z$, we could again include the meet with $\pi_{Y,Z}$ as a middle step. We write
\begin{align*}
I_{acu}(X;Z|Y) & \coloneqq R(\pi_Y) - R(\pi_Y \vee (\pi_Z \wedge \pi_{Y,Z})), \\
I_{rcu}(X;Z|Y) & \coloneqq R(\pi_Y \vee (\pi_Z \wedge \pi_{Y,Z})) - R(\pi_Y \vee \pi_Z),
\end{align*}
for \textit{aligned conditional unique} information $I_{acu}$, capturing the extra gain when adding the common knowledge between $\pi_Z$ and $\pi_{Y,Z}$, and \textit{residual conditional unique} information $I_{rcu}$, capturing the extra knowledge in $\pi_Z$ that is not in $\pi_{Y,Z}$. Then
\begin{equation}
I_{cu}(X;Z|Y) =  I_{acu}(X;Z|Y) + I_{rcu}(X;Z|Y)\, . \label{eq:refined2}
\end{equation}

There are examples that only contain deep redundancy or only pairwise redundancy. For instance, our previous examples $\mathsf{Rdn}$ and $\mathsf{RdnXor}$ only have deep redundancy, whereas one obtains only pairwise redundancy by modifying the latter so that the second $\mathsf{Xor}$ has the values $(0,2,2,0)$ instead of $(2,3,3,2)$.

For the refinement of conditional unique information, we show in Fig.~\ref{fig:moreterms} the knowledge lattice of $\mathsf{And}$, whose PID terms are debated in the literature \cite{Griffith2014,Ay2020}. Our refined PID offers a possible answer: Similarly to $\mathsf{Copy}$, there is no redundant or synergistic information. However, in contrast to $\mathsf{Copy}$, aligned conditional unique information is zero, because there is no common knowledge shared between, say, $\pi_Z$ and $\pi_{Y,Z}$, and so all that is gained when moving from $\pi_Y$ to $\pi_Y\vee \pi_Z$ is residual conditional unique information.

In particular, these examples show that the extra terms in the refined decompositions \eqref{eq:refined1} and \eqref{eq:refined2} are generally non-trivial. Moreover, these are just examples of possible refinements of PIDs. One could also move from $\pi_Y\wedge \pi_Z$ first to $\pi_Y \wedge (\pi_Z \vee \pi_{Y,Z})$ and then to $\pi_Y$, just to name another possible extension. These types of refinements also apply to the case of three sources, as can be seen in the next section.

\begin{figure}
\centering
\includegraphics[width=.48\textwidth]{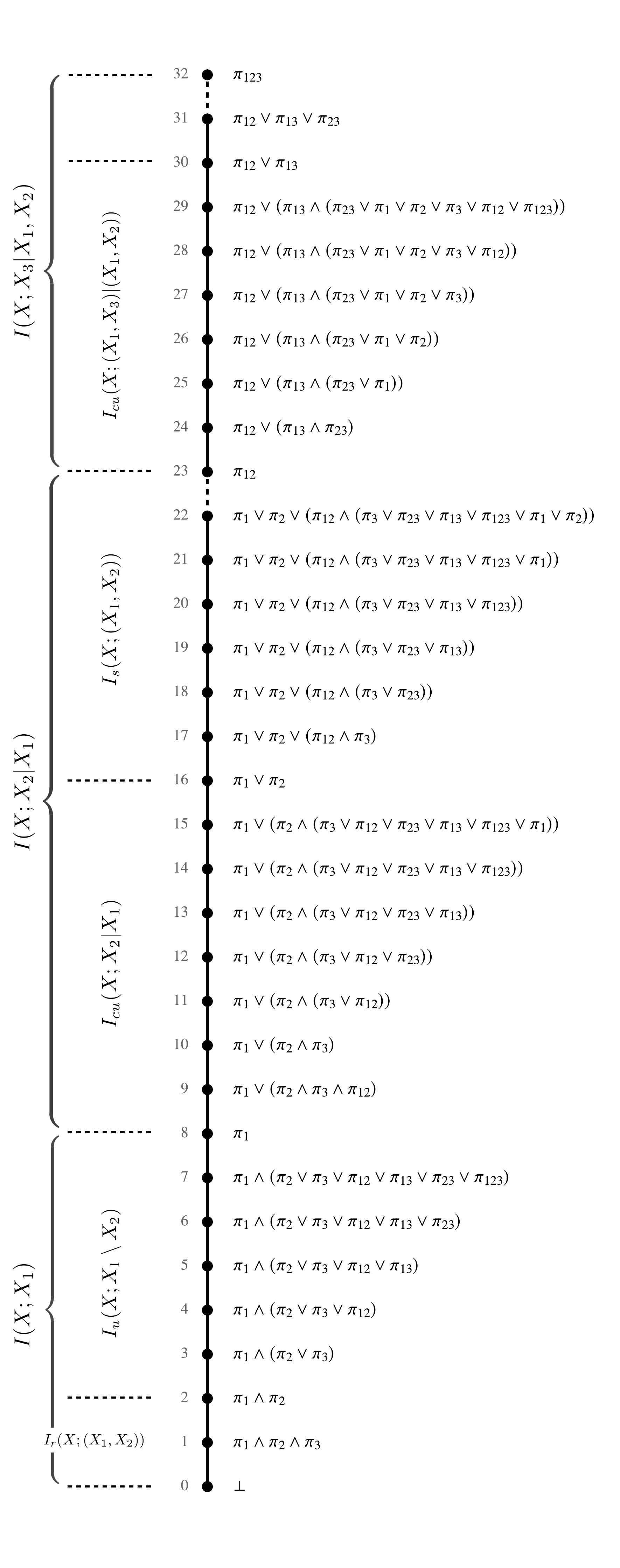}
\vspace{-25pt}
\caption{A possible risk-decreasing path $\gamma = (\bot, \dots, \pi_{123})$, decomposing $I(X;(Y_1,Y_2,Y_3))$ into 32 non-negative terms. We write $\pi_i$ to denote $\pi_{Y_i} = \Pi(\mathbb P_{X|Y_i})$, $\pi_{ij}$ for $\pi_{Y_i,Y_j} = \Pi(\mathbb P_{X|Y_i,Y_j})$, and analogously for more variables. The two synergy steps from node 22 to $\pi_{12}$ and from node 31 to $\pi_{123}$ are the only transitions in this path between incomparable states (indicated by a dashed line). The risk sequence is still non-increasing for any loss, because $R(\pi_1\vee \pi_2 \vee \pi_{12}) = R(\pi_{12})$ by the proof of Lemma \ref{lma:synergygap}, and the final step passes from a type of combined knowledge to joint knowledge as in the two-variable case, so that $R(\pi_{12}\vee \pi_{13}\vee \pi_{23}) \geq R(\pi_{123})$. \label{fig:pid3vars}}
\end{figure}

\begin{figure*}
\centering
\includegraphics[width=\textwidth]{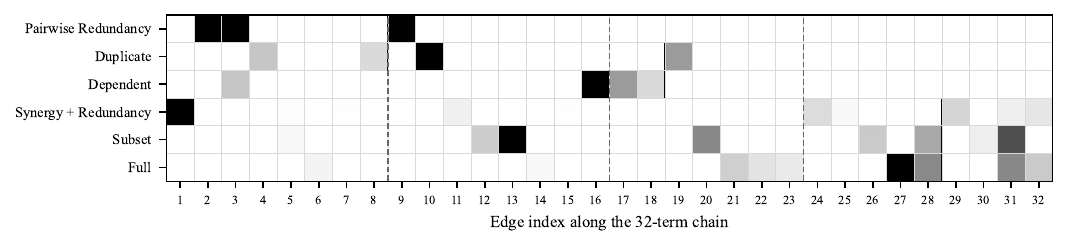}
\caption{Result of the search over the 200,000 examples to find a minimal set of examples covering all 32 terms of the 3-source decomposition of Fig.~\ref{fig:pid3vars}. The grayscale values indicate each term's contribution to the total information. A description of how the six examples were created can be found in Table \ref{tab:pid3_cover_desc}.}
\label{fig:3term-coverage}
\end{figure*}

\begin{table*}[t]
\centering
\small
\caption{Description of how the six examples were sampled, whose non-trivial terms together cover all 32 terms of the decomposition of $I(X;(Y_1,Y_2,Y_3))$ given by the path in Fig.~\ref{fig:pid3vars}. ``Terms'' counts how many of the 32 terms the example makes non-trivial.}
\label{tab:pid3_cover_desc}
\begin{tabular}{l p{0.42\textwidth} c c r}
\toprule
structure & description & $(|Y_1|,|Y_2|,|Y_3|;|X|)$ & $I$ & terms \\
\midrule
Duplicate & $Y_3=Y_2$ (the third source copies the second), and $X$ is then drawn uniformly for each distinct input. & $(3,3,3;4)$  & 1.891 & 4 \\
Pairwise redundancy & $X=(b_1,b_2,b_3)$ with $Y_1=(b_1,b_2)$, $Y_2=(b_2,b_3)$, $Y_3=(b_1,b_3)$, i.e., every pair of sources shares a bit, but none is common to all three. & $(4,4,4;8)$  & 2.985 & 3 \\
Dependent & $Y_3=(Y_1+Y_2)\bmod n_3$ (one source is a deterministic function of the other two), and $X$ is drawn uniformly for each distinct input. & $(3,3,3;4)$  & 1.891 & 4 \\
Full & $X$ is drawn uniformly for every input of the full $n_1\times n_2\times n_3$ grid. & $(3,3,3;3)$  & 1.510 & 9 \\
Synergy $+$ redundancy & A redundant bit copied into all three sources plus a synergistic bit (the XOR of three private bits). & $(4,4,4;4)$  & 1.959 & 10 \\
Subset & $X$ is sampled uniformly on a random subset of the full input grid. & $(3,3,2;2)$ & 0.996 & 8 \\
\bottomrule
\end{tabular}
\end{table*}

\subsection{Partial information decompositions for three sources} \label{app:threesources}

Fig.~\ref{fig:pid3vars} illustrates one possible sequence of partitions of $\Omega$ along a path from $\bot$ to $\pi_{123}$ with decreasing risks, inducing a decomposition of $I(X;(Y_1,Y_2,Y_3))$ through the chain rule in Proposition \ref{prop:properties} $(v)$. The shown risk sequence corresponds to a decomposition with 32 non-trivial terms, which is more fine-grained than the typical assumption of 18 informational terms commonly reported in the PID literature \cite{Griffith2014}. Since, in contrast to current PID formulations, our terms correspond to conditional risk reductions associated with successive transitions between knowledge states, they are guaranteed to be non-negative.

In order to demonstrate that each of the 32 terms is meaningful (i.e., generally non-trivial), we performed a grid search over 200,000 structural variations of a set of examples and found 6 candidates that together show that each term can be non-zero (cf. Fig.~\ref{fig:3term-coverage}). More specifically, we sampled random discrete examples over small alphabets ($|Y_i|\in\{2,3\}$, $|X|\in\{2,3,4\}$ for the randomly drawn targets, whereas the bit-based structured examples have source alphabets of size $4$ and target alphabets of size up to $8$, cf.~Table~\ref{tab:pid3_cover_desc}) drawn from a range of dependence structures: a uniformly random target on the full input grid or on a random subset, structural constraints among the sources (one source duplicating another, one is a deterministic function of the others) with a random target, and structured targets (a shared bit copied into all sources, pairwise-shared bits, nested coarse-grainings, or a redundant bit
combined with a synergistic parity bit). For each example we formed the distribution partitions, evaluated the risk at every node of the chain, and read off each term as the risk reduction $R(\pi)-R(\pi')$ along the corresponding edge (for the log loss).

All $32$ terms are realized as non-zero contributions in
at least one example (the smallest term still reaches $0.34$ bits). We found a minimal cover of six examples, whose non-trivial terms together realize all $32$ terms. These are shown in the coverage map of Fig.~\ref{fig:3term-coverage} and described in
Table~\ref{tab:pid3_cover_desc}.

\bibliographystyle{IEEEtran}

\bibliography{lit_information.bib}

\vfill

\begin{IEEEbiographynophoto}{Sebastian Gottwald}
received his B.Sc. degree in physics from the Ludwig-Maximilians-Universität (LMU) Munich. In 2010, he was selected to enter the elite graduate program Theoretical and Mathematical Physics at the LMU and the Technical University of Munich. From 2010 to 2013 he was a fellow of the Richard-Winter scholarship. He earned his Dr. rer. nat. in mathematics from the LMU Munich in 2016. Currently, he is a Postdoctoral researcher at the Institute of Neural Information Processing at Ulm University, Germany. His current research interests include information theory, abstraction and representation learning in artificial neural networks,  and goal-conditioned learning in dynamic environments.
\end{IEEEbiographynophoto}


\begin{IEEEbiographynophoto}{Daniel A. Braun}
received Diploma degrees in physics and biology from the Albert-Ludwigs-Universität Freiburg, Germany, in 2005, and the Dr. rer. nat. degree in computational neuroscience and the Dr. phil. degree in philosophy of mind from the same university in 2008 and 2011, respectively. He received his habilitation in neural and behavioral biology (2014) and cognitive science (2015) from the Eberhard Karls Universität Tübingen, Germany. From 2006 to 2010, he was with the Computational and Biological Learning Laboratory, University of Cambridge, U.K., and from 2010 to 2011 with the Computational Learning and Motor Control Laboratory, University of Southern California, Los Angeles, CA, USA. In 2011, he received an Emmy Noether Fellowship and established the research group "Sensorimotor Learning and Decision-Making" at the Max Planck Institutes for Biological Cybernetics and for Intelligent Systems, Tübingen. He was awarded an ERC Starting Grant in 2015. Since 2016, he has been a Professor of Learning Systems with Ulm University, Germany. His research interests include cognitive modeling, information-theoretic models of learning and decision-making, abstraction and structural learning, learning robots, and open-ended learning.
\end{IEEEbiographynophoto}

\end{document}